\documentclass{aastex631}
\usepackage{graphicx}

\shorttitle{Calibration of Star Formation Rates in NGC\,628}
\shortauthors{Calzetti et al.}

\graphicspath{{./}{figures/}}

\begin{document}

\title{Feedback in Emerging Extragalactic Star Clusters (JWST--FEAST): Calibration of Star Formation Rates in the Mid--Infrared with NGC\,628}

\author[0000-0002-5189-8004]{Daniela Calzetti}
\affiliation{Department of Astronomy, University of Massachusetts Amherst, 710 North Pleasant Street, Amherst, MA 01003, USA}

\author[0000-0002-8192-8091]{Angela Adamo}
\affiliation{Department of Astronomy, The Oskar Klein Centre, Stockholm University, AlbaNova, SE-10691 Stockholm, Sweden}

\author[0000-0002-1000-6081]{Sean T. Linden}
\affiliation{Department of Astronomy and Steward Observatory, University of Arizona, Tucson, AZ 85721, USA}

\author[0000-0003-4910-8939]{Benjamin Gregg}
\affiliation{Department of Astronomy, 
University of Massachusetts, 710 North Pleasant Street, Amherst, MA 01003, USA}

\author[0000-0003-3893-854X]{Mark R. Krumholz}
\affiliation{Research School of Astronomy and Astrophysics, Australian National University, Camberra, Australia}

\author[0009-0008-4009-3391]{Varun Bajaj}
\affiliation{Space Telescope Science Institute, 3700 San Martin Drive Baltimore, MD 21218, USA}

\author[0000-0001-8068-0891]{Arjan Bik}
\affiliation{Department of Astronomy, The Oskar Klein Centre, Stockholm University, AlbaNova, SE-10691 Stockholm, Sweden}

\author[0000-0001-6291-6813]{Michele Cignoni}
\affiliation{Department of Physics - University of Pisa, Largo B. Pontecorvo 3, 56127 Pisa, Italy }
\affiliation{INFN, Largo B. Pontecorvo 3, 56127 Pisa, Italy }
\affiliation{INAF - Osservatorio di Astrofisica e Scienza dello Spazio di Bologna, Via Gobetti 93/3, I-40129 Bologna, Italy }

\author[0000-0001-6464-3257]{Matteo Correnti}
\affiliation{INAF Osservatorio Astronomico di Roma, Via Frascati 33, 00078, Monteporzio Catone, Rome, Italy}
\affiliation{ASI-Space Science Data Center, Via del Politecnico, I-00133, Rome, Italy}

\author[0000-0002-1723-6330]{Bruce Elmegreen}
\affiliation{Katonah, NY 10598 USA}

\author[0000-0002-2199-0977]{Helena Faustino Vieira}
\affiliation{Cardiff Hub for Astrophysics Research and Technology (CHART), School of Physics \& Astronomy, Cardiff University, The Parade, Cardiff CF24 3AA, UK}

\author[0000-0001-8608-0408]{John S. Gallagher}
\affiliation{Department of Physics and Astronomy, Macalester University, 1600 Grand Avenue, Saint Paul, MN 55105-1899 USA}

\author[0000-0002-3247-5321]{Kathryn~Grasha}
\altaffiliation{ARC DECRA Fellow}
\affiliation{Research School of Astronomy and Astrophysics, Australian National University, Canberra, ACT 2611, Australia}   
\affiliation{ARC Centre of Excellence for All Sky Astrophysics in 3 Dimensions (ASTRO 3D), Australia}   

\author[0000-0002-6447-899X]{Robert A. Gutermuth}
\affiliation{Department of Astronomy,  University of Massachusetts, 710 North Pleasant Street, Amherst, MA 01003, USA}

\author[0000-0001-8348-2671]{Kelsey E. Johnson}
\affiliation{Department of Astronomy, University of Virginia, Charlottesville, VA, USA}

\author[0000-0003-1427-2456]{Matteo Messa}
\affiliation{INAF - Osservatorio di Astrofisica e Scienza dello Spazio di Bologna, Via Gobetti 93/3, I-40129 Bologna, Italy }

\author[0000-0003-0470-8754]{Jens Melinder}
\affiliation{Department of Astronomy, The Oskar Klein Centre, Stockholm University, AlbaNova, SE-10691 Stockholm, Sweden}

\author[0000-0002-3005-1349]{G\"{o}ran \"{Os}tlin}
\affiliation{Department of Astronomy, The Oskar Klein Centre, Stockholm University, AlbaNova, SE-10691 Stockholm, Sweden}

\author[0000-0002-8222-8986]{Alex Pedrini}
\affiliation{Department of Astronomy, The Oskar Klein Centre, Stockholm University, AlbaNova, SE-10691 Stockholm, Sweden}

\author[0000-0003-2954-7643]{Elena Sabbi}
\affiliation{Space Telescope  Science Institute, 3700 San Martin Drive, Baltimore, MD 21218, USA}

\author[0000-0002-0806-168X]{Linda J. Smith}
\affiliation{Space Telescope  Science Institute, 3700 San Martin Drive, Baltimore, MD 21218, USA}

\author[0000-0002-0986-4759]{Monica Tosi}
\affiliation{INAF - Osservatorio di Astrofisica e Scienza dello Spazio di Bologna, Via Gobetti 93/3, I-40129 Bologna, Italy }

\begin{abstract}
New JWST near--infrared imaging of the nearby galaxy NGC\,628 from the Cycle 1 program JWST--FEAST is combined with archival JWST mid--infrared imaging to calibrate the 21~$\mu$m emission as a star formation rate indicator (SFR) at $\sim$120~pc scales. The Pa$\alpha$ ($\lambda$1.8756~$\mu$m) hydrogen recombination emission line targeted by FEAST provides a reference SFR indicator that is relatively insensitive to dust attenuation, as demonstrated by combining this tracer with the HST H$\alpha$ imaging. Our analysis is restricted to regions that appear compact in nebular line emission and are sufficiently bright to mitigate effects of both age and stochastic sampling of the stellar initial mass function. We find that the 21~$\mu$m emission closely correlates with the nebular line emission, with a power--law with exponent=1.07$\pm$0.01, in agreement with past results. We calibrate a hybrid SFR indicator using a combination of H$\alpha$ and 24~$\mu$m (extrapolated from 21~$\mu$m) tracers and derive the proportionality constant between the two tracers $b=0.095\pm0.007$, which is $\sim$ 3--5 times larger than previous derivations using large regions/entire galaxies. We model these discrepancies as an increasing contribution to the dust heating by progressively older stellar populations for increasing spatial scales, in agreement with earlier findings that star formation is hierarchically distributed in galaxies. Thus, use of hybrid SFR indicators requires prior knowledge of the mean age of the stellar populations dominating the dust heating, which makes their application uncertain. Conversely, non--linear calibrations of SFRs from L(24) alone are more robust, with a factor $\lesssim$2.5 variation across the entire range of L(24) luminosities from HII regions to galaxies.
 \end{abstract}


\section{Introduction} \label{sec:intro}
The star formation rate (SFR) is one of the two essential parameters, the other being mass, for tracing the growth of galaxies across cosmic times. The significance of this parameter has spurred vast literature dedicated to calibrating SFR indicators across the full energy spectrum, from the X--ray to radio wavelengths, for use in galaxies and in regions within galaxies \citep[][for a review]{KennicuttEvans2012}. 

Operationally, a SFR is a luminosity, SFR$\propto$ C\, L($\lambda$), calibrated using models of stellar populations or dust emission depending on the wavelength of interest  \citep[e.g.,][]{KennicuttARAA1998, Calzetti2013, Figueira+2022}. The main hurdle is to ensure that the luminosities used in the calibration are dominated by emission from sources with the timescales of 
interest, $\approx$100~Myr for galaxies and somewhat shorter for regions within galaxies. Massive ionizing stars, typically O--type stars with lifetimes shorter than 10~Myr, offer convenient tracers of the timescales of interest for SFRs. However, outside of the light most obviously associated with O--type stars (stellar ultraviolet continuum, UV, and the nebular line and continuum emission from the gas they ionize), most other wavelengths receive contribution from stellar populations that live longer than $\approx$10--100~Myr.  One of the difficulties in deriving reliable SFR indicators is controlling for this contribution.

This problem becomes more acute when leveraging dust emission at infrared wavelengths to measure SFRs, as there is no direct one--to--one mapping between the UV and optical stellar photons that heat the dust and the dust emission in the infrared ($\gtrsim$3~$\mu$m). Numerous authors have recognized and analyzed this problem, as the infrared wavelength range provides a window into the dust--obscured SFR that does not emerge at shorter wavelengths. \citet{Cox+1986}, using infrared surveys of the Milky Way, and \citet{Helou1986}, using IRAS photometry of `normal' (not dominated by AGNs) galaxies, concluded that the dust emission includes contribution from both a  `warm' component, heated by star forming regions, and a `cool' component, also termed `cirrus' and heated by the diffuse interstellar radiation field of the galaxies. \citet{LonsdaleHelou1987}, using 40--120~$\mu$m photometry of nearby galaxy disks from IRAS data, quantified the two contributions: more than 50\% of the infrared emission from galaxies is heated by the non--ionizing stars of the diffuse interstellar radiation field. Subsequent studies reached similar conclusions, finding a dependency of warm--to-cool infrared emission ratio on the star forming--to--diffuse radiation ratio  \citep[e.g.][]{Buat+1988, RowanRobinson+1989, Sauvage+1992, Walterbos+1996, Buat+1996}. More recent observations with the {\em Spitzer Space Telescope} and the {\em Herschel Space Observatory} have further confirmed that warmer dust is associated with higher SFRs in galaxies \citep[e.g.][]{Calzetti+2010, Bendo+2012, Dale+2012, SmithDunne+2012, Magnelli+2014, Gregg+2022}; in the local universe, the infrared emission traces the recent star formation mainly in galaxies where the ratio of the unobscured to obscured SFR, SFR$_{unobs}$/SFR$_{obsc}\lesssim$1/10 \citep[e.g.][]{Rieke+2009}. 

Use of the mid--IR ($\sim$3--40~$\mu$m)  emission to trace SFRs presents more challenges than the longer wavelength emission, since the dust emission in this wavelength range comprises several contributions: non--equilibrium emission composed of continuum emission from stochastically heated small grains and emission features from radiatively excited vibrational and bending modes of Polycyclic Aromatic Hydrocarbons (PAHs), as well as thermal equilibrium emission from hot dust \citep{Greenberg1968,Sellgren+1983,Leger+1984,Desert+1990,Draine+2001,DraineLi2007, Smith+2007, Galliano+2018}. Non--equilibrium emission is heated and excited by non--ionizing UV and optical photons from both the diffuse, evolved stellar population and the recent star formation in a galaxy \citep[e.g.,][]{Draine+2007,Galliano+2018,Draine+2021}, implying that the reliability of mid--IR SFR indicators depends on the relative strength of the heating from these two stellar population components \citep{Boselli+2004,Calzetti+2005,Boquien+2016}. In local galaxies, between 30\% and 80\% of the dust emission at $\sim$8~$\mu$m, which is dominated by PAH features, is due to heating and excitation by evolved stellar populations that are unrelated to recent star formation \citep{Bendo+2008,Crocker+2013,Calapa+2014}. Dust emission becomes progressively more dominated by grains in thermal equilibrium beyond $\sim$20~$\mu$m \citep{Galliano+2018}, which has helped establish SFR indicators at these longer mid--IR wavelengths, including the 24~$\mu$m band of the MIPS instrument on the 
{\em Spitzer Space Telescope} \citep{Rieke+2004}. SFR(24) calibrations have been published by many authors, both for galaxies \citep[e.g.][]{Wu+2005,Zhu+2008,Rieke+2009,Kennicutt+2009} and for regions within galaxies \citep[e.g.][]{Perez+2006,Alonso+2006,Calzetti+2007,Kennicutt+2007,Relano+2007}. Limitations in the angular resolution of pre--JWST space facilities have however constrained analyses to $\sim$0.5--1~kpc regions or larger scales, with the only exception of Local Group Galaxies \citep[e.g.][]{Boquien+2015}. Despite limitations, some of these results have established that the dust emission at 24~$\mu$m is partially due to heating by the diffuse (non--star--forming) stellar population in the galaxy, and a significant portion is from dust thermally heated by recently formed stars \citep{Calzetti+2005, Liu+2011,Leroy+2012, Boquien+2016}. Recent observations with the  {\em James Webb Space Telescope} of nearby galaxies confirm this general picture, assigning about 50\% of the mid--IR emission to dust heated by the diffuse interstellar radiation field \citep{Leroy+2023}. 

The JWST is offering unprecedented resolution at near and mid--IR wavelengths, thus permitting the analysis of mid--IR SFR indicators at the scales of HII regions, i.e., of the most elemental units of star formation. Our main goal is to leverage the high spatial resolution afforded by the JWST to quantify stellar population effects on the dust emission in the mid---IR and their impact on the calibration of dust--emission--based SFR indicators. In this paper, we analyze the Mid-InfraRed Instrument \citep[MIRI, ][]{Rieke+2015}  21~$\mu$m dust emission from the nearby star--forming spiral NGC\,628  at $\sim$100~pc spatial scale, which we combine with the recent observations in the light of the hydrogen recombination line Pa$\alpha$ obtained by the JWST--FEAST program (Adamo et al. 2024, in prep.). Archival {\em Hubble Space Telescope} (HST) observations provide narrow--band imaging centered at the galaxy's H$\alpha$ nebular emission, which we use to correct the Pa$\alpha$ emission line for the effects of dust attenuation. However, the use of the infrared nebular line already provides robustness to the analysis, since the effects of dust attenuation are much reduced at $\approx$2~$\mu$m: a dust column density corresponding to A$_V$=3~mag reduces the Pa$\alpha$ line emission only by a factor 1.8, while it depresses the H$\alpha$ emission by a factor 9.5. Thus, we will adopt the nebular Pa$\alpha$ line emission as our reference SFR indicator.  

Recently, \citet{Belfiore+2023} published a comparison between several JWST mid--IR bands, including 21~$\mu$m, and the dust attenuation--corrected H$\alpha$ luminosity, L(H$\alpha_{corr}$), for a sample of $\sim$20,000 HII regions in 19~galaxies closer than $\sim$20~Mpc. These authors derive SFR calibrations for the dust emission captured by the JWST bands at the $\sim$100~pc scales of HII regions; they find similarities but also differences relative to earlier results, which will be discussed later in this paper. One advantage of the \citet{Belfiore+2023}'s analysis is the sheer number of HII regions considered, which provides statistical strength to their results. Our analysis, although concentrated on a single galaxy, differs in several aspects from the study of  \citet{Belfiore+2023}: we include a nebular emission line long--ward of H$\alpha$, which helps minimize dust attenuation in the hydrogen emission; we remove the diffuse emission from the galaxy in all tracers, thus mitigating contributions from the diffuse (non--star--forming) stellar populations; finally, we limit our analysis to HII regions which are bright enough that the effects of stochastic sampling of the stellar initial mass function (IMF) should be minimized. 

NGC\,628 is located at a distance between 8.6~Mpc and 10.2~Mpc, as determined from the Tip of the Red Giant Branch \citep{Jang+2014,McQuinn+2017,Sabbi+2018,Anand+2021}; we adopt 9.3~Mpc in this paper. The relative proximity, the abundance of HII regions, which yields a global SFR$\sim$3.2~M$_{\odot}$~yr$^{-1}$ \citep{Calzetti+2015a}, and the low inclination \citep[$\sim$9$^o$ from CO and  $\sim$25$^o$ from stellar isophotes,][]{Lang+2020,deVaucouleurs+1991} make this galaxy an excellent study case for SFR calibrations at small spatial scales. The oxygen abundance is about solar in the center\footnote{We adopt a solar oxygen abundance of 12+Log(O/H)=8.69, \citet{Asplund+2009}.} and has a modest gradient \citep{Berg+2020}; the edge of the common region among the JWST mosaics is about 110$^{\prime\prime}$ ($\sim$4.96~kpc) from the galaxy's center, implying that the oxygen abundance has only decreased by 0.14~dex and that metallicity variations will not enter as a parameter in our analysis. 

This paper is organized as follows: Section~\ref{sec:data} presents the data used in this analysis, Section~\ref{sec:selection} describes the source identification and photometry, Section~\ref{sec:results} presents the main results, and Section~\ref{sec:discussion} discusses them, also in comparison with previous results. Conclusions and recommendations on the best mid--IR SFR indicators to use are in Section~\ref{sec:conclusions}.

\section{Imaging Data and Processing} \label{sec:data}

The galaxy NGC\,628 is one of six targets of the Cycle 1 JWST program \# 1783 (Feedback in Emerging extrAgalactic Star clusTers, JWST--FEAST, P.I.: A. Adamo), which is obtaining NIRCam \citep{Rieke+2005} and MIRI  \citep{Rieke+2015} mosaics in 10 bands, covering the wavelength range 1.1--8~$\mu$m. More details on the survey selection and data processing and mosaicing can be found in Adamo et al. (2024, in prep.). For reference, mosaics used in this analysis have been processed through the JWST pipeline version 1.11.4, released at the end of August 2023. At the distance of NGC\,628, the mosaics' coverage subtends  $\sim$2.3$^{\prime}\times$6$^{\prime}$, or 6.2$\times$16.2~kpc$^2$, from the NW to the SE, with the center of the galaxy placed at the center of the mosaics. For this analysis, we employ the Short Wavelength NIRCam mosaics in the light of the Pa$\alpha$  line emission ($\lambda$=1.8797~$\mu$m at the fiducial redshift z=0.00219\footnote{From NED, the NASA Extragalactic Database.}), using the narrow--band F187N filter, and the two adjacent continuum filters, F150W and F200W, for stellar background subtraction. We also employ the Long Wavelength NIRCam mosaic in the F444W filter to verify the amount of stellar light contamination at 21~$\mu$m.  The Full Width at Half Maximum (FWHM) of the Point Spread Function (PSF) for the Short Wavelength mosaics, 0.066$^{\prime\prime}$, subtends 3~pc at the distance of NGC\,628, consistent with the size of an individual star cluster \citep{Ryon+2015,Ryon+2017,Brown+2021}. Exposure time varies along the mosaics, ranging from about 1,000~s to 2,000~s for the narrow--band filter and from 200~s to 1,000~s for the broad band filters. The flux calibration uses the most recent updates, ``jwst\_1126.pmap", from September 2023\footnote{https://jwst-docs.stsci.edu/jwst-near-infrared-camera/nircam-performance/nircam-absolute-flux-calibration-and-zeropoints}, which the FEAST team converts from MJy/sr to Jy/pixel, where each pixel is 0.04$^{\prime\prime}$. 

The new NIRCam observations described above are joined by archival sets from both JWST and HST, retrieved from MAST\footnote{MAST: Mikulski Archive for Space Telescopes at the Space Telescope Science Institute; https://archive.stsci.edu/.} already processed through their respective pipelines. The MIRI F2100W mosaic, centered at 20.563~$\mu$m, is from program \# 2107 (PHANGS--JWST, P.I.: J. Lee), and covers a smaller region than the FEAST mosaics, $\sim$2.2$^{\prime}\times$3.7$^{\prime}$, but with the same orientation and centering, thus maximizing overlap. Pixel scale for the MIRI mosaic is 0.11$^{\prime\prime}$/pix, exposure time is about 300~s, and the MIRI/F2100W PSF FWHM=0.674$^{\prime\prime}$. The mosaic's flux units are MJy/sr. Although the observations were obtained in July 2022, the archival image was re--processed with the more recent pipeline version 1.11.4\footnote{Tests performed on the F2100W mosaic recently released by the PHANGS--JWST collaboration \citep{Williams+2024}, shows that the difference in flux calibration is negligible relative to the MAST mosaic. The released PHANGS mosaic was processed through the pipeline version 1.12.3, which is the most recent available at the time of this writing. We retain the F2100W mosaic processed with the pipeline version 1.11.4 for this analysis for uniformity with the available FEAST mosaics.}. The MAST MIRI mosaic is not background--subtracted, but this does not affect our analysis, which adopts a local background subtraction (Section~\ref{sec:selection}). Archival HST imaging is from the Wide Field of the Advanced Camera for Surveys (ACS/WFC), centered in the light of the H$\alpha$+[NII] doublet  line emission ($\lambda$=0.6562,0.6577,0.6598~$\mu$m at z=0.00219), in the narrow--band F658N, and in the two adjacent broad--band filters F555W and F814W. Field--of--View (FoV) coverage is limited by the F658N images, which only comprise two pointings. The ACS pointings have been mosaicked by the FEAST team and resampled to a pixel scale of 0.04$^{\prime\prime}$/pix. Exposure times in the ACS filters ranges from $\sim$3,000~s to 11,000~s, and the PSF FWHM$\sim$0.07$^{\prime\prime}$. The ACS images are in units of e$^-$/s, and calibration to physical flux is performed by applying the image header keyword PHOTFLAM. Table~\ref{tab:images} lists the telescope/instrument/filters of the mosaics used in this paper, and the programs those data originate from. Although none of the images and mosaics reaches the outskirts of the galaxy, our localized analysis of line-- and dust--emitting regions will not be affected by this limitation.

Emission line mosaics are used to identify HII regions and are obtained by subtracting the stellar continuum from the narrow--band images. The stellar continuum for the F187N (Pa$\alpha$) mosaic is  obtained from the interpolation between the F150W and the F200W. Since the F200W also contains the Pa$\alpha$ line emission, the subtraction is performed iteratively, using the approach described in \citet{Messa+2021}; three iterations are sufficient for convergence to the final continuum--subtracted image, with a final image that differs from the previous iteration by 0.1\% in flux. The stellar continuum for the F658N (H$\alpha+$[NII]) image is derived from the interpolation between the F555W and the F814W. The F555W filter contains the [OIII]($\lambda$ 0.5007~$\mu$m) emission line, which is however weak in metal--rich systems including NGC\,628 \citep[e.g.,][]{Berg+2015}. A small portion of the F555W FoV is also covered by one pointing of the Wide Field Camera 3 in the line--free filter F547M; we use this pointing to measure the impact of the [OIII] emission on  both the F555W and the interpolated stellar continuum for the F658N filter in NGC\,628, verifying that they are at the level of $\lesssim$4\% and $\lesssim$1.5\%, respectively, and will, therefore, not affect our analysis. The optical emission line is further corrected for the [NII] contribution, using the value of [NII]/H$\alpha$=0.4 for the sum of the two [NII] components from \citet{Kennicutt+2008}. Both continuum--subtracted images are then multiplied by the respective filter widths (0.00875~$\mu$m for F658N and 0.024~$\mu$m for F187N\footnote{https://etc.stsci.edu/etcstatic/users\_guide/appendix\_b\_acs.html; https://jwst-docs.stsci.edu/jwst-near-infrared-camera/nircam-instrumentation/nircam-filters.}) and corrected for the filter transmission curve at the galaxy's redshift (z=0.00219), in order to derive the final line flux maps in the light of H$\alpha$ and Pa$\alpha$.

\begin{deluxetable*}{llll}
\tablecaption{Imaging Data Sources\label{tab:images}}
\tablewidth{0pt}
\tablehead{
\colhead{Telescope$^1$} & \colhead{Instrument$^2$} & \colhead{Filters$^3$} & \colhead{Proposal ID$^4$}\\
}
\decimalcolnumbers
\startdata
JWST & NIRCam S+L   &  F150W, F187N, F200W, F444W  & 1783\\
JWST & MIRI   &  F2100W  & 2107 \\
HST  & ACS/WFC   &  F555W, F658N, F814W  & 9796, 10402\\
\enddata
$^1$  JWST=James Webb Space Telescope; HST=Hubble Space Telescope\\
$^2$ NIRCam S+L = Near Infrared Camera, Short and Long Wavelength Channels \citep{Rieke+2005}; MIRI=Mid--Infrared Instrument \citep{Rieke+2015}; ACS/WFC= Advanced Camera for Surveys Wide Field Channel \citep{Sirianni+2005}.\\
$^3$ Filter names.\\
$^4$ Identification of the GO program that obtained the images: JWST/GO--1783 (JWST--FEAST), PI: Adamo; JWST/GO--2107, PI: Lee; HST/GO--9796, PI: Miller; HST/GO--10402, PI: Chandar.\\
\end{deluxetable*}

\section{Source Selection, Photometry, and Derived Quantities} \label{sec:selection}

\begin{figure}
\plotone{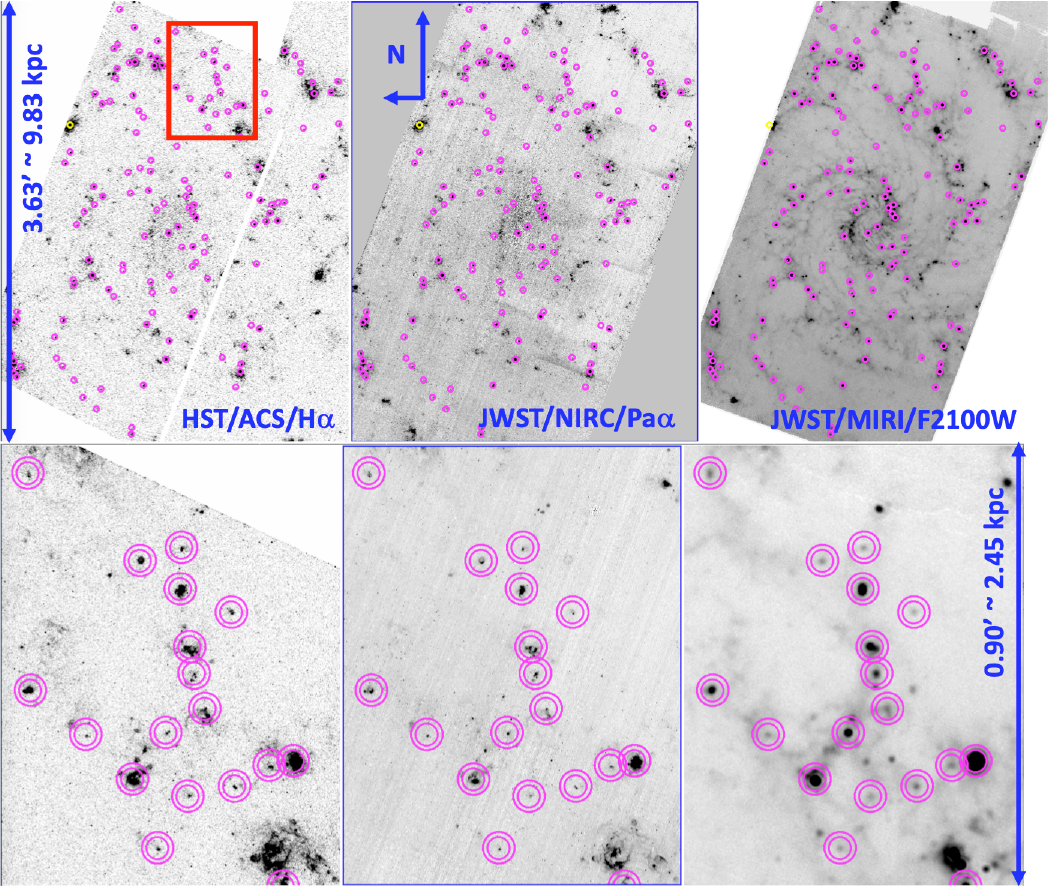}
\caption{{\bf (Top Row:)}: The 143 sources emitting in H$\alpha$, Pa$\alpha$ and 21~$\mu$m are identified with magenta circles on the JWST and HST mosaics in the stellar--continuum subtracted HST/ACS/H$\alpha$ (left), JWST/NIRCam/Pa$\alpha$ (center), and JWST/MIRI/F2100W (right), see Section~\ref{sec:data}. The yellow circle identifies the brightest source in H$\alpha$ and Pa$\alpha$, which is however outside the F2100W footprint. The radius of the circles matches the photometric aperture used in this study, 1.4$^{\prime\prime}$, or $\sim$63~pc.  North  is up, East is left. {\bf (Bottom Row:)} A detail of the images above, in the same order (H$\alpha$, Pa$\alpha$, 21~$\mu$m).  The double magenta circle around each region shows the size of the annulus used for the local background subtraction (0.6$^{\prime\prime}$). The location of this region is drawn on the top--row H$\alpha$ image with a red rectangle. }
 \label{fig:sources}
\end{figure}

Sources are selected by visual inspection, with the following characteristics: (1) they are located within the common FoV of the H$\alpha$, Pa$\alpha$ and 21~$\mu$m mosaics; (2) they are local peaks of emission in {\em both} Pa$\alpha$ and 21~$\mu$m, with the two peaks located within $\sim$0.7$^{\prime\prime}$ ($\sim$1 F2100W PSF FWHM) of each other; (3) they are detected with S/N$\gtrsim$5 at 21~$\mu$m and S/N$\gtrsim$3 in both Pa$\alpha$ and H$\alpha$; (4) they appear morphologically compact in both 21~$\mu$m and Pa$\alpha$; (5) photometric apertures (defined below) of adjacent sources do not overlap more than $\sim$10\% of the aperture's area; (6) the sky annulus used for local background subtraction only contains at most a portion ($<$10\%) of one other source. These criteria are imposed to ensure the following: (1) the sources are emitting both in dust and ionized gas and are, therefore, likely to be HII regions powered by young clusters; (2) we can derive dust attenuation corrections for the nebular lines; (3) we only have one major contributing source to both the line and dust emission within each aperture; (4) the HII regions are sufficiently young that the gas is still coincident with the stellar source, thus mitigating effects of aging \citep{Whitmore+2011}; (5) confusion and cross--contamination in the photometry is minimized, and (6) the level of the local background can be reliably measured. Although only one 21~$\mu$m peak can be present within each aperture, their shape can be round or elongated. Excluded regions of emission generally consist of open (shell--like) HII regions and 21~$\mu$m sources without emission line counterparts (likely red stars: RSGs or AGBs). 
With these criteria, we isolate 143 sources in common between the H$\alpha$, Pa$\alpha$, and 21~$\mu$m mosaics (Figure~\ref{fig:sources}).

The goal of our selection is not to be complete, but to capture all bright ($>$5~$\sigma$) 21~$\mu$m sources in the common FoV among the three mosaics in H$\alpha$, Pa$\alpha$ and 21~$\mu$m. 
We elect to accept line emission sources of any brightness, as faint as 3~$\sigma$, to ensure we include dusty HII regions, which are expected to be faint in the optical and near--IR bands. We separately search for any additional source that may be bright in H$\alpha$, albeit fainter than the set limits in the other two tracers (bright, dust--free sources), finding none. We are missing large numbers of faint sources in all three Pa$\alpha$, H$\alpha$ and 21~$\mu$m; however, as will be discussed in the next section, we will perform a luminosity cut at the faint end to avoid sources that may be affected by stochastic (random) sampling of the stellar IMF. 

Photometry is performed in all filters (F555W, F658N, F814W, F150W, F187N, F200W, F444W, F2100W), using circular apertures with radius 1.4$^{\prime\prime}$ which corresponds to about 63~pc at the distance of the galaxy, and centered on the centroid of the 21~$\mu$m sources. The apertures remain roughly circular also in the plane of the galaxy, given its low inclination. The size of the photometric aperture is chosen to accommodate the mosaics with the broadest PSF, which is the MIRI/F2100W; a radius of 1.4$^{\prime\prime}$  corresponds to 5~$\sigma$ if the PSF of this filter is well represented by a gaussian with FWHM=0.674$^{\prime\prime}$. Growth curves measured on three of the most isolated sources in the FoV of the 21~$\mu$m mosaic indicate that aperture corrections are about 20\%, which we apply to our photometry. Aperture corrections are negligible for all other filters, due to their much narrower PSF FWHMs (0.145$^{\prime\prime}$ for F444W and $\lesssim$0.07$^{\prime\prime}$ for all other filters). 

We elect to remove the local background from our measurements because the diffuse backgrounds in the mid--infrared and in the nebular lines have physically different origins. As discussed in the Introduction, the diffuse emission at 21~$\mu$m is due to the heating of dust by the UV and optical light of the general stellar population in the galaxy, while the diffuse emission in the recombination lines is mostly due to photon leakage from neighboring HII regions as far as 1~kpc from the sources, with a lower--level contribution of scattered light from recombination lines formed elsewhere \citep[e.g.,][]{Reynolds+1984, Reynolds+1990, Ferguson+1996, Hoopes+1996, Hoopes+2003, Voges+2006, Oey+2007, Zhang+2017}. Local background removal also removes the underlying stellar absorption from the nebular line emission, if the composition of the stellar population remains roughly uniform across the area of interest for the photometric measurement. Underlying stellar absorption can be as large as a few \AA, depending on the age of the stellar population \citep{McCall+1985, Calzetti+1997, RosaGonzalez+2002, Hopkins+2003, Li+2005}, which would affect the derived emission line intensity for the fainter regions if not removed. For background removal, we use an annulus of 0.6$^{\prime\prime}$ width around each photometric aperture, to ensure that the background area matches the aperture area in order to mitigate the background noise contribution to the photometric measurements. Thus, the total size of the region used for photometry+background removal is 2$^{\prime\prime}$=90~pc  in radius. A test of the impact of the background is shown in Appendix~\ref{sec:appendixA}.

Nebular emission line fluxes are derived from the photometric measurements using the methodology described in Section~\ref{sec:data}. Stellar continuum is subtracted from the 21~$\mu$m flux density by rescaling the measurements in F444W using a recipe similar to the one by \citet{Helou+2004} and \citet{Calzetti+2007}, adapted to our case: f(21)$_{dust}$=f(21)-0.046 f(444), where the flux densities are in units of Jy. Here we assume the emission in the F444W to be dominated by stellar continuum, which is extrapolated to 21~$\mu$m adopting a Rayleigh-Jeans functional shape for the stellar spectrum. We neglect the small dust attenuation correction for the stellar emission  and the hot dust emission in the F444W filter. As expected, stellar contribution to the 21~$\mu$m emission is at the level of a few \%. The flux density at 21~$\mu$m is then multiplied by the frequency ($\nu$=1.458E+13~Hz) to convert it to flux per octave, which is the standard used in SFR measurements. Fluxes are converted to luminosities, in units of erg~s$^{-1}$, using the adopted galaxy's distance. In addition, we derive the equivalent width (EW) of the Pa$\alpha$ emission, expressed as the ratio of the line luminosity to the luminosity density of the interpolated stellar continuum. The 3~$\sigma$ detection limits for Log[L(H$\alpha$)] and Log[L(Pa$\alpha$)] are 36.14 and 35.91, respectively (after correction for the foreground MW extinction, see below), and the 5~$\sigma$ limit for Log[L(21)] is 37.70, with all luminosities in units of erg~s$^{-1}$.

The color excess E(B$-$V) associated with the dust attenuation is derived from the observed line ratio H$\alpha$/Pa$\alpha$, adopting Case B recombination and an intrinsic ratio of 7.82, which is appropriate for metal rich sources \citep[T$_e$=7,000~K and n$_e$=100~cm$^{-3}$,][]{Osterbrock+2006}. For the extinction curve, we use the extension to JWST filters recently derived by \citet{Fahrion+2023} for 30~Doradus, with 
$\kappa($H$\alpha$)=2.53 and $\kappa$(Pa$\alpha$)=0.678; as our sources are all HII regions, it is likely appropriate to use an extinction curve derived for the largest HII region in the LMC, adopting R$_V$=3.1 as appropriate for the Milky Way and metal--rich galaxies. The resulting color excess contains a small contribution from our own Milky Way's foreground extinction, E(B$-$V)$_{MW}$=0.06~mag \citep{Schlafly+2011}. 

Table~\ref{tab:sources} lists for each source: ID, location on the sky in RA(2000) and DEC(2000), observed luminosity in H$\alpha$, Pa$\alpha$ and 21~$\mu$m (after stellar continuum subtraction), the equivalent width in Pa$\alpha$, and the color excess E(B$-$V) derived from H$\alpha$/Pa$\alpha$. The observed H$\alpha$ and Pa$\alpha$ luminosities are corrected for the Milky Way foreground extinction, and, therefore, the listed color excess has E(B$-$V)$_{MW}$=0.06~mag subtracted. The quantities corrected for foreground MW extinction are those used from now on. Uncertainty in the distance produces a systematic offset in all luminosities between $-$0.06 and $+$0.08 in log scale, depending on the adopted distance. 

\section{Analysis and Results} \label{sec:results}

\begin{figure}
\plottwo{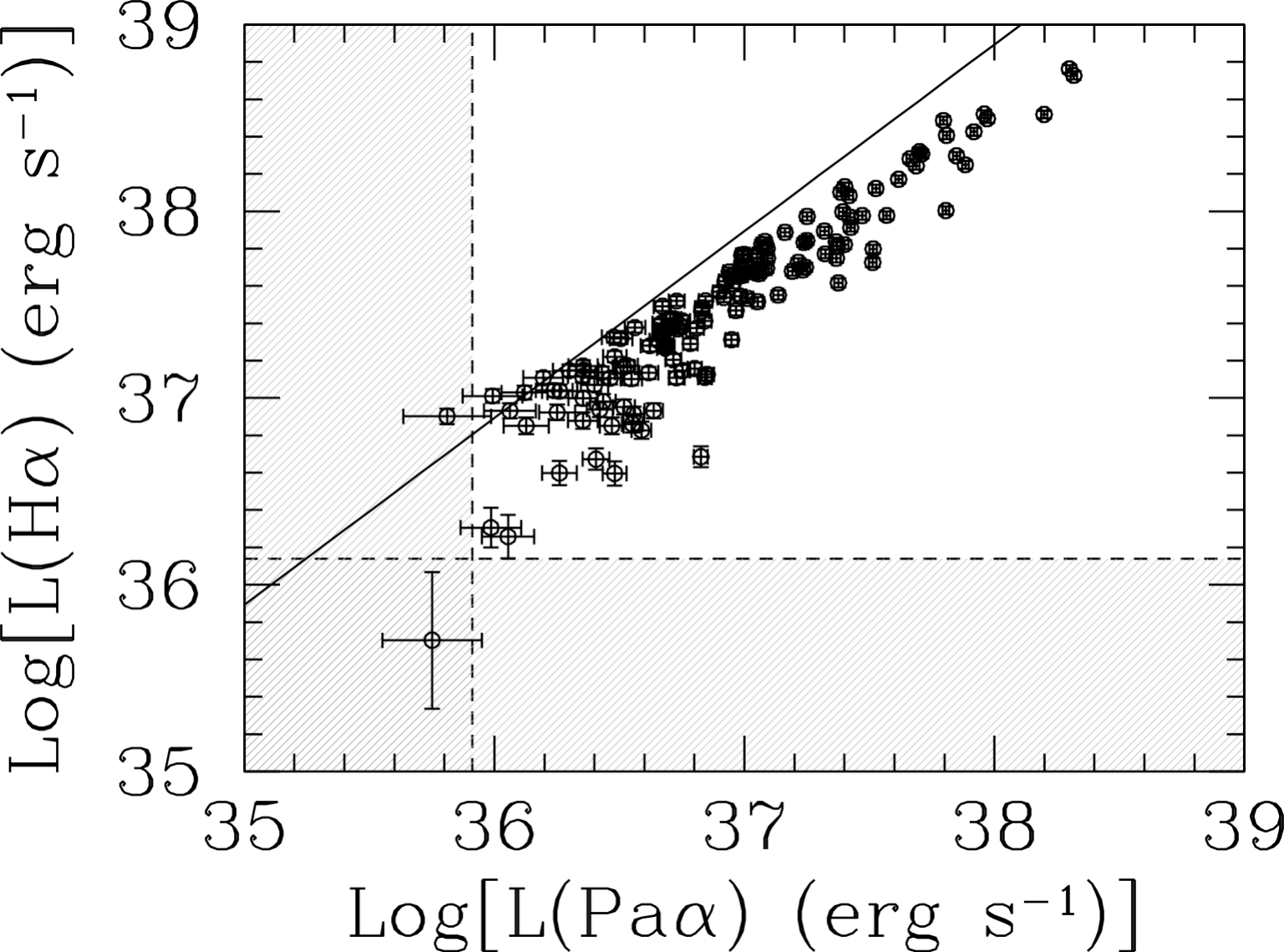}{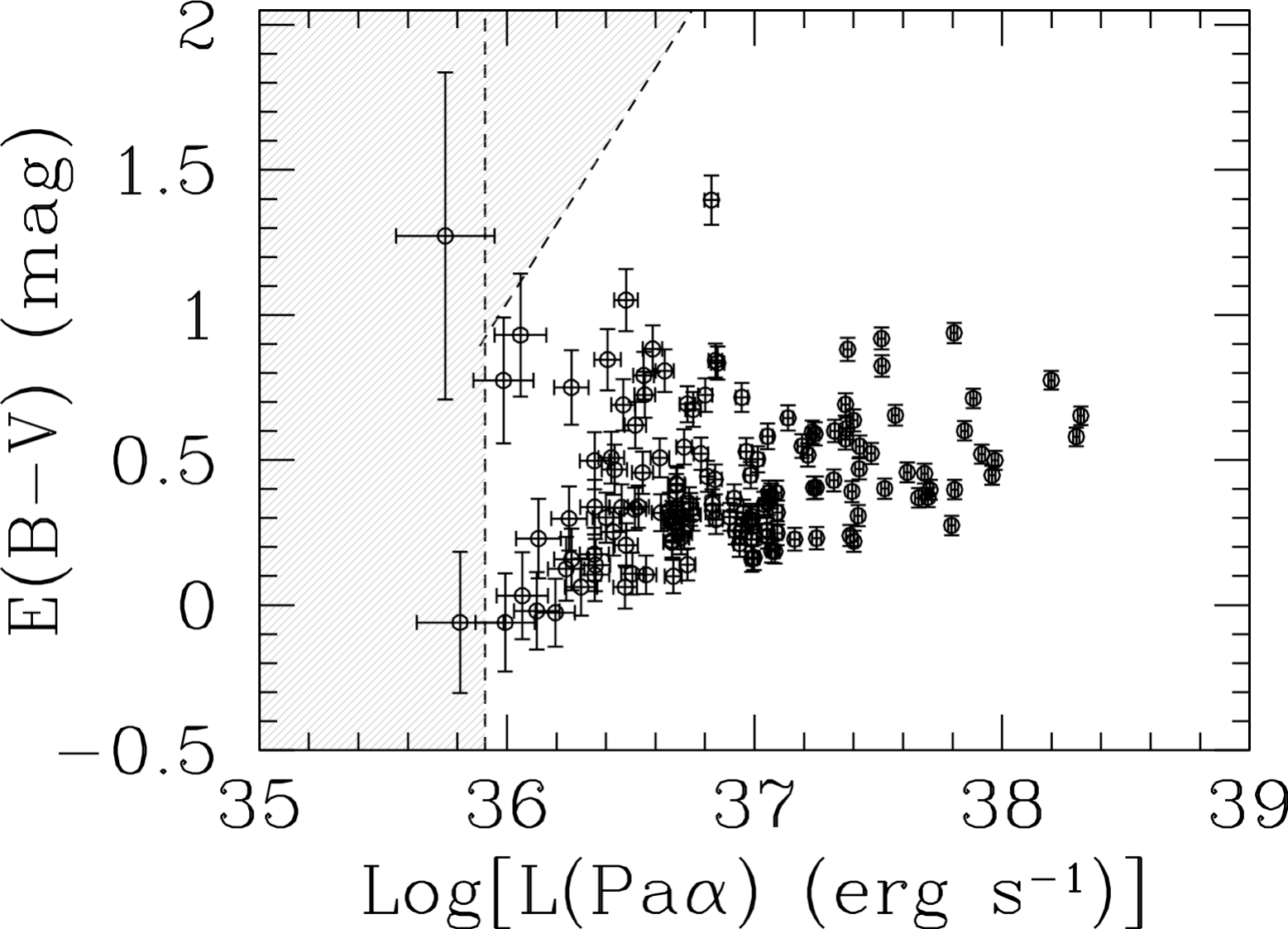}
\caption{The observed luminosity in H$\alpha$ (left panel) and the color excess E(B$-$V) derived from the observed nebular luminosities (right panel) as a function of the observed Pa$\alpha$ luminosity for the 143 regions identified in the JWST+HST mosaics. Luminosities and color excess are corrected for the foreground extinction from the Milky Way (E(B$-$V)$_{MW}$=0.06~mag). The data are shown as black circles with their 1~$\sigma$ uncertainties. The continuous black line in the left panel marks the expected location of luminosities of the two line for our adopted line ratio of 7.82. The grey--shaded regions mark the areas below 3~$\sigma$ detection.} 
\label{fig:obs_lum}
\end{figure}

\begin{figure}
\plottwo{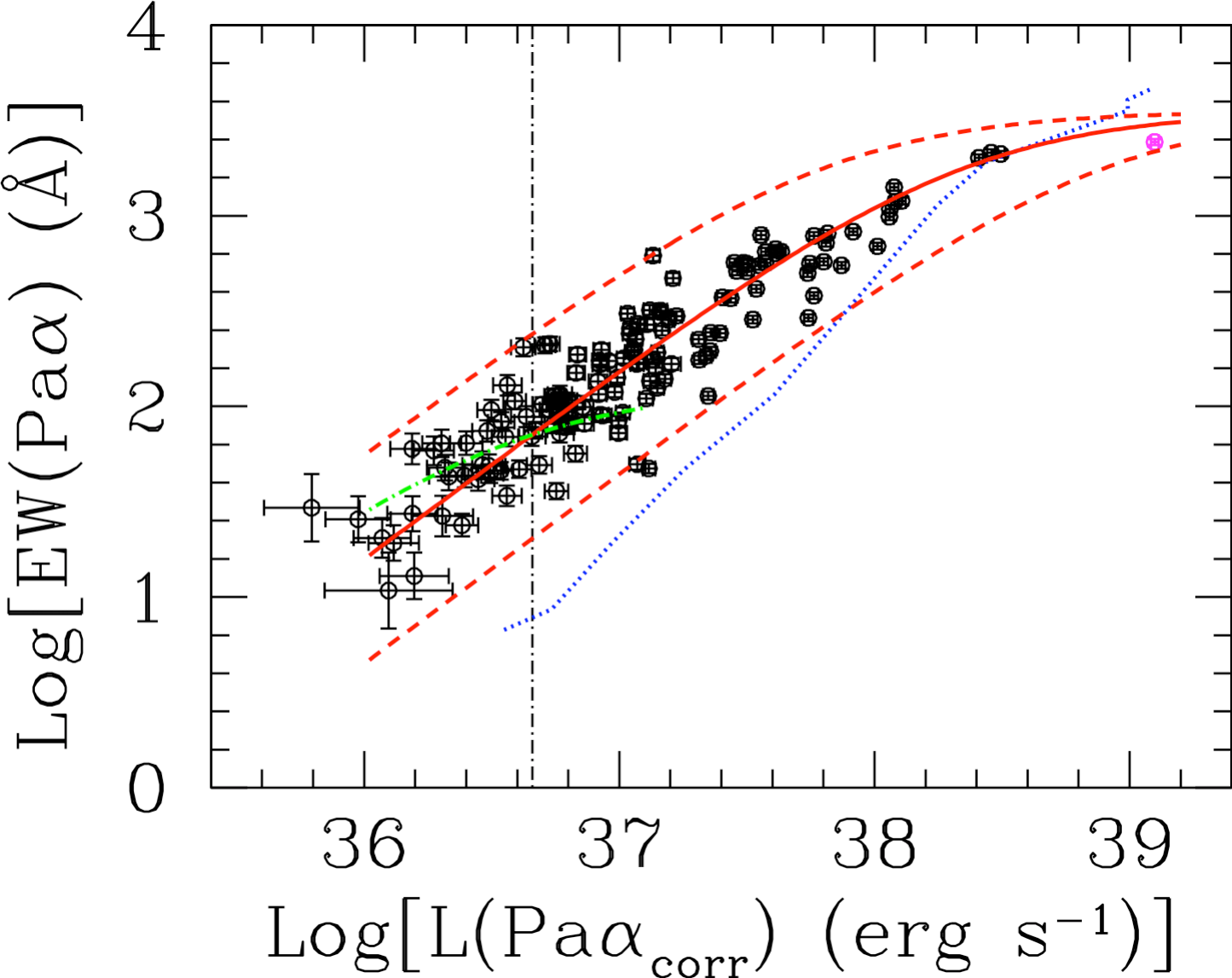}{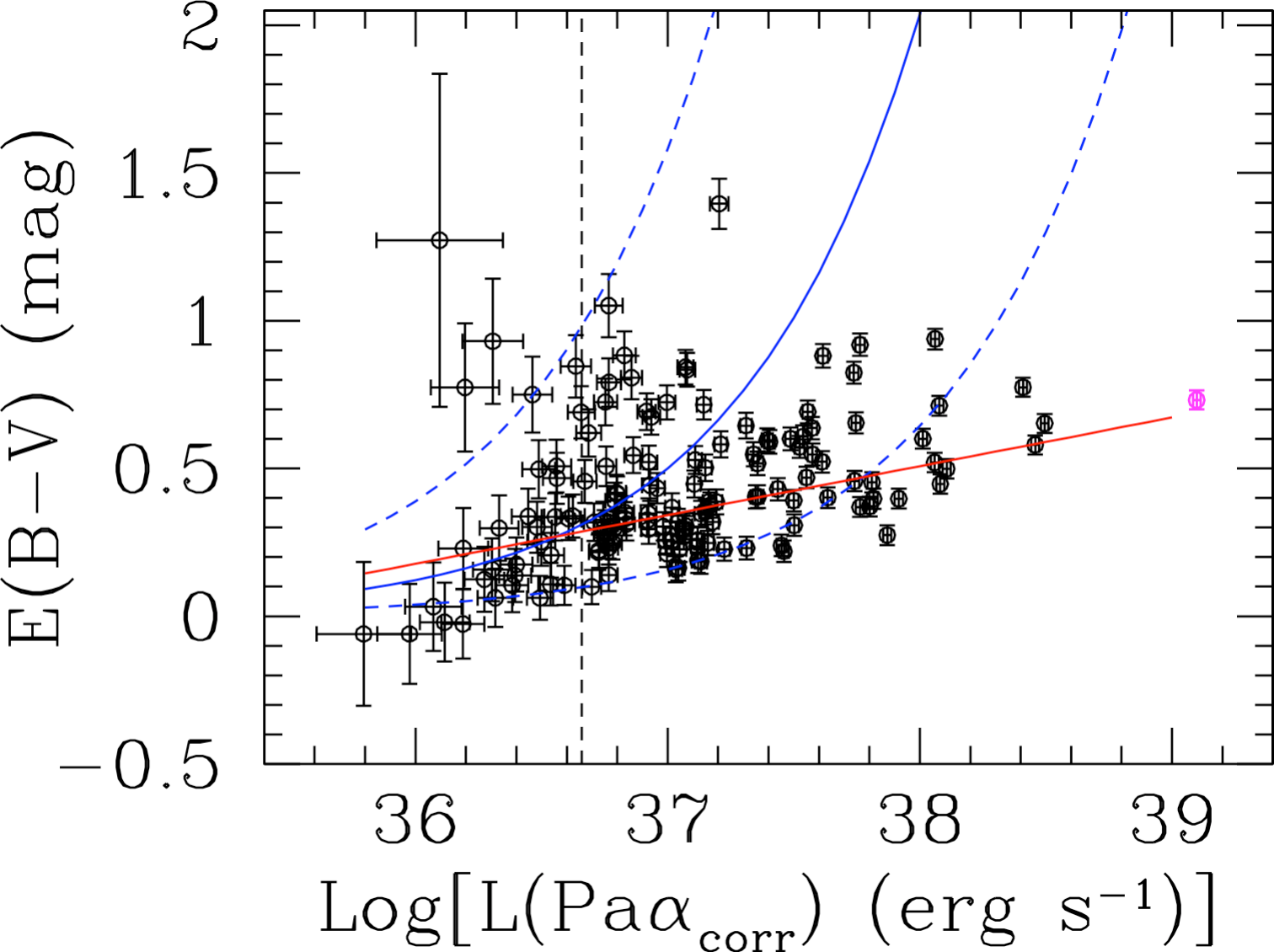}
\caption{The EW(Pa$\alpha$) (left panel) and the color excess E(B$-$V) (right panel) as a function of the attenuation--corrected luminosity in Pa$\alpha$. The data are shown as black circles, with their 1~$\sigma$ uncertainties. The magenta circle represents the highest luminosity Pa$\alpha$ and H$\alpha$ region in the JWST+HST mosaics; this region is not included in our analysis because it is outside of the MIRI footprint. Both panels show as a vertical black dot--dash line the expected Pa$\alpha$ luminosity of a 4~Myr old, 3,000~M$_{\odot}$ cluster, which we take as the lower-end luminosity limit for our analysis to mitigate stochastic IMF sampling \citep{Cervino+2002}. {\bf (Left):} Two sets of models (see text and Table~\ref{tab:ew_model} for more details) are shown: (1) an aging HII region from 1 to 10~Myr at constant mass (dotted blue line, model \# 1 in Table~\ref{tab:ew_model}) and (2) a constant age, 3~Myr, region in the mass range 200--3$\times$10$^5$~M$_{\odot}$ immersed in a constant, non--ionizing stellar background (solid red line, model \# 2 in Table~\ref{tab:ew_model}). The dashed red lines represent a factor $\pm$4 change in the value of the stellar background. The constant--background case is also shown for a 6~Myr region of varying mass (dot--dashed green line, model \# 3 in Table~\ref{tab:ew_model}). {\bf (Right):}  two empirical relations are shown: the E(B$-$V)--versus--L(P$\alpha$) relation derived by \citet{Calzetti+2007} for $\sim$0.5~kpc  star forming regions (blue lines; solid for the mean trend and dashed for the 90--percentile of those authors' data) and the relation by \citet{Garn+2010} for galaxies (continuous red line).} 
\label{fig:corr_lum}
\end{figure}

The trends marked by the observed nebular luminosities are shown in Figure~\ref{fig:obs_lum} (left panel), together with the color excess values E(B$-$V) (right panel) derived from those luminosities. The left panel of Figure~\ref{fig:obs_lum}  shows the location of the expected luminosities for our adopted line ratio of 7.82; almost all values of the H$\alpha$ luminosity are below this line indicating presence of dust attenuation, as supported by the right--hand--side panel. The regions below 3~$\sigma$ in the two tracers are shaded in grey. As expected from our selection criteria, all quantities are above the 3~$\sigma$ thresholds, with only one--two datapoints below those (but still consistent within their 1~$\sigma$ uncertainties). The color excess values are used to correct the nebular lines for the effects of dust attenuation, using an assumption of foreground dust geometry and the standard relation for foreground dust:
\begin{equation}
L(\lambda)_{corr} = L(\lambda) 10^{0.4 E(B-V) \kappa(\lambda)},
\end{equation}
where L($\lambda$) and L($\lambda$)$_{corr}$ are the observed and attenuation--corrected luminosities in units of erg~s$^{-1}$, respectively, and $\kappa$($\lambda$) is the extinction curve described in Section~\ref{sec:selection}.  We assume that the dust attenuation is the same for emission lines and stellar continuum, implying that equivalent widths are not affected by attenuation corrections. Figure~\ref{fig:corr_lum} shows the EW(Pa$\alpha$) and the color excess E(B$-$V) as a function of the attenuation--corrected Pa$\alpha$ luminosity. In these two panels we include the most luminous H$\alpha$ and Pa$\alpha$ source in the JWST+HST frames, located at RA(2000)=1:36:45.2436, DEC(2000)=+15:47:47.586; this source cannot be used for the 21~$\mu$m analysis because it is outside of the MIRI footprint (yellow circle in Figure~\ref{fig:sources}). However, the source is useful to verify trends for the EW and E(B$-$V). Trends of both quantities with L(Pa$\alpha$)$_{corr}$ and their agreement with models or previous results will be discussed in Section~\ref{sec:discussion}. 

We mark the location of the Pa$\alpha$ luminosity expected for a 4~Myr old cluster with mass 3,000~M$_{\odot}$ \citep[derived from the Starburst99 models, see below and][]{Leitherer+1999,Vazquez+2005}  with a vertical line (black dot--dash) in both panels of Figure~\ref{fig:corr_lum}. We take this luminosity, corresponding to Log[L(Pa$\alpha$)]=36.66 and Log[L(H$\alpha$)]=37.55, as the lowest limit we use in our analysis and fits in order to mitigate effects of stochastic sampling of the stellar IMF \citep{Cervino+2002}. Including regions below this luminosity limit can artificially affect trends: \citet{Fumagalli+2011} show that, below M$\approx$3$\times$10$^3$~M$_{\odot}$, the ratio L(UV)/L(H$\alpha$) (a proxy for L(21)/L(Pa$\alpha$)) is a factor $\sim$3 or more overluminous relative to higher masses. These stochastic IMF sampling limits are a factor 25.7 and 5.6 above the 3~$\sigma$ thresholds for the H$\alpha$ and Pa$\alpha$, respectively. Tests run using more conservative mass limits, up to M=10$^4$~M$_{\odot}$, do not significantly change the results presented in this paper, but reduce the dynamical range of the data, thus increasing uncertainties. For instance, for a limit of M=10$^4$~M$_{\odot}$, uncertainties increase by a factor $\sim$4 for both the slope and intercept of the best fit line between the mid--IR emission and L(P$\alpha_{corr}$) (see later in the section).  The insensitivity of our results to the choice of stellar mass above 3,000~M$_{\odot}$ will be further discussed later in this section.

To derive relations between attenuation--corrected luminosities and EWs, we adopt the Starburst99 population synthesis models for stellar populations \citep{Leitherer+1999,Vazquez+2005}. We use the instantaneous burst outputs, under the assumption that a $\sim$120~pc diameter region dominated by a compact line emission source is likely to be well described by a single or, at most, a few coeval cluster(s). We adopt the Padova AGB evolutionary tracks \citep{Girardi+2000} with metallicity Z=0.02 (solar) and a \citet{Kroupa2001} IMF in the stellar mass range 0.1--120~M$_{\odot}$. Standard outputs from Starburst99 include lists of several nebular lines luminosities and EWs as a function of age (at the fixed mass of 10$^6$~M$_{\odot}$), including H$\alpha$, but excluding Pa$\alpha$. We derive L(Pa$\alpha$) from the models by simply dividing L(H$\alpha$) by 7.82 (see section~\ref{sec:selection}). We derive the EW(Pa$\alpha$) as a function of age by dividing the model line luminosity by the continuum level around the Pa$\alpha$ wavelength in the model spectra, measured in a manner that mimics our continuum subtraction method for the observations. Table~\ref{tab:Paa} lists the Pa$\alpha$ luminosities and EWs for a 10$^6$~M$_{\odot}$ instantaneous burst population in the age range 1--10~Myr, as derived from the Starburst99 models. Using the Geneva tracks with no rotation \citep{Ekstrom+2012} yields emission line fluxes that are typically $\sim$30\%--40\% higher than those of the Padova AGB tracks, and similarly higher EWs, for the first $\sim$4~Myr, and converging to comparable values at older ages.

{We use the single stellar stellar population models above to introduce composite population models that can be used to explain the trend between EW(Pa$\alpha$) and the attenuation--corrected Pa$\alpha$ luminosity (left panel of Figure~\ref{fig:corr_lum}). The implications of these models will be discussed in Section~\ref{sec:discussion}. We consider two basic model tracks for the EW--luminosity relation: the track expected for an aging, constant--mass, instantaneous--burst population from Starburst99 (dotted blue line in Figure~\ref{fig:corr_lum}, left panel) and the track expected for a constant--age, decreasing mass instantaneous--burst stellar population immersed in a constant--level non--ionizing stellar population (solid red and dot--dashed green lines in Figure~\ref{fig:corr_lum}, left panel). The parameters we adopt for these tracks are listed in Table~\ref{tab:ew_model}. For the aging HII region (model \# 1 in Table~\ref{tab:ew_model} and dotted blue line in Figure~\ref{fig:corr_lum}, left), the model ranges from 1~Myr to 10~Myr, with constant mass$\sim$2$\times$10$^5$~M$_{\odot}$; the mass is chosen to have the youngest region match the largest Pa$\alpha$ luminosity reported. The model with constant age and decreasing mass is constructed by choosing the largest mass for the HII region to match the largest P$\alpha$ luminosity in Figure~\ref{fig:corr_lum}, left, that also has Log[EW(Pa$\alpha$)]=3.56; from Table~\ref{tab:Paa}, this corresponds to a 3~Myr old stellar population with stellar mass$\sim$3$\times$10$^5$~M$_{\odot}$ (model \# 2 in Table~\ref{tab:ew_model} and solid red line in Figure~\ref{fig:corr_lum}, left). This HII region is surrounded by non--ionizing stellar light that matches the typical values found in our images and is kept constant. Conversely, we allow the Pa$\alpha$--emitting population's mass (and luminosity) to decrease by a little over 3~orders of magnitude, to match the range of the Pa$\alpha$ luminosity in the Figure. Thus, the non--ionizing stellar light's intensity ranges from 1/7 to 220 times the HII region continuum intensity, producing the observed trend. The same Figure shows the expected range (dashed red lines) if the non--ionizing population intensity is changed by a factor $\pm$4 relative to its default value. For comparison, we also show the track with varying mass of HII regions with age 6~Myr (model \# 3 in Table~\ref{tab:ew_model}, dot--dashed green line in Figure~\ref{fig:corr_lum}, left). In all cases, model parameters are chosen to maximize overlap between data and models.

\begin{deluxetable}{rrr}
\tablecolumns{3}
\tabletypesize{\footnotesize}
\tablecaption{Pa$\alpha$ Emission Line Model Properties.\label{tab:Paa}}
\tablewidth{0pt}
\tablehead{
\colhead{Age (Myr)} &  \colhead{Log[L(Pa$\alpha$)]\tablenotemark{a}} & \colhead{Log[EW(Pa$\alpha$)]\tablenotemark{b}} 
\\
}
\startdata
\hline
1  & 39.828 & 3.663\\
2  & 39.741 & 3.608\\
3  & 39.744 & 3.553\\
4  & 39.187 & 3.282\\
5  & 38.997 & 3.057\\
6  & 38.357 & 2.062\\
7  & 38.003 & 1.667\\
8  & 37.723 & 1.285\\
9  & 37.485 & 0.936\\
10  & 37.285 & 0.820\\
\hline
\enddata
\tablenotetext{a}{Logarithm (base 10) of the luminosity at Pa$\alpha$ ($\lambda_{rest}$ 1.8756~$\mu$m) for a 10$^6$~M$_{\odot}$ instantaneous burst stellar population, in units of erg~s$^{-1}$.}
\tablenotetext{b}{Logarithm (base 10) of the equivalent width of Pa$\alpha$, in units of \AA. The continuum underlying the emission line is measured from the interpolation between the SED values obtained from the model
spectra convolved with  the F150W and F200W filter bandpasses, using the same method used for the observational data.} 
\tablecomments{Based on instantaneous burst stellar populations from Starburst99 with metallicity Z=0.02, Padova AGB tracks \citep{Girardi+2000}, and \citet{Kroupa2001} stellar IMF between 0.1 and 120 M$_{\odot}$ \citep{Leitherer+1999,Vazquez+2005}.}
 \end{deluxetable}

\begin{deluxetable*}{llcccccc}
\tablecaption{Models for the Equivalent Width versus Luminosity in the Pa$\alpha$ Line\label{tab:ew_model}}
\tablewidth{0pt}
\tablehead{
\colhead{ID $^1$} & \colhead{Model$^2$} & \colhead{Age$^3$} & \colhead{Mass$^3$} & \colhead{Log(L$_{stars}$)$^4$} &  \colhead{Log[L(Pa$\alpha$)]$^5$} &  \colhead{Log[L$_{c,Pa\alpha}$]$^5$} & \colhead{Log[EW(Pa$\alpha$)]$^6$} \\
\colhead{} & \colhead{} & \colhead{(Myr)} & \colhead{(M$_{\odot}$)} & \colhead{(erg~s$^{-1}$~\AA$^{-1}$)} & \colhead{(erg~s$^{-1}$)} & \colhead{(erg~s$^{-1}$~\AA$^{-1}$)}  & \colhead{(\AA)} \\
}
\decimalcolnumbers
\startdata
\# 1 & Const. Mass & 10--1 & 2.0E5  & 0. & 36.59--39.13 &35.77--35.47 & 0.82--3.66\\
\# 2 & Const. Age    &  3     & 2.0E2--3.0E5& 34.80 & 36.00--39.22 & 32.45--35.67 & 1.20--3.56\\
\# 3 & Const. Age   &   6     & 4.4E3--5.5E5 & 34.45 & 36.00--37.10 & 33.90--35.00  & 1.44--2.00\\
\enddata
$^1$  Model ID\\
$^2$  Description of the stellar population model associated with the HII region: we consider both the constant mass with varying age (\# 1) and the constant age with varying mass (\# 2 and \# 3) cases; \# 2 is for a population with 3~Myr age and \# 3 is for 6~Myr.\\
$^3$ Age (or age range) and mass (or mass range) of the stellar population associated with the HII region.\\
$^4$ Logarithm  (base 10) of the flux density from the stellar population {\em unrelated} to the HII region; this stellar population is assumed to be non--ionizing.\\ 
$^5$ Logarithm  (base 10) of the emission line flux at Pa$\alpha$, L(Pa$\alpha$), and  of the underlying stellar continuum, L$_{c,Pa\alpha}$, of the stellar population associated with the HII region, rescaled from the outputs of Starburst99 \citep{Leitherer+1999,Vazquez+2005} listed in Table~\ref{tab:Paa}.\\ 
$^6$ Logarithm  (base 10) of the equivalent width (EW) of the Pa$\alpha$ emission resulting from the ratio of the emission line intensity divided by the combined continuum, L$_{c,Pa\alpha}$+L$_{stars}$, of the two stellar populations: the one responsible for the HII region and the one unrelated to it.\\ 
\tablecomments{Quantities are ordered from smallest EW(Pa$\alpha$) (largest age or smallest mass) to largest EW(Pa$\alpha$) (smallest age or largest mass). Model parameters are chosen  to ensure that the model lines overlap with the data.}
\end{deluxetable*}

The relation between L(21) and L(Pa$\alpha$)$_{corr}$  is shown in Figure~\ref{fig:l21_vs_lpa} (left panel), with its equivalent in surface brightness at 24~$\mu$m, $\Sigma$(24), shown in the right panel. The change from luminosity to surface brightness is a constant for all regions ($+$1.90 dex in log10 scale) as they are all located at the same distance. All surface brightnesses are expressed in units of erg~s$^{-1}$~kpc$^{-2}$. 

We perform the conversion from L(21) to L(24) in order to relate our measurements to the vast literature using L(24) from the {\em Spitzer Space Telescope} and discussed in the next section. The relation between L(21) and L(24) for NGC\,628 is determined by first convolving the JWST/MIRI/F2100W image to the PSF FWHM$\sim$6$^{\prime\prime}$.4 of Spitzer/MIPS24 \citep{Aniano+2011}, and then measuring common 
bright regions in the two bands\footnote{The Spitzer MIPS24 image of NGC\,628 was retrieved from NED, the NASA Extragalactic Database.}. We choose an aperture of 10$^{\prime\prime}$, with a background annulus of 3$^{\prime\prime}$ width, to perform background--subtracted photometry; the aperture size is a 
compromise between measuring regions that are large enough to mitigate the effects of small mismatches in the PSFs and are small enough to mitigate the low sensitivity of MIRI to diffuse emission. With these choices, we determine that:
\begin{equation}
Log L(24) = Log L(21) - 0.057.
\label{equa:l24}
\end{equation}
In other words, L(24) is fainter, by 14\%, than L(21). Reasons for this can be at least two--fold:  the luminosity L$\propto\nu$f($\nu$) is about 14\% lower at 24~$\mu$m than at 21~$\mu$m if the spectrum is relatively flat in f($\nu$) \citep[e.g.,][]{DraineLi2007}, and/or there may still be small calibration mismatches between the two bands. The observed mid--IR spectrum of the central $\approx$30$^{\prime\prime}$ of NGC\,628 is flat in $\nu$f($\nu$) between 21 and 24~$\mu$m \citep{Smith+2007}, which may support the presence of some residual calibration mismatch\footnote{However, see the most recent calibration releases at: https://jwst-docs.stsci.edu/jwst-data-calibration-considerations/jwst-data-absolute-flux-calibration}. For the purpose of this analysis, the origin of the difference is of secondary importance, and the difference itself is small enough that applying equation~\ref{equa:l24} to our photometry will not affect conclusions.

Figure~\ref{fig:l21_vs_lpa} also shows the best linear fits through the datapoints. The two relations, one connecting luminosities and the other surface brightnesses, are identical between the left and right panels, except for the small offset in the intercept, which reflects the shift between L(21) and L(24) from equation~\ref{equa:l24}. The best fit for all 143 regions is:
\begin{equation}
Log[\Sigma(24])=(1.02 \pm 0.01) Log[\Sigma(P\alpha)_{corr}] + (1.02 \pm 0.41),
\end{equation}
while the best fit through the 111 regions above the stochastic IMF sampling limit (the `censored' data) is:
\begin{equation}
Log[\Sigma(24)]=(1.07 \pm 0.01) Log[\Sigma(P\alpha)_{corr}] - (1.11 \pm 0.46).
\label{equa:bestfit}
\end{equation}
The fits reported use a bi--regression algorithm, which includes uncertainties in both x and y coordinates. The average of the x--vs--y and y--vs--x fits yields identical results to those given above. Using a least square bisector fitting routine does not change the basic results: this algorithm yields a slope for the uncensored data (all 143 regions) of 0.97$\pm$0.03 and for the censored data (111 regions) of 1.10$\pm$0.03, thus virtually identical to the slopes yielded by the bi--regression algorithm within the uncertainties. The scatter about the mean fit line is $\pm$0.3~dex for the censored data, which is captured by the uncertainty in the best fit intercept value.

\begin{figure}
\plottwo{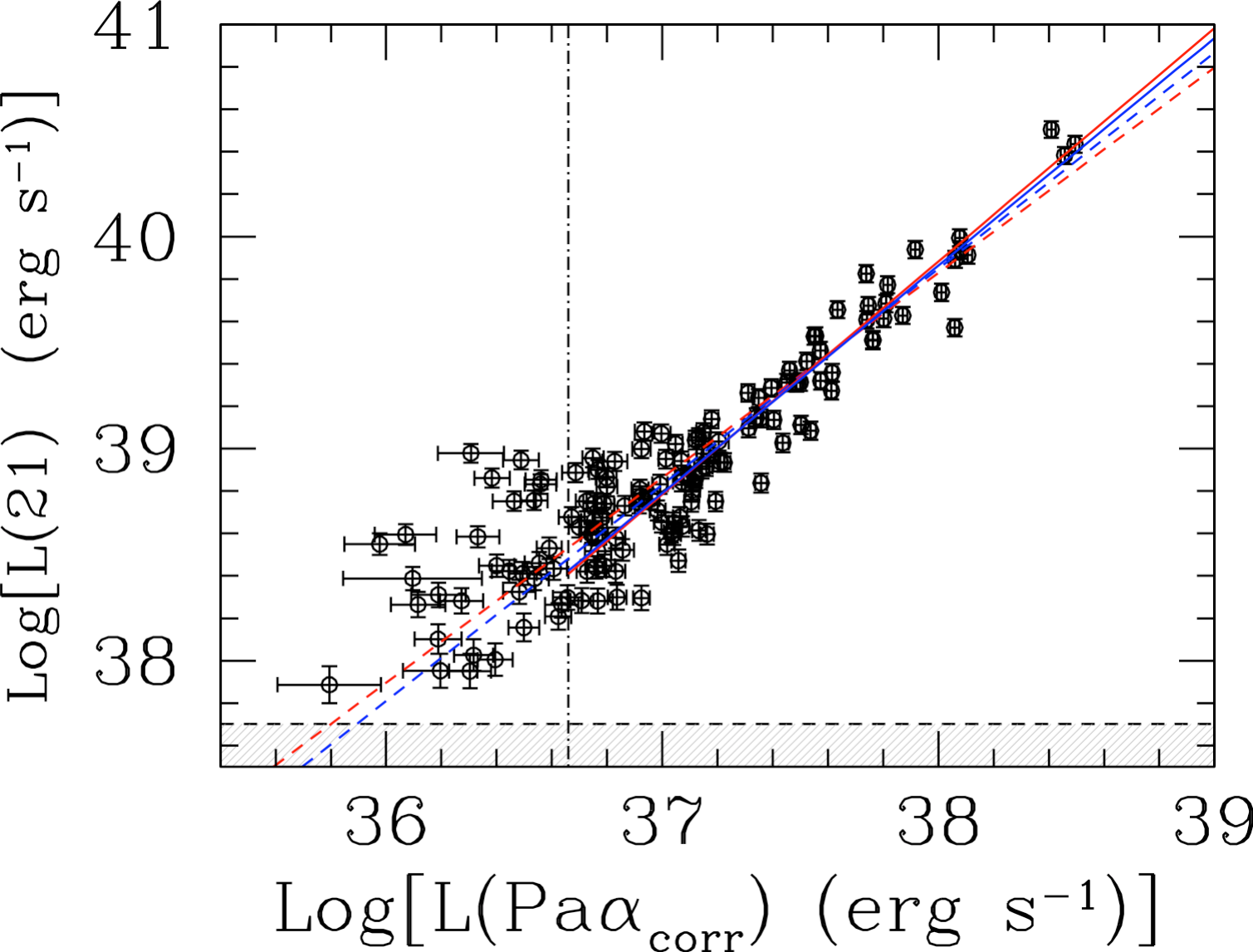}{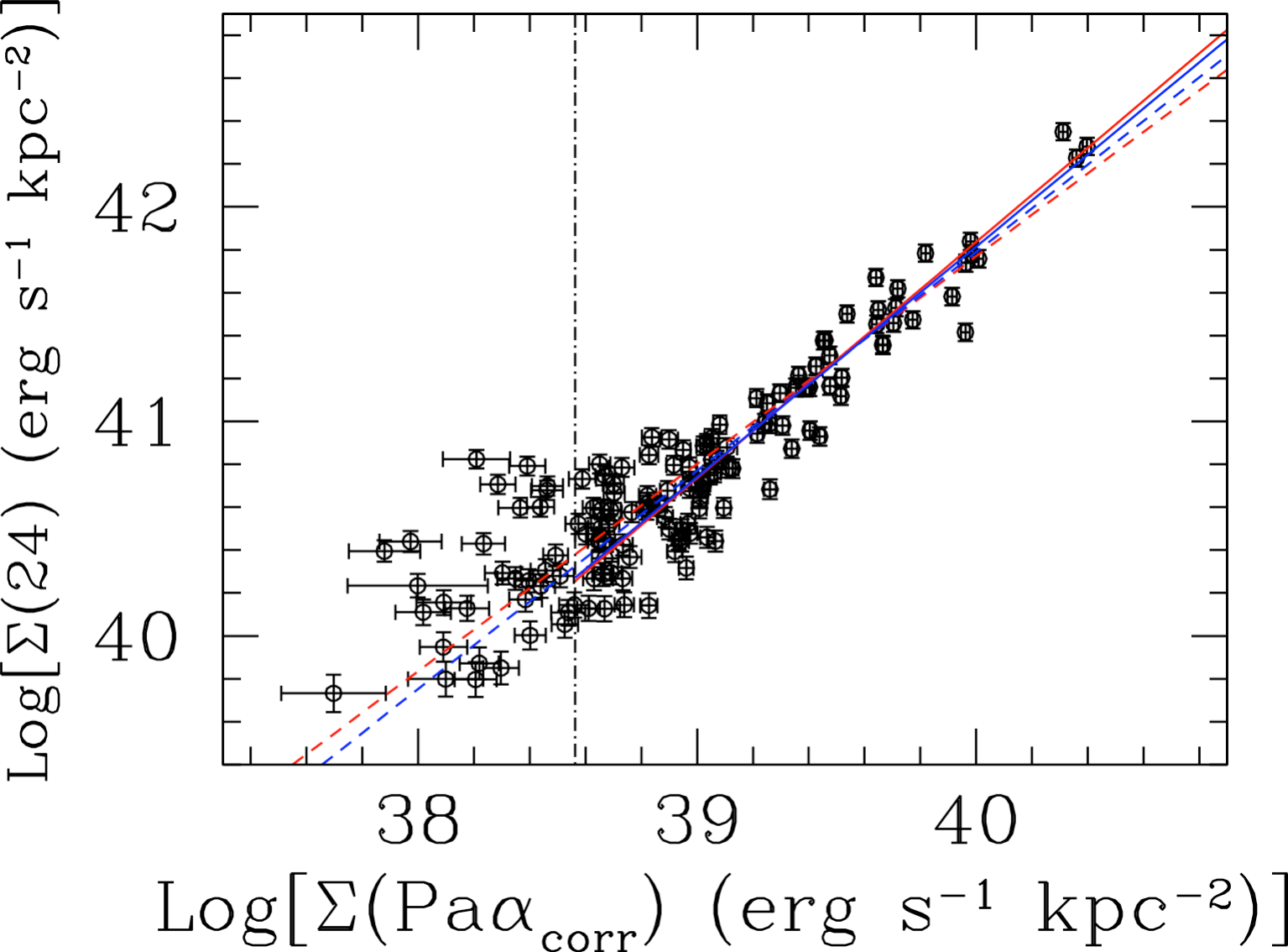}
\caption{(Left:) The luminosity at 21~$\mu$m, L(21), as a function of the attenuation--corrected luminosity at Pa$\alpha$ for the 143 line--emitting regions in this sample (black circles), with 1~$\sigma$ uncertainties. The horizontal dash line and grey region show the location of the 5~$\sigma$ threshold for the 21~$\mu$m detections. (Right:) The surface brightness at 24~$\mu$m, $\Sigma$(24), as a function of the attenuation--corrected surface brightness at Pa$\alpha$ for the same regions (black circles), also with 1~$\sigma$ uncertainties. The luminosity at 21~$\mu$m is converted to the one at 24~$\mu$m using equation~\ref{equa:l24}. In both panels, the vertical line (black dot-dash) marks the transition luminosity above which stochastic sampling of the IMF is mitigated. Color lines are for the best fits through the data: dash lines for the full sample and solid lines for the 111 regions above the stochastic IMF sampling limit. Red for fits with the least square bisector and blue for the bi--regression.}
\label{fig:l21_vs_lpa}
\end{figure}

Within the formal uncertainty, the relation between $\Sigma$(24) and $\Sigma$(P$\alpha$)$_{corr}$ for the regions that are more luminous than the stochastic IMF sampling limit has an exponent above unity. The significance of this result is difficult to ascertain, given the scatter in the data, but it may reflect the observation that more luminous regions tend also to be dustier, with a higher average color excess than lower luminosity regions (Figure~\ref{fig:corr_lum}; discussion in the next section);  this implies a higher infrared emission per unit of attenuation--corrected Pa$\alpha$ emission at higher luminosity \citep[e.g.,][]{Calzetti+2007}. The increase in $\Sigma$(24) for increasing $\Sigma$(P$\alpha_{corr}$)  is due to two effects: the linear growth expected for increasing luminosity at constant dust attenuation and an additional contribution to $\Sigma$(24) if attenuation increases at higher luminosities. This is visualized in Figure~\ref{fig:ebv_vs_l24ratio}, where the color excess E(B$-$V) is shown as a function of the ratio Log[$\Sigma(24)/\Sigma(Pa\alpha_{corr})$], for the 111 HII regions with luminosity above the stochastic IMF sampling limit. We observe a mild increase of the median E(B$-$V) for increasing Log[$\Sigma(24)/\Sigma(Pa\alpha_{corr})$], although the scatter in the data is large; the increase is comparable when bins in Log[$\Sigma(24)/\Sigma(Pa\alpha_{corr})$] are chosen by equal size or equal number of datapoints. A similar result is obtained when binning in E(B$-$V) instead of Log[$\Sigma(24)/\Sigma(Pa\alpha_{corr})$]. Also shown in Figure~\ref{fig:ebv_vs_l24ratio} are the best fit lines obtained by using a bi--regression (continuous line) and a bisector (dot--dash line) algorithm; the results from the fits are different, supporting the visual impression that the two variables are weakly correlated. 

There is a separate physical mechanism that can produce a super--linear $\Sigma(24)$--vs.--$\Sigma(Pa\alpha_{corr})$ correlation, similar to the effect of increasing attenuation. In the presence of dust, a fraction of ionizing photons are lost to direct dust absorption and do not contribute to the recombination cascade, although they 
contribute to the infrared luminosity. If this fraction increases with HII region luminosity, the super--linear slope would be due to a smaller proportion of the ionizing photons being reprocessed into recombination lines, rather than a larger proportion of the starlight being reprocessed into IR in brighter HII regions.  Models show that even modest amounts of dust in HII regions, E(B$-$)$\lesssim$0.3~mag, can decrease by 10\% the number of ionizing photons available for processing into recombination lines \citep{Draine+2011}. Using the models of \citet{Krumholz+2009} and going from  0\% to the maximum theoretical value of 50\% ionizing photons directly absorbed by dust from the lowest to the highest HII region luminosities, the slope of the Log[$\Sigma$(24)]--vs--Log[$\Sigma$(P$\alpha$$_{corr}$)] would change from 1 to $\sim$1.2, slightly steeper than observed, but not inconsistent with it. Thus, the observed trend in Figure~\ref{fig:l21_vs_lpa} can be due to multiple mechanisms at play. 

\begin{figure}
\plotone{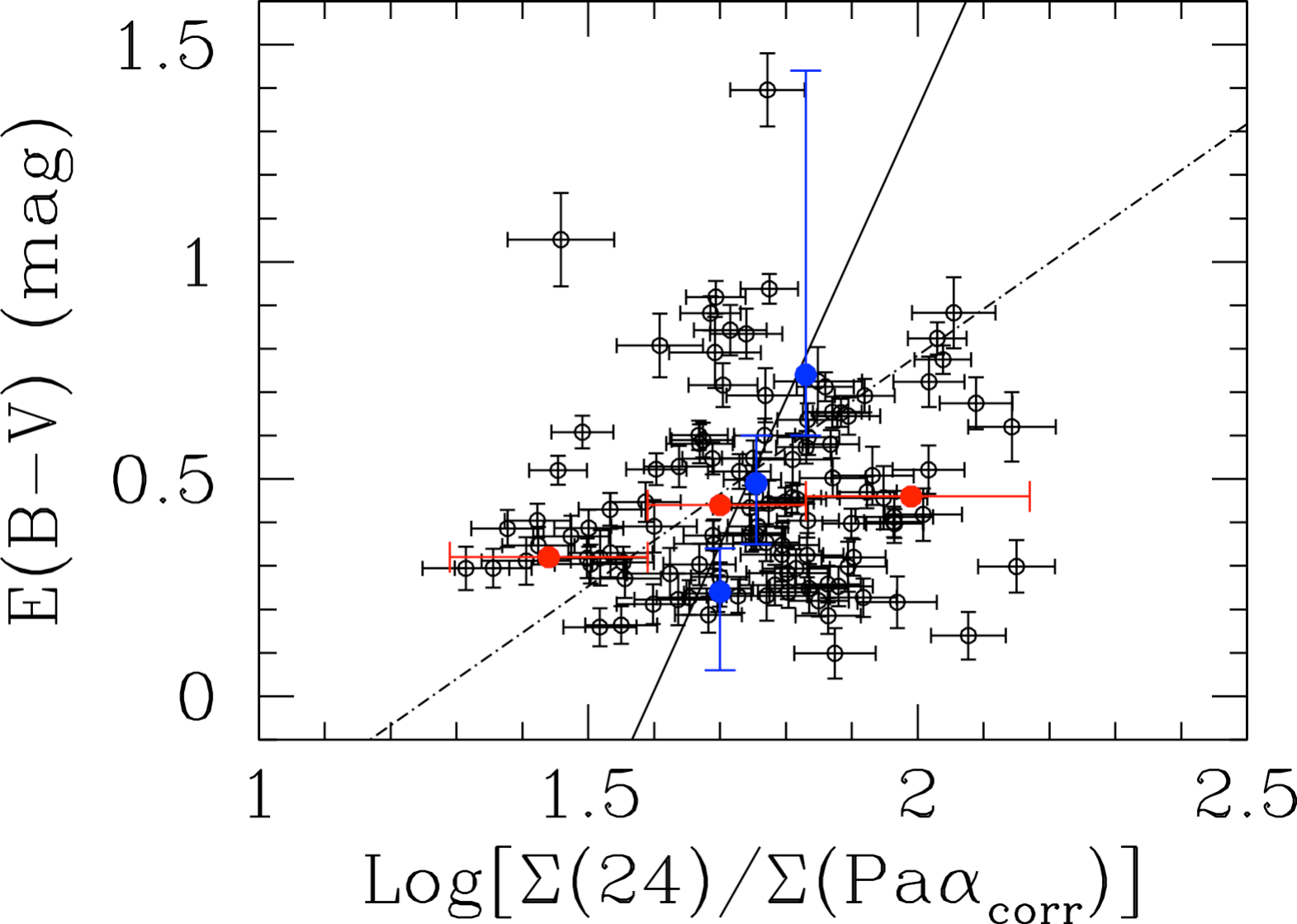}
\caption{The color excess E(B$-$V) for the 111 HII regions with luminosity above the stochastic IMF sampling limit as a function of the ratio Log[$\Sigma(24)/\Sigma(Pa\alpha_{corr})$]. Data are shown as black circles with 1~$\sigma$ uncertainty, and the median E(B$-$V) for three bins of increasing Log[$\Sigma(24)/\Sigma(Pa\alpha_{corr})$ are shown as filled red circles, with the bin widths shown by horizontal bars. The bins are centered at Log[$\Sigma(24)/\Sigma(Pa\alpha_{corr})$=1.44,1.70,2.00, with the median E(B$-$V)=0.38,0.50,0.52. The bins are equal size, but the results are virtually identical for bins of equal numbers of datapoints per bin. The case where the data are binned in E(B$-$V), for equal number bins and centered at 0.24,0.49,0.74~mag, are also shown as filled blue circles with vertical blue bars to show the width of the bins. The two black lines show the results for linear fits through the data, performed using the bi--regression (continuous line) and the least square bisector (dot--dash line) algorithms. }
 \label{fig:ebv_vs_l24ratio}
\end{figure}

\citet{Wang+1996} introduced the idea of combining a tracer of the unattenuated stellar light in the UV/optical with a tracer of the dust--reprocessed stellar light in the infrared to capture all the light associated with recent star formation. This was later expanded by several authors to calibrate the UV+IR as a SFR indicator \citep[e.g.,][]{Buat+1999, Meurer+1999, Hirashita+2003, Treyer+2010, Hao+2011, Liu+2011} and by \citet{Calzetti+2007},  \citet{Kennicutt+2007}, and \citet{Kennicutt+2009} to combine nebular emission and IR into SFR tracers. Following the technique presented in \citet{Calzetti+2007}, we combine the observed H$\alpha$ surface density with the 24~$\mu$m surface density to obtain a linear relation which reproduces the attenuation--corrected H$\alpha$ surface density:
\begin{equation}
\Sigma(SFR) \propto \Sigma(H\alpha_{corr}) = \Sigma(H\alpha) + b  \Sigma(24),
\label{equa:mixsigma}
\end{equation}
where $\Sigma(H\alpha)$ and $\Sigma(24)$ are the observed quantities and $\Sigma(H\alpha_{corr}$) is the attenuation-corrected one. The difference $\Sigma(H\alpha_{corr}) -  \Sigma(H\alpha)$ is the fraction of nebular emission absorbed by dust, which can be related to the fraction of massive starlight attenuated by dust; $\Sigma(24)$ is a measure of the emission from the dust heated by those stars for a simplistic assumption of foreground dust. The ratio between these two quantities:
\begin{equation}
b = {\Sigma(H\alpha_{corr}) - \Sigma(H\alpha) \over  \Sigma(24)},
\label{equa:b_definition}
\end{equation}
provides the scaling for the dust emission to be translated to a SFR \citep{KennicuttARAA1998, KennicuttEvans2012}. In Figure~\ref{fig:b_vs_Pa}, we plot $b$ as a function of both $\Sigma(Pa\alpha_{corr})$ \citep{Calzetti+2007} and EW(Pa$\alpha$) \citep{Belfiore+2023}, and provide fits for these quantities using data that are above the stochastic IMF sampling limit. For the EW(Pa$\alpha$), we determine the lower limit  from the left panel of Figure~\ref{fig:corr_lum} to be Log[EW(P$\alpha$)]$\simeq$1.8. 
In both cases, $b$ is almost independent (within 2~$\sigma$) of either variable, $\Sigma(Pa\alpha_{corr}$) or EW(P$\alpha$), for the censored data. The quantity $b$ is, therefore, well described by a constant which corresponds to the median value through the data:
\begin{equation}
\Sigma(H\alpha_{corr}) = \Sigma(H\alpha) + (0.095\pm0.007)  \Sigma(24).
\label{equa:mix}
\end{equation}

\begin{figure}
\plottwo{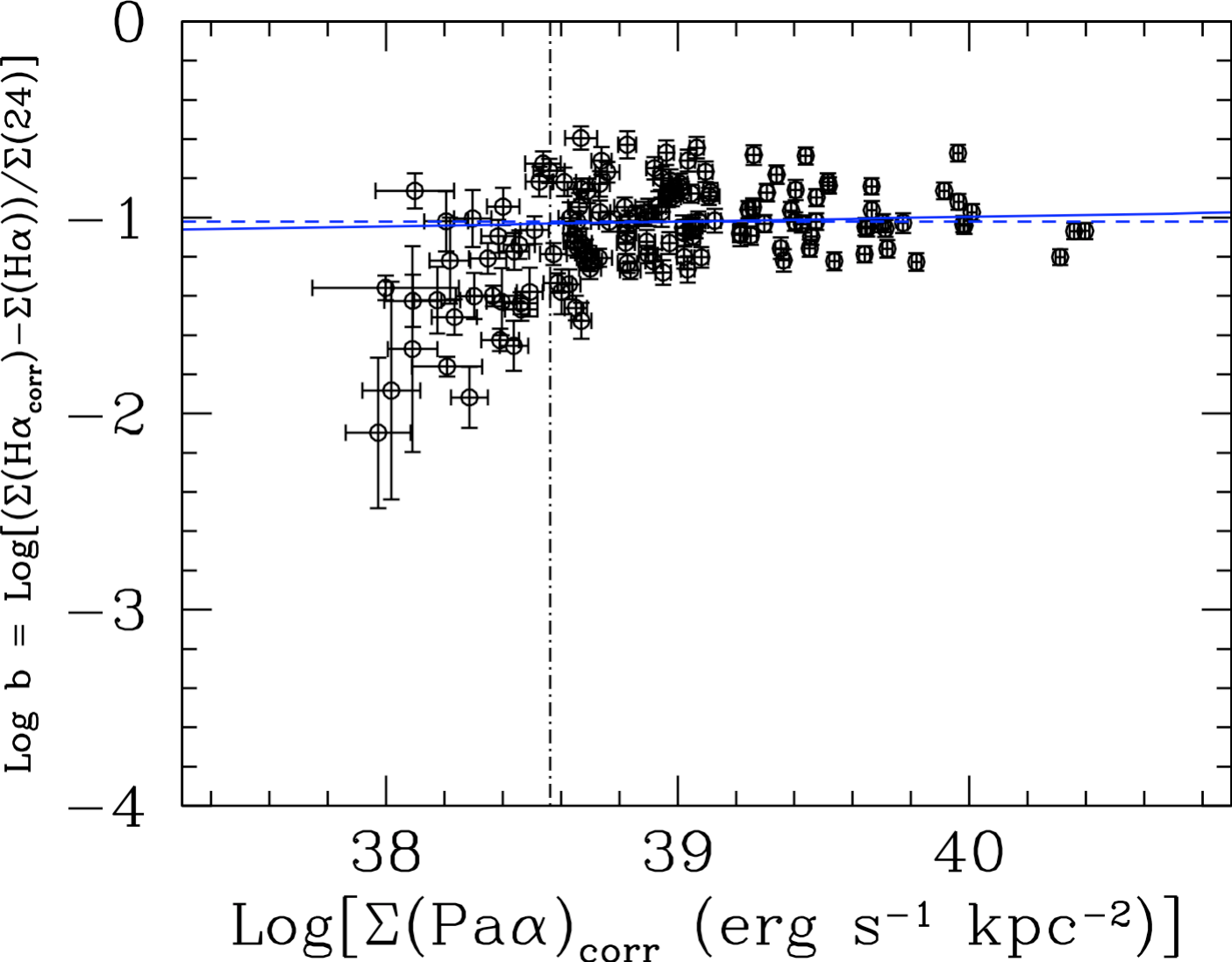}{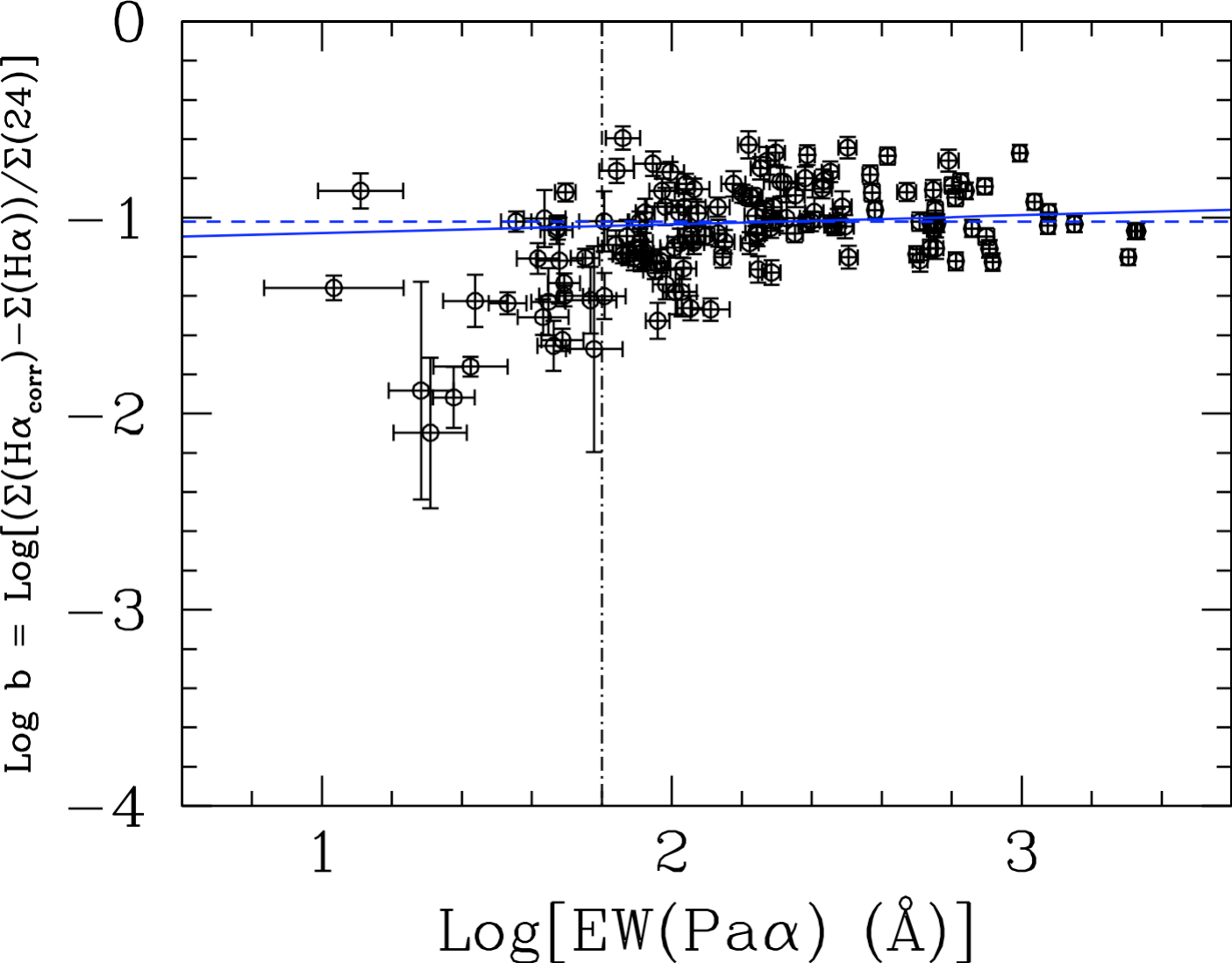}
\caption{The ratio of the nebular hydrogen emission absorbed by dust at H$\alpha$ and the dust emission at 24~$\mu$m as a function of the surface brightness of Pa$\alpha$ (attenuation corrected, left panel) and of the EW(Pa$\alpha$) (right panel). The best linear fits, in log--log scale, and  the best constant value fits  for regions with luminosity and EW above the stochastic IMF sampling limit (vertical dot--dash black lines) are shown on the plots as continuous and dashed blue lines, respectively; the value of $b$ in this regime is practically independent of either variable.} 
\label{fig:b_vs_Pa}
\end{figure}

The mostly downward scatter of the points in Figure~\ref{fig:b_vs_Pa} for low P$\alpha$ luminosity or EW is consistent with stochastic IMF sampling. The quantity $\Sigma(H\alpha)_{corr})$-$\Sigma(H\alpha)$ traces the gas ionized by the dust--attenuated massive stars while $\Sigma(24)\approx \Sigma(UV)$, i.e., the UV emission from those stars (under assumption of foreground dust geometry). The O--type stars producing ionizing photons are more sensitive to the effects of stochastic sampling of the IMF than the B--type stars that contribute to the UV emission,  because they are less numerous. This also means that high values of H$\alpha$ emission (or of its dust--absorbed component) will be less common than low values, producing the type of downward scatter of points seen in Figure~\ref{fig:b_vs_Pa} \citep[resembling Figure~1 of ][]{Fumagalli+2011}. Conversely, the independence, within the uncertainties, of $b$ on both $\Sigma(Pa\alpha_{corr}$) and EW(P$\alpha$) for values above the stochastic IMF sampling limit lends support to our choice of a stellar mass M=3,000~M$_{\odot}$ to set this limit.

\section{Discussion} \label{sec:discussion}

\subsection{Characteristics of the HII regions}

Our aim has been to isolate HII regions or small clusters of HII regions in NGC\,628 which are relatively uniform in terms of age. We have done this by requiring that the selected $\lesssim$120~pc--size regions are dominated by one compact nebular emitting source. In this section, we further investigate the properties of these regions.

The EW(Pa$\alpha$) shows a clear trend as a function of the attenuation--corrected Pa$\alpha$ luminosity (left panel of Figure~\ref{fig:corr_lum}), and the models introduced in the previous section and in Table~\ref{tab:ew_model} attempt to reproduce that trend. The main conclusion from these models is that, although there are areas where both the aging model and the decreasing--mass model are consistent with the data, only the model with constant age and decreasing HII region mass closely follows the trend marked by the data (red solid line and model \# 2 in Table ~\ref{tab:ew_model}). In other words, the observed trend of EW(Pa$\alpha$) with Pa$\alpha$ luminosity is well accounted for by constant--age HII regions whose intensity (and mass) is not constant, but varies relative to the intensity (mass) of the surrounding stellar population. The scatter of about an order of magnitude in the data relative to the mean trend is consistent with the observed variations in stellar background levels across the footprint of the galaxy imaged by MIRI. The age that best accounts for the data is $\approx$3~Myr, although different ages can be accommodated by the scatter in the data; older ages, for instance, will generate smaller EWs (Table~\ref{tab:Paa} and model \# 3 in Table~\ref{tab:ew_model}), and can easily be masked by younger sources with higher stellar backgrounds. We may also have cases where multiple ages (multiple clusters) are present in a single region, but the luminosities of the older--age systems will be sub--dominant relative to the youngest ones. The regions that are brighter than the stochastic IMF sampling limit are likely dominated by sources younger than $\sim$5~Myr, as demonstrated by the location of the 6~Myr model (dot-dash green line). Thus, the original selection of Pa$\alpha$--compact regions has netted a sample of truly young HII regions, in agreement with \citet{Whitmore+2011}. 

Our sample regions display an interesting characteristic: we do not find luminous Pa$\alpha$ regions that have low values of E(B$-$V) (Figure~\ref{fig:corr_lum}, right panel), and the higher the luminosity the higher the minimum color excess we derive. This trend is physical, because no selection effect can explain it: low--E(B$-$V), luminous HII regions will be luminous at all wavelengths and will not be missed by our visual inspection and our selection criteria. 
We note that while we cannot find low--attenuation bright HII regions, fainter regions are found with a wide range of E(B$-$V) values. At faint Pa$\alpha$ luminosities, Log(Pa$\alpha_{corr}$)$<$36.5, the values of E(B$-$V) appear to bifurcate; at this stage, we cannot establish whether the bifurcation is due to incompleteness in our visual selection or other effects. However this will not affect our results which concentrate on brighter regions. A trend of higher attenuations for higher Pa$\alpha$ surface brightnesses was found by \citet{Calzetti+2007} for their 180  $\sim$0.5--kpc--size star--forming regions in a sample of 23 $\sim$solar--metallicity local galaxies. Conversely, \citet{Emsellem+2022} do not find a clear correlation between E(B$-$V) and the observed H$\alpha$ surface brightness in their PHANGS--MUSE sample, but this result may be affected by the high sensitivity of the H$\alpha$ emission to dust attenuation. Similarly to our result,  trends between E(B$-$V) and SFRs are also commonly found in galaxies, with explicit relations published by several authors \citep[e.g.,][]{Hopkins+2001, Calzetti2001, Garn+2010, Xiao+2012}. Although these relations are often described as a consequence of the mass--metallicity relation (SFR$\propto$mass and color excess$\propto$ metallicity), \citet{Xiao+2012} still find positive E(B$-$V)--SFR relations after dividing their sample of star forming galaxies in bins of constant metallicity. Figure~\ref{fig:corr_lum} (right panel) shows the trends marked by the relation derived by \citet{Calzetti+2007} for their $\sim$solar metallicity star--forming regions and the relation by \citet{Garn+2010} for galaxies, after converting observed H$\alpha$ luminosity to attenuation--corrected Pa$\alpha$ luminosity for the latter. For the relation by \citet{Calzetti+2007}, both the mean trend (continuous blue line) and the region enclosing 90\% of their data points (dash blue lines) are shown. The spread in both the data from this work and the relations from the literature indicates that the correlation between E(B$-$V) and nebular line luminosity is weak. 

We observe a mild trend for increasing E(B$-$V) with increasing $\Sigma(24)/\Sigma(Pa\alpha_{corr})$ (or, equivalently, L(21)/L(Pa$\alpha_{corr}$)), which is preserved when binning either along the x--axis or the y--axis (Figure~\ref{fig:ebv_vs_l24ratio}). Dustier regions are expected to have both higher  E(B$-$V) and mid--IR/line ratio values, but the scatter in the observed trend is larger than the trend itself. This may be due to the different nature of the two dust tracers. E(B$-$V) traces the foreground dust absorption in the nebular gas, while the mid--IR emission traces the dust emission associated with both the ionizing photons and the non--ionizing stellar continuum and is generally dominated by the latter. Thus, small variations in the dust attenuation of the gas and the stellar continuum, as well as complex geometries for the distribution of the dust relative to both the gas and the stars, will affect and increase the scatter of the mid--IR/line intensity ratio. Furthermore, the L(21)/L(Pa$\alpha_{corr}$) can be affected by small variations in the ages of the regions: over the first 5~Myr of life of an HII region, the nebular line emission decreases by almost a factor of 7 (Table~\ref{tab:Paa}), while the UV--optical stellar continuum only changes by a factor of 2.1 at most. This translates into a scatter of 0.5~dex in Log[$\Sigma(24)/\Sigma(Pa\alpha_{corr})$], perfectly adequate to conceal an intrinsic relation between this quantity and E(B$-$V).

In summary, the 120~pc--size regions identified in this work are consistent with being dominated by young (age$\approx$3~Myr and $<$5~Myr) line--emitting stellar populations surrounded by stellar populations that contribute to the continuum but not to the line emission, and are, therefore, older than the primary line--emitting population.

\subsection{Calibration of L(24) as a SFR tracer in NGC\,628}

An HII region powered by an instantaneous--burst stellar population has, by definition, SFR=0. While the size of our regions, $\sim$120~pc, is larger than the typical HII region and may contain more than one such regions, the area is still sufficiently small that single--age populations may dominate the emission, as discussed in the previous sub--section. Thus, the SFR calibrations presented here should be taken with caution.

The conversion from surface brightness to luminosity is trivial in a single galaxy analysis, and this sub--section will revert to luminosities in order to present SFR calibrations. We assume that the attenuation--corrected Pa$\alpha$ luminosity is a tracer of SFR via:
\begin{equation}
SFR(line) (M_{\odot} yr^{-1}) = 5.45\times 10^{-42} L(H\alpha) = 4.26\times 10^{-41} L(Pa\alpha),
\label{equa:SFRHa}
\end{equation}
appropriate for our choice of stellar IMF \citep{Calzetti+2010}, and with all luminosities expressed in units of erg~s$^{-1}$. With this, we can re--write equation~\ref{equa:bestfit} as:
\begin{equation}
SFR(24) (M_{\odot} yr^{-1}) = 3.06^{+3.04}_{-1.91}\times 10^{-40} L(24)^{0.9318\pm 0.0087} \ \ \ \ \ \ \ \ \ \ \ \ \ \ \ \ \ \ \ \ \ \ for \ \ \ \ \ 10^{38}\lesssim L(24) \lesssim 10^{41},
\label{equa:SFR24}
\end{equation}
and the hybrid optical--IR calibration from equation~\ref{equa:mix} as:
\begin{equation}
SFR(H\alpha + 24) (M_{\odot} yr^{-1}) = 5.35\times 10^{-42} [L(H\alpha) + (0.095\pm0.007) L(24)].
\label{equa:SFRmix}
\end{equation}

The equations above assume that  the 120~pc regions capture all the light in each waveband. This is likely a correct assumption for the mid--IR (e.g., Lawton+2010), but incorrect for the nebular emission, as HII regions are known to leak between 1/3 and 1/2 of the ionizing photons they produce into the interstellar medium out to about 1~kpc distance \citep{Ferguson+1996, Oey+2007,Pellegrini+2012,DellaBruna+2021}. Assuming the most extreme case that about 1/2 of the ionizing photons leak out of HII regions and are not recovered by our 120~pc photometric apertures, we obtain the following updated coefficients for our equations: the proportionality factor in equation~\ref{equa:SFR24} changes from $3.060\times 10^{-40} $ to $6.120\times 10^{-40} $ and $b$ in equation~\ref{equa:SFRmix} changes from $(0.095\pm0.007)$ to $(0.19\pm0.01)$. 

The same equations as \ref{equa:SFR24} and \ref{equa:SFRmix} expressed for the 21~$\mu$m emission are:
\begin{equation}
SFR(21) (M_{\odot} yr^{-1}) = 2.71^{+2.69}_{-1.69}\times 10^{-40} L(21)^{0.9318\pm 0.0087} \ \ \ \ \ \ \ \ \ \ \ \ \ \ \ \ \ \ \ \ \ \ for \ \ \ \ \ 10^{38}\lesssim L(21) \lesssim 10^{41},
\label{equa:SFR21}
\end{equation}
and:
\begin{equation}
SFR(H\alpha + 21) (M_{\odot} yr^{-1}) = 5.35\times 10^{-42} [L(H\alpha) + (0.083\pm0.006) L(21)].
\label{equa:SFR21mix}
\end{equation}

\subsection{Comparison with Previous Work}

The calibration of the 24~$\mu$m emission as a SFR indicator has a long history, with many such calibrations available in the literature. Although the ones quoted here are mainly from observations with the {\em Spitzer Space Telescope}, the calibration of monochromatic SFR indicators dates back to the {\em Infrared Space Observatory} \citep[e.g.,][]{Roussel+2001, Forster+2004, Boselli+2004}. Figure~\ref{fig:sfr24} compares the calibration derived in this work from the MIRI data of a single galaxy, NGC\,628 (equation~\ref{equa:SFR24}) with the HII regions calibrations from Spitzer published by several authors, usually from samples of several or many galaxies \citep{Perez+2006, Alonso+2006, Calzetti+2007, Relano+2007}. All calibrations are rescaled to a common stellar IMF, as discussed in \citet{Calzetti+2010}. All HII region calibrations are sub--linear in log--log scale, with slopes in the range 0.768--0.885; for comparison, we find a slope of 0.932. The calibrations are relatively consistent among themselves, within a factor $\sim$2.5, in the luminosity range spanned by L(24) for HII regions in this work ($\sim$10$^{38 - 41}$~erg~s$^{-1}$). The main exception is the calibration we obtain when correcting for leakage of ionizing photons out of HII regions; this pushes the overall relation to higher values than the other calibrations (solid grey line in Figure~\ref{fig:sfr24}), making it discrepant with other calibrations across most of the 24~$\mu$m luminosity range; however, we consider this calibration as an extreme case, as stated in the previous sub--section. Previous works do not correct for the effects of leakage of ionizing photons out of HII regions, but these works, based on Spitzer data, generally consider physically larger regions  ($\gtrsim$0.5~kpc) than those analyzed here, and are therefore less affected by leakage.

In the luminosity range of galaxies, L(24)$\gtrsim$10$^{40}$~erg~s$^{-1}$, the non--linear HII region calibrations remain consistent among themselves still within a factor $\sim$2.5, except for the one by \citet{Perez+2006} (solid cyan line in Figure~\ref{fig:sfr24}), which has been derived only for L(24)$<$10$^{42}$~erg~s$^{-1}$ and under--predicts the SFR above this luminosity value. 

Figure~\ref{fig:sfr24} also shows published calibrations for galaxies, both non--linear \citep{Wu+2005, Zhu+2008} and linear \citep{Wu+2005, Zhu+2008, Rieke+2009}, although it should be noted that the galaxy calibrations are extrapolated to several orders of magnitude lower luminosities than the range where they have been derived, typically L(24)$\sim$10$^{41}$ -- 3$\times$10$^{44}$~erg~s$^{-1}$ . At low luminosities, the galaxy linear calibrations are more discrepant than the non--linear ones with the HII region calibrations, being a factor $\sim$4 lower in predicted SFR at L(24)$\sim$10$^{38}$~erg~s$^{-1}$.  However, both linear and non--linear SFR(24) calibrations for galaxies agree with the HII region calibration, again within a factor $\sim$2.5, for L(24)$>$10$^{41.5}$~erg~s$^{-1}$. 

In summary, non--linear calibrations between SFR and L(24) remain consistent with each other over most of the luminosity range of HII regions and galaxies in the local universe and can be used, within an accuracy of $\sim$2.5X, across the full range L(24)$\sim$10$^{38 - 44}$~erg~s$^{-1}$.  Linear calibrations, derived for galaxies, are also consistent with each other, and become consistent with the non--linear ones in the regime of galaxy luminosities. 
 The observed variations in slope and intercept between different non--linear calibrations of the SFR--L(24) relation are briefly discussed in the next subsection.

Calibrations of hybrid SFR indicators involving nebular lines and mid--IR emission were obtained by \citet{Calzetti+2007} \citep[who expanded on the calibration by][]{Kennicutt+2007} for galaxy regions $\sim$500~pc across and by \citet{Kennicutt+2009} for entire galaxies, obtaining $b=0.031\pm0.006$ and $b=0.020\pm0.005$, respectively. The samples those authors analyze are dominated by metal--rich galaxies ($\sim$solar metallicity), similar to NGC\,628. In both cases, the calibration parameters these authors obtain are between a factor 3 and 4.8 lower than the one derived in this work. This discrepancy is discussed in the next subsection.

\citet{Belfiore+2023} derive several calibrations for hybrid optical+mid--IR tracers, using JWST mid--IR observations in the wavelength range 3.3--21~$\mu$m; they determine that the proportionality factor $b$ in equation~\ref{equa:mixsigma} (and in similar equations with other mid--IR bands) is a function of EW(H$\alpha$), becoming constant at high EWs. We obtain a similar result for our uncensored data (Figure~\ref{fig:b_vs_Pa}, right panel), although we only analyze the data above the stochastic IMF sampling limit. The median value of $b$ derived by \citet{Belfiore+2023} ranges between 0.031 and 0.051 for L(21), corresponding to 0.035 and 0.058 for L(24), lower than our value of 0.095. However, at least part of the difference may be due to our approach of only fitting the censored data. When using  those authors' high EW(H$\alpha$) asymptote, which is $\sim$0.2~dex above the median, their background--subtracted regions yield b$\sim$0.08--0.09 for L(24), which is closer to our value. The residual difference could be due to those authors' use of  H$\alpha$/H$\beta$ to perform attenuation corrections, which may be insufficient to recover the intrinsic value of the nebular line intensity and lead to an underestimate of $b$. The sample of \citet{Belfiore+2023} includes galaxies with a range of metallicities, but lower--than--solar metallicities increase $b$, since metal--poor regions will have proportionally fainter L(24), thus metallicity effects cannot cause the observed discrepancy. Leakage of ionizing photons may or may not contribute to the difference: the regions analyzed by \citet{Belfiore+2023} are reported to be around 100~pc, thus slightly smaller than ours; leakage of ionizing photons may impact their nebular line photometry a little more than in our case leading to further underestimation of $b$. In what follows, we will include the higher of the two values reported by \citet{Belfiore+2023}, corresponding to the median value of their background--subtracted regions. 

\begin{figure}
\plottwo{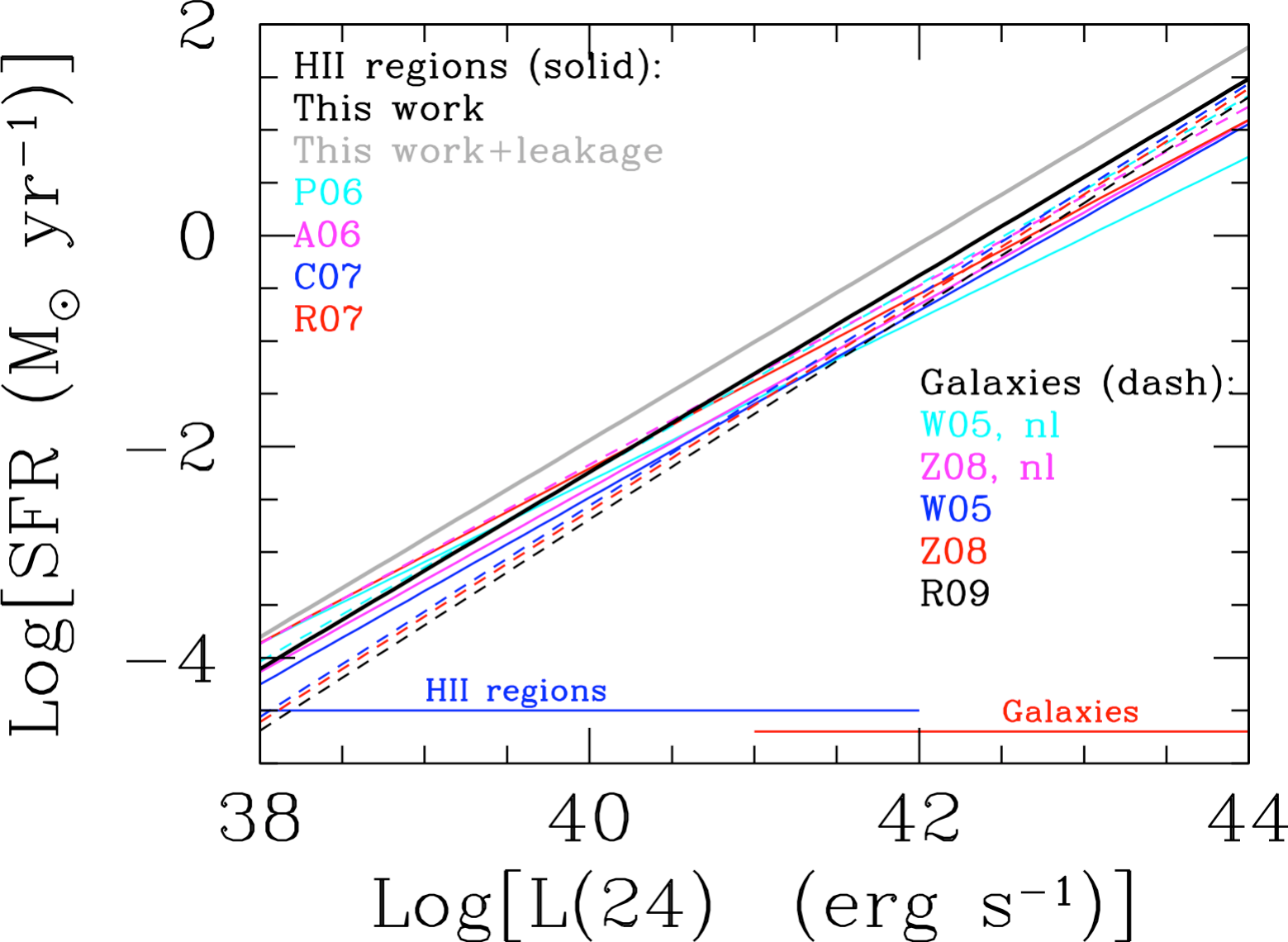}{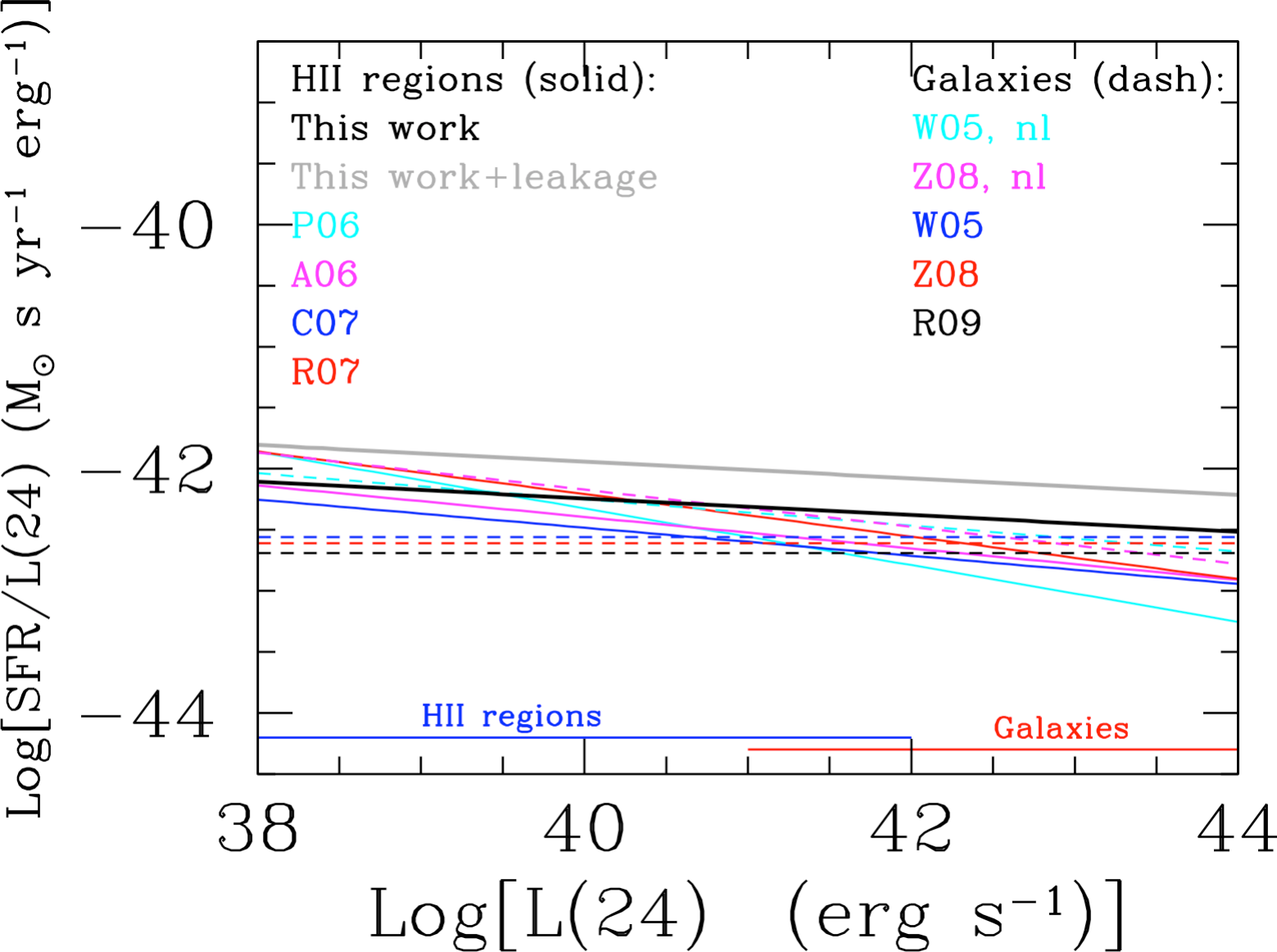}
\caption{{\bf (Left):} A collection of calibrations between SFR and the luminosity at 24~$\mu$m from the {\em Spitzer Space Telescope} derived by several authors both for HII regions and galaxies is shown in comparison with the current calibration for NGC\,628 (equation~\ref{equa:SFR24}, black solid line;  grey solid line for the calibration corrected for ionizing photon leakage). For HII regions/starbursts (all solid lines): \citet[][P06, cyan]{Perez+2006}; \citet[][A06, magenta]{Alonso+2006}; \citet[][C07, blue]{Calzetti+2007}; \citet[][R07, red]{Relano+2007}. All relations for HII regions/starbursts are non--linear, and with a slope$<$1 in log--log scale. For star forming galaxies (all dash lines): non--linear and linear relations from  \citet[][W05 nl, cyan, and W05, blue, respectively,]{Wu+2005}; non--linear and linear relations from  \citet[][Z08 nl, magenta, and Z08, red,respectively,]{Zhu+2008}; the linear relation from \citet[][R09, black]{Rieke+2009}.  The horizontal blue and red bars mark the approximate range in luminosity of HII regions and galaxies, respectively. {\bf (Right):} The same as the left panel, but using the SFR normalized by L(24) on the vertical axis, to emphasize both similarities and disagreements between the different calibrations. All relations for galaxies are extrapolated to much lower luminosities than their calibration range.}
 \label{fig:sfr24}
\end{figure}

\subsection{The Role of the Dust--Heating Stellar Population in L(24)}

As already mentioned in the Introduction, the 24~$\mu$m (and 21~$\mu$m) emission is due to both stochastic and thermal dust heating by UV and optical photons, implying that old stellar populations also contribute to the heating of the dust \citep{Greenberg1968,Sellgren+1983,Leger+1984,Desert+1990,Draine+2001,DraineLi2007, Smith+2007, Galliano+2018, Draine+2021}. In addition, since star formation is hierarchically clustered in galaxies and larger spatial scales correspond to older mean ages of the stellar populations \citep[e.g.,][]{Efremov+1998,Delafuente+2009,Elmegreen+2010,Grasha+2017}, galaxy regions of increasing sizes, when locally background--subtracted, will capture increasingly older mixes of stellar populations, all contributing in different ways to the dust emission. 

This concept was used by \citet{Li+2013} to model the decreasing SFR calibration coefficient at 70~$\mu$m for increasing region size: in small  galactic regions, one expects the 70~$\mu$m emission to be contributed mainly by recently formed stars, with the contribution from progressively older stellar populations increasing for increasing galactic region size; with this model, the smaller SFR(70) calibration coefficient at larger scales is an outcome of the smaller fraction of 70~$\mu$m emission due to recent star formation in larger regions. 

We expand and further develop the model from \citet{Li+2013}, to attempt a description of the trend in $b$, which decreases for increasing spatial scale, as stated in the previous sub--section, from $\sim$0.1 at 120~pc scales \citep[$\sim$0.06 at 100~pc in the derivation of][]{Belfiore+2023}  to $\sim$0.03 at $\sim$500~pc scales to $\sim$0.02 at galaxy scales \citep{Calzetti+2007, Kennicutt+2009}. For the purpose of our toy model, we set:
\begin{equation}
L(H\alpha_{corr})-L(H\alpha) = L(H\alpha_{corr}) 10^{-0.4 E(B-V) \kappa(H\alpha)},
\label{equa:first_term}
\end{equation}
and:
\begin{equation}
L(24) = f L(IR) = f L_{bol} 10^{-0.4 E(B-V) \kappa_{eff}},
\label{equa:second_term}
\end{equation}
where E(B$-$V) is the dust column density of the region, under the simplistic assumption of foreground dust; $\kappa(H\alpha)$ and $\kappa_{eff}$ are the attenuation curve at H$\alpha$ and its effective value across the entire stellar wavelength range; $f$ is the fraction of the bolometric IR emission that is captured at 24~$\mu$m; L$_{bol}$ is the stellar bolometric emission. Technically, equation~\ref{equa:second_term} should be the integral of the luminosity density multiplied by the attenuation as a function of wavelength; we use the central limit theorem to replace this integral with the bolometric luminosity multiplied by the effective value of the attenuation. We obtain L(H$\alpha_{corr}$) and L$_{bol}$ from the Starburst99 models for both an instantaneous burst and a constant SFR population \citep{Leitherer+1999}. We adopt $f=0.14$ from the median value of L(24)/L(IR) reported in \citet{Calzetti+2010} for high--SFR galaxies; this median value has a 1~$\sigma$ scatter of at least 0.1~dex. 

We are interested in the ratio of equation~\ref{equa:first_term} to equation~\ref{equa:second_term}, meaning that, while we don't know the exact values of E(B$-$V), $\kappa(H\alpha)$ and $\kappa_{eff}$, we can use other information to constrain the ratios of the attenuation terms in those two equations. We note that star--forming galaxies have been found to have A$_V\sim$1~mag \citep{Kennicutt+1983}, corresponding to about 1/2 of the luminosity at the H$\alpha$ wavelength to be absorbed by dust. We also know that about 1/3--to--1/2 of the bolometric luminosity of local galaxies is absorbed by dust and re--emitted in the infrared \citep[e.g.,][]{Dale+2007}. These considerations allow us to adopt:
\begin{equation}
{10^{-0.4 E(B-V) \kappa(H\alpha)} \over 10^{-0.4 E(B-V) \kappa_{eff}}} \sim 1.
\end{equation}
With this, the ratio of equation~\ref{equa:first_term} to equation~\ref{equa:second_term} simplifies to (see, also, equation~\ref{equa:b_definition}):
\begin{equation}
b = {L(H\alpha_{corr})-L(H\alpha) \over L(24)} \sim {L(H\alpha_{corr}) \over f L_{bol}},
\label{equa:b_mod}
\end{equation}
which is plotted in Figure~\ref{fig:b_model}. The $b$ values derived in this work, \citet{Calzetti+2007} and \citet{Kennicutt+2009} mark a sequence for increasing duration of the star formation with increasing spatial scale. The value $b=0.095$ derived in this work is consistent with either constant or instantaneous star formation for a few$\times$10$^6$~yr; the $b\sim0.03$ from \citet{Calzetti+2007} and $b\sim0.02$ from \citet{Kennicutt+2009} are consistent with constant star formation over timescales of a few$\times$10$^8$~yr and $\approx$10$^{10}$~yr, respectively. These durations are not inconsistent with the spatial scales sampled by those authors. For comparison, \citet{Efremov+1998} and \citet{Grasha+2017} find that typical age separations for star clusters are around 10~Myr and 100~Myr over scales of 10--20~pc and 0.5~kpc, respectively, indicating a trend in the same direction as the one found here. Thus, the decreasing $b$ values  for increasing spatial scale are consistent with the dust being heated by progressively older stellar populations. 

The value $b\sim0.06$ by \citet{Belfiore+2023} is consistent with either an instantaneous, $\sim$3$\times$10$^6$~yr, burst or constant star formation over $\sim$10$^7$~yr in duration. However, neither option is consistent with the relatively small EW(H$\alpha$), with mean value $\sim$30~\AA, these authors recover. This, together with the small spatial regions sampled by the authors which are likely to include a small number of star clusters per region, reinforces the possibility that the inclusion of HII regions heavily affected by stochastic sampling of the IMF has impacted the results by \citet{Belfiore+2023}, by having included low--mass young objects with H$\alpha$ suppressed by their relative paucity of massive stars. Overall, the $b$ value derived by \citet{Belfiore+2023} is out of order relative to the age--size trend, and is highlighted in Figure~\ref{fig:b_model} with a left--pointing arrow. It is also important to consider that the  results for the larger regions are affected by the adopted star formation history: instantaneous bursts or even constant star formation are dramatic simplifications, as present day star--forming galaxies show evidence for decreasing SFR over cosmic times. Thus the values in Figure~\ref{fig:b_model} that map the correspondence between region size and duration of the star formation should be considered indicative only, and within the context of this toy model.  

It is tempting to try to interpret the observed variations in the non--linear SFR--L(24) relation, specifically slope and intercept, in light of the model above. However, we do not observe a systematic trend between region size and change in the slope and/or intercept for this relation. This may be due to the fact that the relation includes two effects: (1) the contribution of the diffuse, stochastically heated, dust emission to the L(24) luminosity discussed above; and (2) the decrease in the mean dust attenuation for decreasing luminosity, i.e., SFR, discussed in section~\ref{sec:results}. In the hybrid calibration (equation~\ref{equa:mixsigma}), the second effect is captured by the observed L(H$\alpha$), while in the SFR--L(24) calibration this effect is captured by the non--linearity between the two quantities, as already remarked by  \citet{Kennicutt+2009}. The two effects are both present and are not easy to separate, which may explain the lack of a systematic trend with region size for the SFR--L(24) relation.

\begin{figure}
\plotone{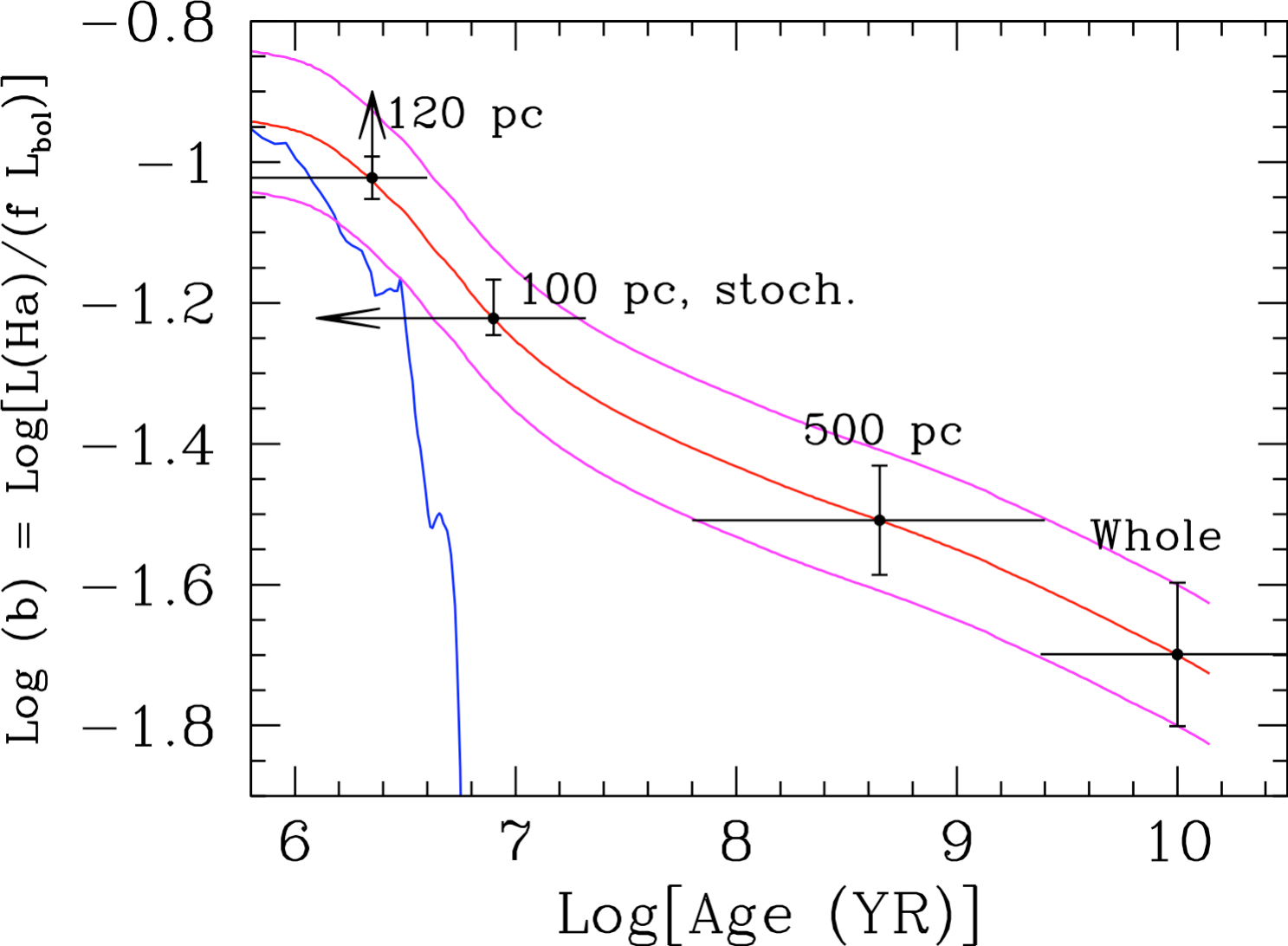}
\caption{ The model ratio L(H$\alpha_{corr}$)/[f L$_{bol}$], which is approximately equal to $b$, as a function of age/duration for the two cases of instantaneous burst population (blue line) and constant star formation (red line). Uncertainties in $f$, the fraction of L(IR) that is captured at 24~$\mu$m (see text), are shown as magenta lines for the constant SFR case only. The values of $b\sim$L(H$\alpha_{corr}$)/[f L$_{bol}$] derived from data are shown as representative points on the constant star formation model with their 1~$\sigma$ uncertainties, for the following results:  from this work (marked `120~pc'), from \citet{Belfiore+2023} (marked `100~pc, stoch' and with a left--pointing arrow to remind that the value of $b$ includes regions that are below the stochastic IMF sampling limit), from \citet{Calzetti+2007} (marked `500~pc') and from \citet{Kennicutt+2009} (for whole galaxies, marked `Whole'). The location of the $b$ values along the horizontal axis is forced to be within the range marked by the constant star formation model, to show the approximate age range each region size ($b$--value) corresponds to. The horizontal black lines span the full age range of the models at the level of the observational $b$ values. The upward arrow at the smallest scale indicates the direction of the data if leakage of ionizing photons in these regions is corrected for.}
 \label{fig:b_model}
\end{figure}

\newpage

\section{Summary and Conclusions} \label{sec:conclusions}
We have combined new JWST/NIRCam observations in the light of the Pa$\alpha$ nebular line for the nearby galaxy NGC\,628 with archival JWST mosaics in the MIRI/21~$\mu$m band and archival HST images in the light of the H$\alpha$ nebular line, to isolate line--emitting and dust--emitting regions at the scale of 120~pc. We identify 143 such regions, 111 of which are luminous enough to be considered above the limit for which stochastic sampling of the stellar IMF is a concern; we term these 111 sources our `censored data'. We calibrate L(21) (converted to the Spitzer L(24)) as a SFR indicator, using the dust-attenuation--corrected P$\alpha$ luminosity as a reference SFR; we find that the relation between L(24) and L(P$\alpha_{corr}$) is non--linear for the censored data, with an exponent$\sim$1.07, although the scatter of the data around the mean trend is large, $\approx$0.3~dex. The non--linear trend is consistent with what previous authors have found with Spitzer on larger--scale regions, and indicates that fainter regions are proportionally less dust attenuated or have a smaller fraction of ionizing photons directly absorbed by dust than brighter regions. 

The above results translate into a relation SFR$\propto$L(24)$^{\alpha}$ with the exponent $\alpha<1$. We determine the exponent to be 0.9318$\pm$0.087 for the HII regions in NGC\,628, with previous results in the range 0.768--0.885 for HII regions across a wide variety of galaxies. We find that the non--linear calibrations of SFR as a function of L(24) agree with each other within a factor $\sim$2.5 across about 6 orders of magnitude in luminosity, and agree with both linear and non--linear calibrations for galaxies in the luminosity range of the latter. Thus, we conclude that deriving SFRs using L(24) with non--linear calibrations provides robust results across the full range of luminosities for HII regions and galaxies. 

We also present a calibration for the hybrid [L(H$\alpha$)+b L(24)] SFR indicator at the same spatial scale of 120~pc, finding that the proportionality constant $b$ is larger by a factor between 3 and $\sim$5 than previous derivations at larger spatial scales. We model these results as an effect of the increasing mean age of the stellar population contributing to the 24~$\mu$m emission for increasing spatial scales, finding that at larger scales the dust is heated by older stellar populations than at small scales. While this is not a new result, as prior investigations of this effect date back to IRAS observations, this work provides a quantification of its effect on the hybrid SFR indicators and a possible interpretation in light of the hierarchical distribution of star formation in galaxies. Thus, the use of hybrid SFR indicators requires prior knowledge of or an educated guess on the mean age of the stellar population dominating the dust heating. 

Based on the results summarized above, the main recommendation from this study is to derive SFRs from L(24) alone using non--linear calibrations. However, this investigation is limited to the HII regions in the  one galaxy for which we have been able to secure the three key observations required for the analysis: P$\alpha$, H$\alpha$ and 21~$\mu$m emission. As the JWST and HST archives continue to be populated with new observations, this type of study can be extended to other galaxies to test the interpretation presented in this work. 

\begin{acknowledgments}

The authors would like to thank the anonymous referee, whose comments have improved the content and presentation of this paper.

This work is based in part on observations made with the NASA/ESA/CSA James Webb Space Telescope, which is operated by the Association of Universities 
for Research in Astronomy, Inc., under NASA contract NAS 5-03127. 
These observations are associated with program \# 1783. Support for program \# 1783 was provided by NASA through a grant from the Space Telescope Science Institute, 
which is operated by the Association of Universities for Research in Astronomy, Inc., under NASA contract NAS 5-03127.

The authors acknowledge the team of the `JWST-HST-VLT/MUSE-ALMA Treasury of Star Formation in Nearby Galaxies',  led by coPIs Lee, Larson, Leroy, Sandstrom, Schinnerer, and Thilker, 
for developing the JWST observing program \# 2107 with a zero-exclusive-access period.

Based also on observations made with the NASA/ESA Hubble Space Telescope, and obtained from the Hubble Legacy Archive, 
which is a collaboration between the Space Telescope Science Institute (STScI/NASA), the Space Telescope European Coordinating 
Facility (ST-ECF/ESA) and the Canadian Astronomy Data Centre (CADC/NRC/CSA).

Archival data presented in this paper were obtained from the Mikulski Archive for Space Telescopes (MAST)
at the Space Telescope Science Institute. The specific observations analyzed can be accessed via 
\dataset[10.17909/q3rh-mk45] .

This research has made use of the NASA/IPAC Extragalactic Database (NED) which is operated by the Jet
Propulsion Laboratory, California Institute of Technology, under contract with the National Aeronautics and Space
Administration.

KG is supported by the Australian Research Council through the Discovery Early Career Researcher Award (DECRA) Fellowship (project number DE220100766) funded by the Australian Government and by the Australian Research Council Centre of Excellence for All Sky Astrophysics in 3 Dimensions (ASTRO~3D), through project number CE170100013. 

AA and AP acknowledge support from the Swedish National Space Agency (SNSA) through the grant 2021- 00108.

MRK acknowledges support from the Australian Research Council through Laureate Fellowship FL220100020.

MM acknowledges the financial support through grant PRIN-MIUR 2020SKSTHZ. 

\end{acknowledgments}

\vspace{5mm}
\facilities{James Webb Space Telescope (NIRCam, MIRI)}
\software{JWST Calibration Pipeline \citep[][]{Bushouse+2022, Greenfield+2016}, Drizzlepac \citep[][and the STSCI Development Team]{Gonzaga+2012}, IRAF \citep{Tody1986, Tody1993}, SAOImage DS9 \citep{Joye+2003}, Fortran.}

\appendix
\section{Impact of the galaxy's diffuse emission on source photometry}\label{sec:appendixA}

We investigate the impact  of the local background removal, which we have adopted as the default approach in this paper, on the photometry of the sources in the 21~$\mu$m mosaic. For this analysis, we use the mosaic released by the 
PHANGS--JWST collaboration, which has been processed through a custom pipeline to remove the sky background \citep{Williams+2024} for the project's specific observing approach \citep{Lee+2023}. 
Thus, we assume that any smooth emission left in the mosaic is due to the galaxy's diffuse emission. Figure~\ref{fig:phot_PHANGS} (left) shows the ratio of the diffuse to the source's emission as a function 
of the source's 21~$\mu$m luminosity. The luminosity of the diffuse emission is measured in the same photometric aperture used for the sources. The vertical line in the figure marks the approximate location of the luminosity above which sources can be considered minimally affected by stochastic sampling of the stellar IMF (section~\ref{sec:results}), as inferred from Figure~\ref{fig:l21_vs_lpa}. While the galaxy's diffuse emission contributes at most 10\% of the source flux at the highest 21~$\mu$m luminosity in our sample, it represents an equal amount of flux as the source at the stochastic `boundary', with a smooth trend from high to low luminosity.

Figure~\ref{fig:phot_PHANGS} (right) shows $\Sigma(24)$ as a function of the attenuation corrected $\Sigma$(Pa$\alpha$) for the 143 regions in our sample, using on both axes surface brightnesses that include the contribution of the diffuse emission. The Pa$\alpha$ diffuse emission scatters around a constant, with both positive and negative values, and represents a small fraction of the sources luminosity. The positive and negative values are due to slight misalignments in background levels of the tiles of the FEAST mosaics, and correspond to a (peak--to--peak) 2.5$\sigma$ scatter in the background levels. In order to avoid biasing the Pa$\alpha$ surface brightness towards very low values, thus artificially flattening the $\Sigma(24)$--$\Sigma$(Pa$\alpha$) relation, we elect to set to zero negative values of the diffuse emission in the Pa$\alpha$ mosaic. This approach provides a lower limit to the impact of the diffuse emission in the mid--IR. The 24~$\mu$m surface brightness is derived from the 21~$\mu$m one as discussed in the main text. Best linear fits to the data obtained with both the bi--regression and the least square bisector algorithms yield values for the slopes that are within 1~$\sigma$ of each other, with values: $\sim$0.92 for the 117 datapoints that are above the  stochastic IMF sampling limit and $\sim$0.85 for the entire sample of 143 datapoints. Our derived slope of 0.85 for the entire sample is indistinguishable from the slope of 0.86 derived by \citet{Belfiore+2023} for their larger sample using the attenuation corrected H$\alpha$ surface brightness. This agreement lends robustness to our analysis, suggesting that even if our results are derived for a single galaxy, they may be representative of trends across many galaxies. 

\begin{figure}
\plottwo{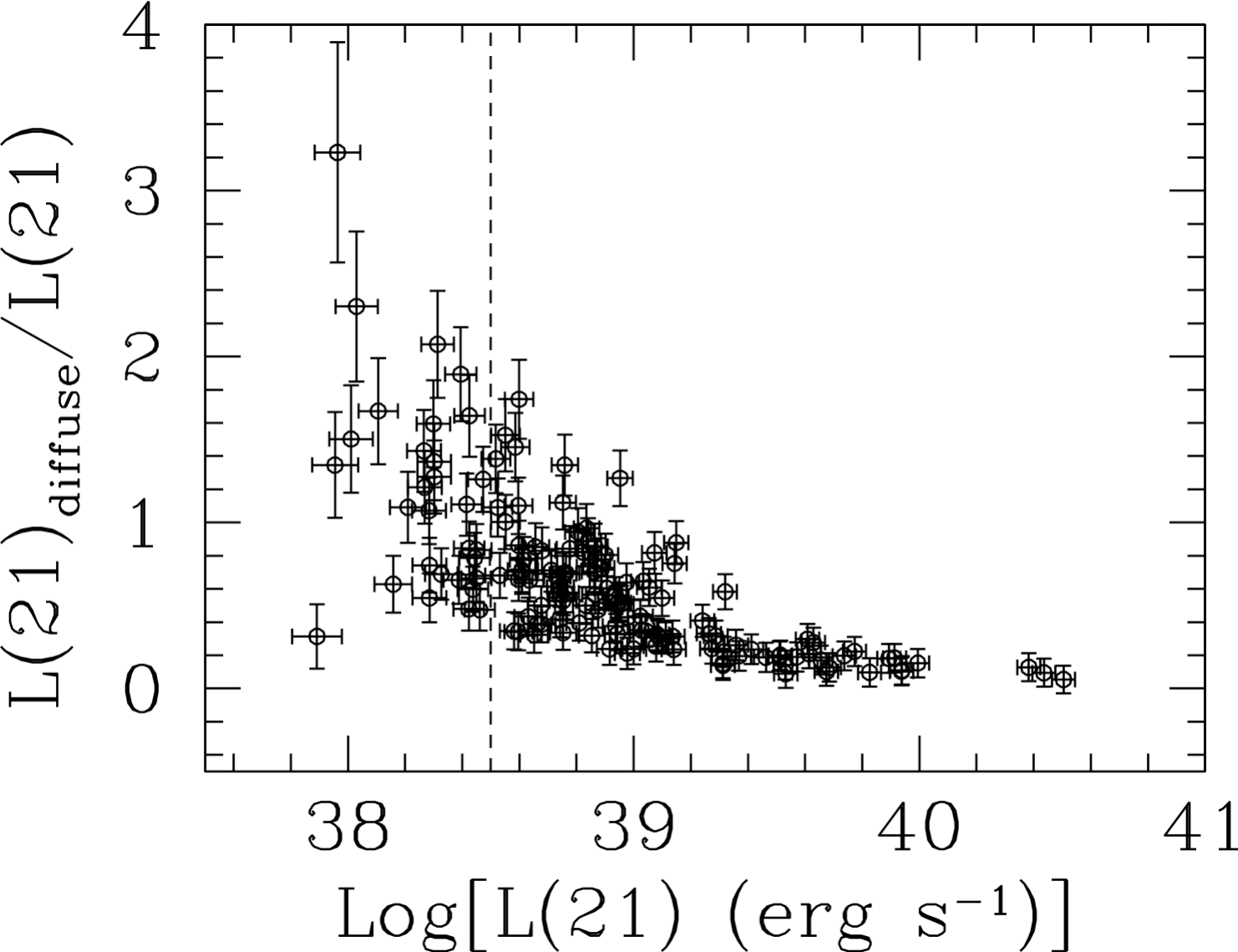}{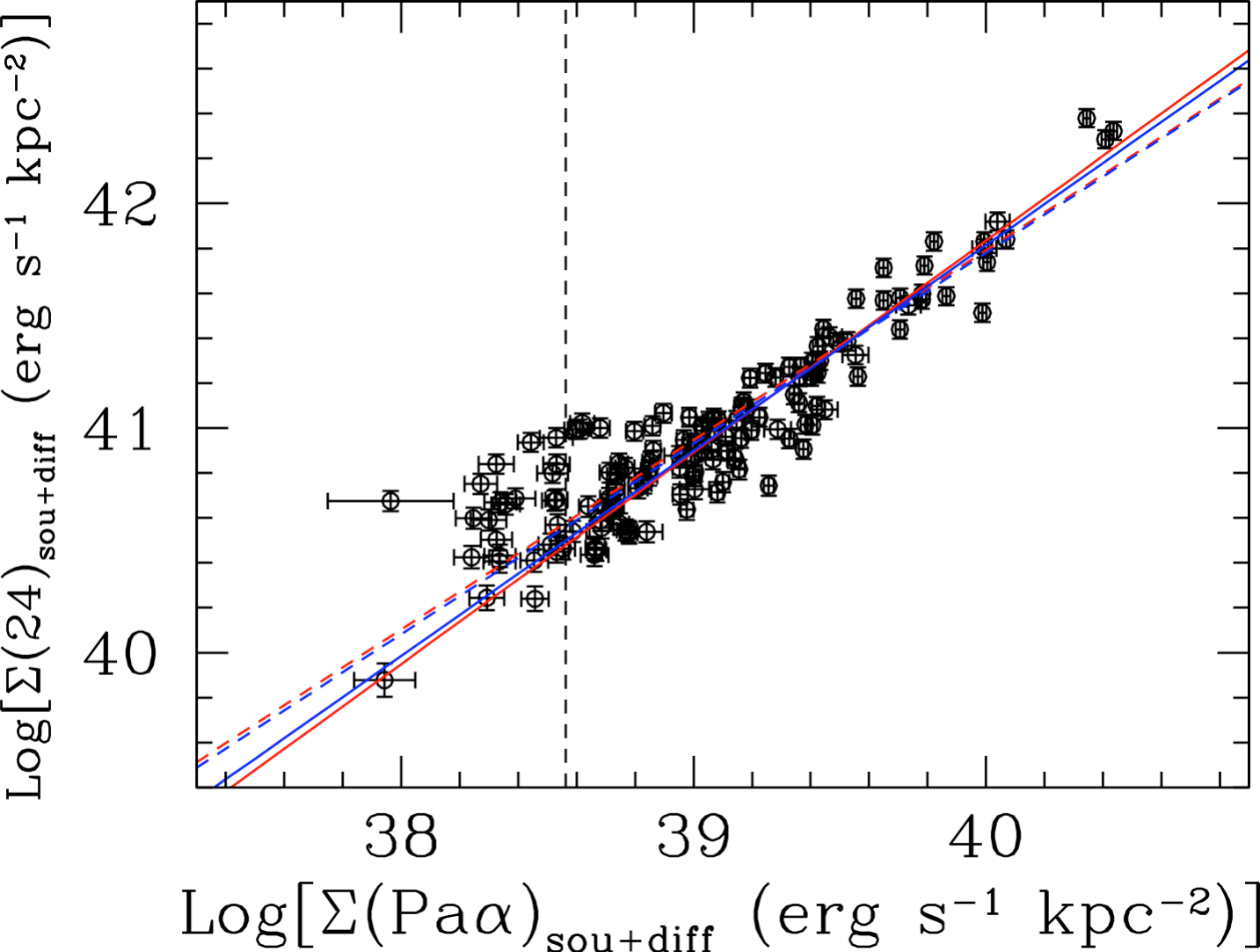}
\caption{{\bf (Left:)} The diffuse--to--source luminosity ratio at 21~$\mu$m as a function of the source luminosity L(21). The sky background has been removed from the mosaic, and only the diffuse emission attributable to the galaxy is captured in L(21)$_{diffuse}$. The source's luminosity L(21) has been corrected for the light outside the photometric aperture (see section~\ref{sec:selection}), while L(21)$_{diffuse}$ has not been corrected under the assumption that it is smoothly distributed. The vertical line marks the approximate location of the luminosity above which the effects of stochastic sampling of the stellar IMF are minimized (see text). The diffuse luminosity dominates over the source luminosity, by as much as a factor $>$3, at the low luminosities but only represents $\lesssim$10\% of the source emission at the high luminosities. {\bf (Right:)} The surface brightness at 24~$\mu$m  as a function of the attenuation--corrected surface brightness at Pa$\alpha$ for the 143 regions in our sample. Both surface brightnesses include the contribution from the galaxy's diffuse emission. The vertical line marks the value in the Pa$\alpha$ surface brightness above which the IMF stochastic sampling is mitigated. The blue line is for the best linear fit using bi--regression and red for the least square bisector; continuous lines for the censored data (117 points above the stochastic IMF sampling limit) and dash lines for the uncensored data. }
 \label{fig:phot_PHANGS}
\end{figure}

\startlongtable
\begin{deluxetable}{lrrrrrr}
\tablecolumns{7}
\tabletypesize{\small}
\tablecaption{Source Location, Luminosity and Derived Quantities \label{tab:sources}}
\tablewidth{100pt}
\tablehead{
\colhead{ID} & \colhead{RA(2000),DEC(2000)}  & \colhead{Log[L(H$\alpha$)]} & \colhead{Log[L(Pa$\alpha$)]} & \colhead{Log[L(21)]} & \colhead{Log[EW(Pa$\alpha$)]} & \colhead{E(B$-$V)} 
\\
\colhead{(1)} & \colhead{(2)} & \colhead{(3)} & \colhead{(4)} & \colhead{(5)}  & \colhead{(6)} & \colhead{(7)} 
\\
}
\startdata
\hline
    1 &1:36:47.1993, +15:45:50.101 &38.495$\pm$0.021 &37.971$\pm$0.013 &39.912$\pm$0.040 &3.077$\pm$0.018 &0.498$\pm$0.033\\
    2 &1:36:47.2742, +15:45:53.541 &37.515$\pm$0.025 &37.053$\pm$0.021 &38.936$\pm$0.044 &2.672$\pm$0.027 &0.581$\pm$0.045\\
    3 &1:36:46.9416, +15:45:47.061 &37.653$\pm$0.024 &36.979$\pm$0.023 &38.471$\pm$0.052 &2.297$\pm$0.027 &0.295$\pm$0.045\\
    4 &1:36:47.2880, +15:45:45.781 &37.546$\pm$0.025 &36.984$\pm$0.023 &38.749$\pm$0.046 &2.041$\pm$0.026 &0.447$\pm$0.046\\
    5 &1:36:47.1133, +15:45:42.021 &37.538$\pm$0.025 &36.918$\pm$0.025 &38.549$\pm$0.050 &2.253$\pm$0.028 &0.368$\pm$0.048\\
    6 &1:36:47.1881, +15:45:44.061 &37.665$\pm$0.024 &37.058$\pm$0.021 &38.597$\pm$0.049 &2.502$\pm$0.026 &0.386$\pm$0.043\\
    7 &1:36:45.5063, +15:45:47.464 &37.312$\pm$0.029 &36.949$\pm$0.024 &38.905$\pm$0.045 &2.202$\pm$0.027 &0.716$\pm$0.050\\
    8 &1:36:45.7113, +15:45:51.463 &36.598$\pm$0.064 &36.260$\pm$0.070 &38.752$\pm$0.046 &1.693$\pm$0.071 &0.750$\pm$0.128\\
    9 &1:36:45.7668, +15:46:02.023 &37.415$\pm$0.027 &36.752$\pm$0.031 &38.299$\pm$0.058 &2.272$\pm$0.034 &0.311$\pm$0.055\\
   10 &1:36:45.4842, +15:46:10.344 &37.977$\pm$0.022 &37.471$\pm$0.016 &39.273$\pm$0.042 &2.826$\pm$0.021 &0.523$\pm$0.036\\
   11 &1:36:45.1019, +15:46:27.784 &37.824$\pm$0.023 &37.069$\pm$0.021 &38.860$\pm$0.045 &2.133$\pm$0.024 &0.188$\pm$0.042\\
   12 &1:36:45.1241, +15:46:30.264 &37.683$\pm$0.024 &36.991$\pm$0.023 &38.677$\pm$0.048 &2.353$\pm$0.027 &0.271$\pm$0.044\\
   13 &1:36:45.0189, +15:46:57.944 &37.291$\pm$0.029 &36.784$\pm$0.029 &38.998$\pm$0.044 &1.958$\pm$0.032 &0.521$\pm$0.056\\
   14 &1:36:44.6032, +15:46:40.344 &38.282$\pm$0.021 &37.663$\pm$0.014 &39.509$\pm$0.041 &2.581$\pm$0.018 &0.370$\pm$0.034\\
   15 &1:36:44.4757, +15:46:32.505 &38.297$\pm$0.021 &37.849$\pm$0.013 &39.736$\pm$0.041 &2.840$\pm$0.018 &0.601$\pm$0.034\\
   16 &1:36:43.9159, +15:46:25.625 &37.968$\pm$0.022 &37.423$\pm$0.016 &39.530$\pm$0.041 &2.738$\pm$0.021 &0.470$\pm$0.037\\
   17 &1:36:44.7529, +15:47:04.344 &37.425$\pm$0.027 &36.715$\pm$0.032 &38.675$\pm$0.048 &1.914$\pm$0.035 &0.248$\pm$0.057\\
   18 &1:36:43.4227, +15:46:37.865 &37.367$\pm$0.027 &36.683$\pm$0.034 &38.440$\pm$0.053 &2.035$\pm$0.037 &0.282$\pm$0.059\\
   19 &1:36:43.4227, +15:46:35.145 &37.149$\pm$0.032 &36.357$\pm$0.058 &38.005$\pm$0.076 &1.636$\pm$0.060 &0.138$\pm$0.090\\
   20 &1:36:44.4246, +15:47:16.815 &37.685$\pm$0.024 &37.054$\pm$0.021 &38.976$\pm$0.044 &2.283$\pm$0.025 &0.354$\pm$0.043\\
   21 &1:36:44.1142, +15:47:12.015 &37.837$\pm$0.023 &37.368$\pm$0.017 &39.411$\pm$0.041 &2.456$\pm$0.020 &0.573$\pm$0.038\\
   22 &1:36:44.4080, +15:47:07.695 &37.340$\pm$0.028 &36.668$\pm$0.035 &38.956$\pm$0.044 &2.054$\pm$0.037 &0.299$\pm$0.060\\
   23 &1:36:43.1499, +15:47:13.696 &37.206$\pm$0.031 &36.716$\pm$0.032 &38.731$\pm$0.047 &1.908$\pm$0.035 &0.545$\pm$0.060\\
   24 &1:36:42.5291, +15:47:14.816 &37.534$\pm$0.025 &37.013$\pm$0.022 &39.077$\pm$0.043 &2.096$\pm$0.025 &0.502$\pm$0.045\\
   25 &1:36:43.2441, +15:47:23.696 &36.924$\pm$0.041 &36.251$\pm$0.071 &38.584$\pm$0.049 &1.632$\pm$0.073 &0.298$\pm$0.111\\
   26 &1:36:42.4737, +15:47:24.416 &37.770$\pm$0.023 &37.052$\pm$0.021 &38.870$\pm$0.045 &1.674$\pm$0.024 &0.236$\pm$0.042\\
   27 &1:36:42.6676, +15:46:52.256 &37.519$\pm$0.025 &36.729$\pm$0.032 &38.901$\pm$0.045 &1.960$\pm$0.034 &0.139$\pm$0.055\\
   28 &1:36:42.8284, +15:47:25.936 &37.029$\pm$0.036 &36.121$\pm$0.092 &38.265$\pm$0.060 &1.283$\pm$0.093 &-.020$\pm$0.134\\
   29 &1:36:42.2562, +15:47:05.387 &36.828$\pm$0.046 &36.588$\pm$0.039 &38.939$\pm$0.044 &1.752$\pm$0.041 &0.883$\pm$0.082\\
   30 &1:36:42.4580, +15:47:29.811 &37.380$\pm$0.027 &36.708$\pm$0.033 &38.740$\pm$0.047 &1.891$\pm$0.035 &0.298$\pm$0.058\\
   31 &1:36:45.1181, +15:45:40.597 &36.998$\pm$0.038 &36.355$\pm$0.059 &38.422$\pm$0.054 &1.617$\pm$0.060 &0.337$\pm$0.094\\
   32 &1:36:44.1704, +15:45:36.118 &37.040$\pm$0.036 &36.239$\pm$0.073 &38.282$\pm$0.059 &1.768$\pm$0.075 &0.124$\pm$0.110\\
   33 &1:36:42.7075, +15:45:37.799 &37.772$\pm$0.023 &37.324$\pm$0.017 &39.311$\pm$0.042 &2.753$\pm$0.022 &0.601$\pm$0.039\\
   34 &1:36:41.9427, +15:45:50.759 &38.244$\pm$0.021 &37.687$\pm$0.014 &39.684$\pm$0.041 &2.858$\pm$0.019 &0.454$\pm$0.034\\
   35 &1:36:44.7911, +15:45:31.877 &37.108$\pm$0.034 &36.195$\pm$0.079 &38.102$\pm$0.069 &1.778$\pm$0.081 &-.026$\pm$0.116\\
   36 &1:36:41.8264, +15:46:00.039 &37.770$\pm$0.023 &36.998$\pm$0.023 &38.650$\pm$0.048 &2.407$\pm$0.027 &0.164$\pm$0.044\\
   37 &1:36:43.1452, +15:45:13.719 &37.657$\pm$0.024 &36.990$\pm$0.023 &38.633$\pm$0.048 &2.433$\pm$0.027 &0.304$\pm$0.045\\
   38 &1:36:43.1064, +15:45:16.679 &36.913$\pm$0.041 &36.557$\pm$0.041 &38.659$\pm$0.048 &1.555$\pm$0.043 &0.725$\pm$0.079\\
   39 &1:36:44.3256, +15:45:26.198 &37.071$\pm$0.035 &36.401$\pm$0.054 &38.324$\pm$0.057 &1.871$\pm$0.056 &0.301$\pm$0.087\\
   40 &1:36:43.7384, +15:46:22.118 &37.146$\pm$0.032 &36.752$\pm$0.031 &39.081$\pm$0.043 &1.952$\pm$0.033 &0.674$\pm$0.060\\
   41 &1:36:42.4748, +15:46:29.559 &37.484$\pm$0.026 &36.831$\pm$0.028 &38.808$\pm$0.046 &2.063$\pm$0.030 &0.325$\pm$0.051\\
   42 &1:36:42.4083, +15:46:23.879 &37.103$\pm$0.034 &36.548$\pm$0.042 &38.677$\pm$0.048 &1.875$\pm$0.044 &0.457$\pm$0.073\\
   43 &1:36:41.6823, +15:46:25.399 &36.931$\pm$0.040 &36.062$\pm$0.104 &38.595$\pm$0.049 &1.310$\pm$0.104 &0.032$\pm$0.150\\
   44 &1:36:42.3972, +15:46:14.119 &37.135$\pm$0.033 &36.618$\pm$0.037 &38.744$\pm$0.047 &2.007$\pm$0.040 &0.508$\pm$0.067\\
   45 &1:36:42.6022, +15:46:08.199 &37.895$\pm$0.022 &37.320$\pm$0.017 &39.027$\pm$0.043 &2.567$\pm$0.021 &0.430$\pm$0.038\\
   46 &1:36:41.0837, +15:46:12.279 &38.084$\pm$0.021 &37.419$\pm$0.016 &39.112$\pm$0.043 &2.747$\pm$0.021 &0.308$\pm$0.036\\
   47 &1:36:40.9618, +15:46:14.279 &37.400$\pm$0.027 &36.667$\pm$0.035 &38.751$\pm$0.046 &1.983$\pm$0.037 &0.216$\pm$0.059\\
   48 &1:36:39.5210, +15:45:36.919 &36.982$\pm$0.038 &36.435$\pm$0.051 &38.853$\pm$0.045 &2.111$\pm$0.054 &0.467$\pm$0.086\\
   49 &1:36:39.3603, +15:45:44.679 &38.426$\pm$0.021 &37.918$\pm$0.013 &39.570$\pm$0.041 &2.994$\pm$0.018 &0.520$\pm$0.033\\
   50 &1:36:39.2272, +15:45:57.319 &38.407$\pm$0.021 &37.808$\pm$0.014 &39.938$\pm$0.040 &2.917$\pm$0.018 &0.398$\pm$0.033\\
   51 &1:36:39.9920, +15:45:51.639 &37.107$\pm$0.034 &36.463$\pm$0.048 &38.459$\pm$0.053 &1.840$\pm$0.050 &0.336$\pm$0.080\\
   52 &1:36:39.3381, +15:45:48.359 &37.764$\pm$0.023 &36.989$\pm$0.023 &38.607$\pm$0.049 &2.487$\pm$0.027 &0.159$\pm$0.044\\
   53 &1:36:38.9834, +15:46:07.718 &35.702$\pm$0.367 &35.751$\pm$0.199 &38.388$\pm$0.055 &1.034$\pm$0.200 &1.272$\pm$0.563\\
   54 &1:36:41.0172, +15:46:47.079 &37.749$\pm$0.023 &37.092$\pm$0.021 &39.139$\pm$0.043 &2.145$\pm$0.024 &0.319$\pm$0.042\\
   55 &1:36:41.3775, +15:46:46.679 &37.171$\pm$0.032 &36.355$\pm$0.059 &38.861$\pm$0.045 &1.377$\pm$0.060 &0.104$\pm$0.090\\
   56 &1:36:41.0228, +15:46:37.159 &37.551$\pm$0.025 &37.136$\pm$0.020 &39.263$\pm$0.042 &2.352$\pm$0.023 &0.645$\pm$0.043\\
   57 &1:36:40.8953, +15:46:38.919 &36.860$\pm$0.044 &36.553$\pm$0.042 &38.516$\pm$0.051 &2.063$\pm$0.044 &0.791$\pm$0.082\\
   58 &1:36:40.6182, +15:46:44.279 &36.259$\pm$0.116 &36.055$\pm$0.105 &38.978$\pm$0.044 &1.425$\pm$0.106 &0.930$\pm$0.212\\
   59 &1:36:40.5018, +15:46:31.079 &37.711$\pm$0.023 &36.986$\pm$0.023 &39.022$\pm$0.043 &2.285$\pm$0.026 &0.227$\pm$0.044\\
   60 &1:36:41.8430, +15:46:53.799 &37.973$\pm$0.022 &37.250$\pm$0.018 &39.097$\pm$0.043 &2.244$\pm$0.021 &0.231$\pm$0.038\\
   61 &1:36:41.5271, +15:46:56.839 &37.111$\pm$0.034 &36.842$\pm$0.027 &38.843$\pm$0.045 &1.696$\pm$0.029 &0.843$\pm$0.058\\
   62 &1:36:41.9982, +15:47:08.359 &37.319$\pm$0.028 &36.506$\pm$0.045 &38.756$\pm$0.046 &1.662$\pm$0.047 &0.108$\pm$0.072\\
   63 &1:36:40.8953, +15:47:01.319 &38.307$\pm$0.021 &37.709$\pm$0.014 &39.772$\pm$0.041 &2.907$\pm$0.019 &0.398$\pm$0.034\\
   64 &1:36:41.0061, +15:47:03.399 &37.651$\pm$0.024 &36.945$\pm$0.024 &38.951$\pm$0.044 &1.968$\pm$0.027 &0.253$\pm$0.046\\
   65 &1:36:41.1447, +15:47:06.359 &37.679$\pm$0.024 &37.192$\pm$0.019 &39.148$\pm$0.043 &2.264$\pm$0.022 &0.548$\pm$0.041\\
   66 &1:36:40.9673, +15:47:07.959 &37.726$\pm$0.023 &37.216$\pm$0.018 &39.143$\pm$0.043 &2.291$\pm$0.022 &0.517$\pm$0.040\\
   67 &1:36:38.4733, +15:46:59.558 &38.132$\pm$0.021 &37.402$\pm$0.016 &39.368$\pm$0.042 &2.710$\pm$0.021 &0.220$\pm$0.036\\
   68 &1:36:38.9001, +15:46:58.358 &37.425$\pm$0.027 &36.703$\pm$0.033 &38.594$\pm$0.049 &2.024$\pm$0.035 &0.231$\pm$0.057\\
   69 &1:36:38.8226, +15:46:38.438 &37.492$\pm$0.026 &36.672$\pm$0.035 &38.630$\pm$0.049 &2.011$\pm$0.037 &0.099$\pm$0.058\\
   70 &1:36:40.5960, +15:46:51.559 &37.165$\pm$0.032 &36.517$\pm$0.044 &38.434$\pm$0.053 &1.671$\pm$0.046 &0.331$\pm$0.074\\
   71 &1:36:41.8153, +15:46:32.519 &37.378$\pm$0.027 &36.562$\pm$0.041 &38.530$\pm$0.051 &2.028$\pm$0.043 &0.104$\pm$0.066\\
   72 &1:36:38.7670, +15:47:26.598 &38.172$\pm$0.021 &37.617$\pm$0.015 &39.607$\pm$0.041 &2.464$\pm$0.018 &0.457$\pm$0.035\\
   73 &1:36:39.7979, +15:47:22.439 &37.370$\pm$0.027 &36.736$\pm$0.031 &38.579$\pm$0.050 &2.020$\pm$0.034 &0.350$\pm$0.056\\
   74 &1:36:39.0552, +15:47:12.598 &37.325$\pm$0.028 &36.477$\pm$0.047 &38.411$\pm$0.054 &1.647$\pm$0.049 &0.062$\pm$0.074\\
   75 &1:36:37.2039, +15:47:45.797 &36.774$\pm$0.049 &35.811$\pm$0.175 &37.887$\pm$0.087 &1.468$\pm$0.176 &-.060$\pm$0.246\\
   76 &1:36:38.4011, +15:47:54.678 &37.222$\pm$0.030 &36.481$\pm$0.047 &38.387$\pm$0.055 &1.920$\pm$0.049 &0.206$\pm$0.076\\
   77 &1:36:39.2823, +15:47:57.398 &38.521$\pm$0.021 &37.959$\pm$0.013 &39.937$\pm$0.040 &3.075$\pm$0.018 &0.448$\pm$0.033\\
   78 &1:36:39.3711, +15:47:41.479 &37.838$\pm$0.023 &37.082$\pm$0.021 &39.053$\pm$0.043 &2.247$\pm$0.024 &0.185$\pm$0.041\\
   79 &1:36:36.8158, +15:48:03.076 &38.764$\pm$0.020 &38.300$\pm$0.013 &40.382$\pm$0.040 &3.330$\pm$0.017 &0.580$\pm$0.032\\
   80 &1:36:37.2924, +15:48:13.957 &37.127$\pm$0.033 &36.851$\pm$0.027 &38.874$\pm$0.045 &2.229$\pm$0.030 &0.834$\pm$0.058\\
   81 &1:36:38.5341, +15:48:03.078 &37.038$\pm$0.036 &36.261$\pm$0.070 &37.951$\pm$0.081 &1.807$\pm$0.072 &0.157$\pm$0.106\\
   82 &1:36:40.6958, +15:47:55.079 &38.122$\pm$0.021 &37.526$\pm$0.015 &39.655$\pm$0.041 &2.812$\pm$0.020 &0.401$\pm$0.035\\
   83 &1:36:40.2080, +15:47:52.919 &36.851$\pm$0.045 &36.127$\pm$0.091 &38.310$\pm$0.058 &1.438$\pm$0.092 &0.229$\pm$0.137\\
   84 &1:36:41.5937, +15:48:06.359 &37.665$\pm$0.024 &36.990$\pm$0.023 &38.942$\pm$0.044 &2.222$\pm$0.026 &0.294$\pm$0.044\\
   85 &1:36:40.4740, +15:47:46.359 &37.147$\pm$0.032 &36.300$\pm$0.065 &38.026$\pm$0.074 &1.679$\pm$0.066 &0.062$\pm$0.098\\
   86 &1:36:40.4020, +15:48:00.999 &36.685$\pm$0.056 &36.826$\pm$0.028 &39.033$\pm$0.043 &2.223$\pm$0.031 &1.396$\pm$0.084\\
   87 &1:36:36.8767, +15:48:08.596 &37.682$\pm$0.024 &37.067$\pm$0.021 &38.973$\pm$0.044 &2.397$\pm$0.025 &0.375$\pm$0.043\\
   88 &1:36:37.8356, +15:48:24.517 &38.518$\pm$0.021 &38.199$\pm$0.013 &40.504$\pm$0.040 &3.304$\pm$0.017 &0.775$\pm$0.033\\
   89 &1:36:37.4919, +15:48:21.957 &37.375$\pm$0.027 &36.810$\pm$0.028 &38.761$\pm$0.046 &2.297$\pm$0.032 &0.444$\pm$0.053\\
   90 &1:36:37.2758, +15:48:18.277 &37.352$\pm$0.028 &36.669$\pm$0.035 &38.605$\pm$0.049 &1.909$\pm$0.037 &0.283$\pm$0.060\\
   91 &1:36:39.8255, +15:48:16.278 &37.137$\pm$0.033 &36.430$\pm$0.051 &38.157$\pm$0.065 &1.981$\pm$0.054 &0.252$\pm$0.082\\
   92 &1:36:40.2745, +15:48:19.239 &38.102$\pm$0.021 &37.386$\pm$0.016 &39.312$\pm$0.042 &2.755$\pm$0.021 &0.239$\pm$0.036\\
   93 &1:36:40.1913, +15:48:11.879 &37.700$\pm$0.023 &37.244$\pm$0.018 &39.135$\pm$0.043 &2.572$\pm$0.022 &0.590$\pm$0.040\\
   94 &1:36:39.7978, +15:47:54.199 &37.116$\pm$0.033 &36.353$\pm$0.059 &38.448$\pm$0.053 &1.807$\pm$0.061 &0.175$\pm$0.091\\
   95 &1:36:42.2311, +15:48:29.479 &37.396$\pm$0.027 &36.668$\pm$0.035 &38.421$\pm$0.054 &2.328$\pm$0.039 &0.223$\pm$0.059\\
   96 &1:36:42.4750, +15:48:22.199 &37.695$\pm$0.024 &37.088$\pm$0.021 &38.751$\pm$0.046 &2.453$\pm$0.025 &0.387$\pm$0.042\\
   97 &1:36:42.3309, +15:48:16.999 &38.728$\pm$0.020 &38.318$\pm$0.013 &40.436$\pm$0.040 &3.323$\pm$0.017 &0.652$\pm$0.032\\
   98 &1:36:42.0703, +15:48:17.239 &37.833$\pm$0.023 &37.239$\pm$0.018 &39.239$\pm$0.042 &2.055$\pm$0.021 &0.404$\pm$0.039\\
   99 &1:36:42.2588, +15:48:20.759 &37.913$\pm$0.022 &37.425$\pm$0.016 &39.319$\pm$0.042 &2.811$\pm$0.021 &0.547$\pm$0.037\\
  100 &1:36:40.2634, +15:48:24.519 &37.173$\pm$0.032 &36.531$\pm$0.043 &38.208$\pm$0.063 &2.310$\pm$0.047 &0.339$\pm$0.072\\
  101 &1:36:40.6292, +15:48:22.839 &37.390$\pm$0.027 &36.741$\pm$0.031 &38.422$\pm$0.054 &2.176$\pm$0.034 &0.330$\pm$0.056\\
  102 &1:36:41.1059, +15:48:00.759 &36.306$\pm$0.106 &35.986$\pm$0.121 &37.953$\pm$0.081 &1.111$\pm$0.122 &0.774$\pm$0.218\\
  103 &1:36:39.4930, +15:47:56.838 &36.941$\pm$0.040 &35.993$\pm$0.119 &38.550$\pm$0.050 &1.408$\pm$0.120 &-.060$\pm$0.170\\
  104 &1:36:40.1525, +15:48:08.519 &37.109$\pm$0.034 &36.728$\pm$0.032 &38.742$\pm$0.047 &2.131$\pm$0.035 &0.692$\pm$0.062\\
  105 &1:36:40.0583, +15:48:04.039 &37.569$\pm$0.025 &36.900$\pm$0.025 &38.708$\pm$0.047 &2.075$\pm$0.028 &0.304$\pm$0.048\\
  106 &1:36:42.7244, +15:47:51.359 &36.852$\pm$0.045 &36.470$\pm$0.048 &38.298$\pm$0.058 &1.842$\pm$0.050 &0.691$\pm$0.088\\
  107 &1:36:43.7332, +15:48:25.718 &37.798$\pm$0.023 &37.515$\pm$0.015 &39.825$\pm$0.041 &2.700$\pm$0.020 &0.824$\pm$0.037\\
  108 &1:36:43.5254, +15:48:45.078 &37.280$\pm$0.029 &36.622$\pm$0.037 &38.284$\pm$0.059 &2.320$\pm$0.041 &0.318$\pm$0.064\\
  109 &1:36:43.0349, +15:48:36.118 &37.675$\pm$0.024 &37.039$\pm$0.022 &38.614$\pm$0.049 &2.791$\pm$0.029 &0.347$\pm$0.043\\
  110 &1:36:42.8575, +15:48:24.198 &36.673$\pm$0.057 &36.406$\pm$0.053 &38.263$\pm$0.060 &1.946$\pm$0.056 &0.846$\pm$0.106\\
  111 &1:36:41.6158, +15:48:33.959 &36.933$\pm$0.040 &36.637$\pm$0.036 &38.522$\pm$0.051 &1.996$\pm$0.039 &0.807$\pm$0.073\\
  112 &1:36:43.6002, +15:48:27.558 &37.465$\pm$0.026 &36.829$\pm$0.028 &38.775$\pm$0.046 &2.241$\pm$0.031 &0.348$\pm$0.051\\
  113 &1:36:43.1956, +15:48:18.678 &37.822$\pm$0.023 &37.400$\pm$0.016 &39.462$\pm$0.041 &2.753$\pm$0.021 &0.637$\pm$0.038\\
  114 &1:36:43.6223, +15:48:15.638 &37.468$\pm$0.026 &36.966$\pm$0.023 &38.804$\pm$0.046 &2.430$\pm$0.028 &0.529$\pm$0.047\\
  115 &1:36:44.1433, +15:47:58.118 &37.685$\pm$0.024 &37.233$\pm$0.018 &39.286$\pm$0.042 &2.384$\pm$0.022 &0.596$\pm$0.040\\
  116 &1:36:44.3705, +15:48:03.157 &37.726$\pm$0.023 &37.513$\pm$0.015 &39.513$\pm$0.041 &2.894$\pm$0.021 &0.919$\pm$0.038\\
  117 &1:36:43.0736, +15:48:02.118 &37.978$\pm$0.022 &37.569$\pm$0.015 &39.674$\pm$0.041 &2.749$\pm$0.019 &0.654$\pm$0.036\\
  118 &1:36:43.0126, +15:48:18.838 &38.486$\pm$0.021 &37.796$\pm$0.014 &39.628$\pm$0.041 &2.740$\pm$0.018 &0.274$\pm$0.033\\
  119 &1:36:43.6611, +15:48:10.918 &37.752$\pm$0.023 &37.050$\pm$0.021 &39.040$\pm$0.043 &2.505$\pm$0.026 &0.258$\pm$0.043\\
  120 &1:36:43.8606, +15:47:59.398 &37.275$\pm$0.029 &36.688$\pm$0.034 &38.821$\pm$0.045 &1.912$\pm$0.036 &0.413$\pm$0.060\\
  121 &1:36:42.8685, +15:47:54.518 &37.413$\pm$0.027 &36.841$\pm$0.027 &38.761$\pm$0.046 &2.239$\pm$0.031 &0.433$\pm$0.052\\
  122 &1:36:42.9516, +15:47:43.718 &37.288$\pm$0.029 &36.684$\pm$0.034 &38.447$\pm$0.053 &1.972$\pm$0.036 &0.391$\pm$0.060\\
  123 &1:36:41.5604, +15:47:36.919 &36.596$\pm$0.064 &36.481$\pm$0.047 &38.281$\pm$0.059 &1.860$\pm$0.049 &1.051$\pm$0.107\\
  124 &1:36:38.6730, +15:47:10.186 &37.888$\pm$0.022 &37.163$\pm$0.019 &38.934$\pm$0.044 &2.474$\pm$0.023 &0.227$\pm$0.040\\
  125 &1:36:37.8638, +15:47:09.145 &37.308$\pm$0.029 &36.656$\pm$0.035 &38.595$\pm$0.049 &2.055$\pm$0.038 &0.326$\pm$0.061\\
  126 &1:36:38.0523, +15:47:08.666 &36.879$\pm$0.043 &36.354$\pm$0.059 &38.946$\pm$0.044 &1.688$\pm$0.060 &0.497$\pm$0.098\\
  127 &1:36:38.2906, +15:47:07.786 &37.267$\pm$0.029 &36.684$\pm$0.034 &38.863$\pm$0.045 &2.033$\pm$0.036 &0.418$\pm$0.061\\
  128 &1:36:38.0856, +15:46:59.706 &37.997$\pm$0.022 &37.394$\pm$0.016 &39.314$\pm$0.042 &2.708$\pm$0.021 &0.391$\pm$0.037\\
  129 &1:36:38.3460, +15:47:03.786 &37.676$\pm$0.024 &36.940$\pm$0.024 &38.653$\pm$0.048 &1.924$\pm$0.027 &0.212$\pm$0.046\\
  130 &1:36:41.1449, +15:47:15.787 &36.952$\pm$0.039 &36.518$\pm$0.044 &38.886$\pm$0.045 &1.692$\pm$0.046 &0.620$\pm$0.080\\
  131 &1:36:41.3001, +15:47:19.947 &36.939$\pm$0.040 &36.422$\pm$0.052 &38.833$\pm$0.045 &1.531$\pm$0.053 &0.508$\pm$0.089\\
  132 &1:36:41.2558, +15:47:12.107 &37.158$\pm$0.032 &36.801$\pm$0.029 &39.071$\pm$0.043 &1.861$\pm$0.031 &0.724$\pm$0.058\\
  133 &1:36:47.2022, +15:46:09.381 &38.249$\pm$0.021 &37.884$\pm$0.013 &39.994$\pm$0.040 &3.150$\pm$0.019 &0.712$\pm$0.034\\
  134 &1:36:47.0748, +15:46:10.981 &38.005$\pm$0.022 &37.806$\pm$0.014 &39.892$\pm$0.040 &3.036$\pm$0.019 &0.938$\pm$0.035\\
  135 &1:36:47.0748, +15:46:14.261 &37.617$\pm$0.024 &37.376$\pm$0.016 &39.358$\pm$0.042 &2.801$\pm$0.022 &0.882$\pm$0.040\\
  136 &1:36:45.4233, +15:47:28.689 &37.801$\pm$0.023 &37.091$\pm$0.021 &38.915$\pm$0.044 &2.495$\pm$0.025 &0.248$\pm$0.041\\
  137 &1:36:45.2348, +15:47:34.129 &37.613$\pm$0.024 &36.955$\pm$0.024 &38.596$\pm$0.049 &2.381$\pm$0.028 &0.317$\pm$0.046\\
  138 &1:36:38.7063, +15:46:05.958 &37.815$\pm$0.023 &37.372$\pm$0.016 &39.085$\pm$0.043 &2.617$\pm$0.021 &0.608$\pm$0.038\\
  139 &1:36:36.5556, +15:47:21.796 &38.319$\pm$0.021 &37.701$\pm$0.014 &39.613$\pm$0.041 &2.759$\pm$0.018 &0.371$\pm$0.034\\
  140 &1:36:36.7828, +15:47:16.756 &37.626$\pm$0.024 &36.922$\pm$0.025 &38.832$\pm$0.045 &2.152$\pm$0.028 &0.256$\pm$0.047\\
  141 &1:36:35.9789, +15:47:48.115 &37.749$\pm$0.023 &37.368$\pm$0.017 &39.532$\pm$0.041 &2.898$\pm$0.022 &0.692$\pm$0.038\\
  142 &1:36:36.2504, +15:48:03.875 &37.842$\pm$0.023 &37.249$\pm$0.018 &38.838$\pm$0.045 &2.387$\pm$0.022 &0.405$\pm$0.039\\
  143 &1:36:36.0010, +15:48:09.155 &37.520$\pm$0.025 &36.845$\pm$0.027 &38.296$\pm$0.058 &2.219$\pm$0.030 &0.294$\pm$0.050\\
\enddata
\tablenotetext{}{(1) The identification number of the source.}
\tablenotetext{}{(2) Right Ascension  and Declination in J2000 coordinates.}
\tablenotetext{}{(3)--(5) Logarithm of the luminosity  of each source at the indicated wavelength, in units of erg~s$^{-1}$. The photometry is measured in circular apertures with 1$^{\prime\prime}$.4 radius on the plane of the sky. The H$\alpha$ and Pa$\alpha$ luminosities are corrected for the Milky Way foreground extinction, E(B$-$V)$_{MW}$=0.06~mag. See text for more details.}
\tablenotetext{}{(6) The logarithm of the equivalent width (EW) of Pa$\alpha$, in \AA, calculated from the ratio of the emission line flux  to the stellar continuum flux density. }
\tablenotetext{}{(7) The color excess, E(B$-$V), derived from the H$\alpha$/Pa$\alpha$ luminosity ratio.}
\end{deluxetable}

\bibliographystyle{aasjournal}
\bibliography{bibliography_ngc628}{}

\begin{thebibliography}{}
\expandafter\ifx\csname natexlab\endcsname\relax\def\natexlab#1{#1}\fi

\bibitem[{{Alonso-Herrero} {et~al.}(2006){Alonso-Herrero}, {Rieke}, {Rieke},
  {Colina}, {P{\'e}rez-Gonz{\'a}lez}, \& {Ryder}}]{Alonso+2006}
{Alonso-Herrero}, A., {Rieke}, G.~H., {Rieke}, M.~J., {et~al.} 2006, \apj, 650,
  835

\bibitem[{{Anand} {et~al.}(2021){Anand}, {Lee}, {Van Dyk}, {Leroy},
  {Rosolowsky}, {Schinnerer}, {Larson}, {Kourkchi}, {Kreckel}, {Scheuermann},
  {Rizzi}, {Thilker}, {Tully}, {Bigiel}, {Blanc}, {Boquien}, {Chandar}, {Dale},
  {Emsellem}, {Deger}, {Glover}, {Grasha}, {Groves}, {S. Klessen}, {Kruijssen},
  {Querejeta}, {S{\'a}nchez-Bl{\'a}zquez}, {Schruba}, {Turner}, {Ubeda},
  {Williams}, \& {Whitmore}}]{Anand+2021}
{Anand}, G.~S., {Lee}, J.~C., {Van Dyk}, S.~D., {et~al.} 2021, \mnras, 501,
  3621

\bibitem[{{Aniano} {et~al.}(2011){Aniano}, {Draine}, {Gordon}, \&
  {Sandstrom}}]{Aniano+2011}
{Aniano}, G., {Draine}, B.~T., {Gordon}, K.~D., \& {Sandstrom}, K. 2011, \pasp,
  123, 1218

\bibitem[{{Asplund} {et~al.}(2009){Asplund}, {Grevesse}, {Sauval}, \&
  {Scott}}]{Asplund+2009}
{Asplund}, M., {Grevesse}, N., {Sauval}, A.~J., \& {Scott}, P. 2009, \araa, 47,
  481

\bibitem[{{Belfiore} {et~al.}(2023){Belfiore}, {Leroy}, {Williams}, {Barnes},
  {Bigiel}, {Boquien}, {Cao}, {Chastenet}, {Congiu}, {Dale}, {Egorov},
  {Eibensteiner}, {Emsellem}, {Glover}, {Groves}, {Hassani}, {Klessen},
  {Kreckel}, {Neumann}, {Neumann}, {Querejeta}, {Rosolowsky},
  {Sanchez-Blazquez}, {Sandstrom}, {Schinnerer}, {Sun}, {Sutter}, \&
  {Watkins}}]{Belfiore+2023}
{Belfiore}, F., {Leroy}, A.~K., {Williams}, T.~G., {et~al.} 2023, \aap, 678,
  A129

\bibitem[{{Bendo} {et~al.}(2008){Bendo}, {Draine}, {Engelbracht}, {Helou},
  {Thornley}, {Bot}, {Buckalew}, {Calzetti}, {Dale}, {Hollenbach}, {Li}, \&
  {Moustakas}}]{Bendo+2008}
{Bendo}, G.~J., {Draine}, B.~T., {Engelbracht}, C.~W., {et~al.} 2008, \mnras,
  389, 629

\bibitem[{{Bendo} {et~al.}(2012){Bendo}, {Boselli}, {Dariush}, {Pohlen},
  {Roussel}, {Sauvage}, {Smith}, {Wilson}, {Baes}, {Cooray}, {Clements},
  {Cortese}, {Foyle}, {Galametz}, {Gomez}, {Lebouteiller}, {Lu}, {Madden},
  {Mentuch}, {O'Halloran}, {Page}, {Remy}, {Schulz}, \&
  {Spinoglio}}]{Bendo+2012}
{Bendo}, G.~J., {Boselli}, A., {Dariush}, A., {et~al.} 2012, \mnras, 419, 1833

\bibitem[{{Berg} {et~al.}(2020){Berg}, {Pogge}, {Skillman}, {Croxall},
  {Moustakas}, {Rogers}, \& {Sun}}]{Berg+2020}
{Berg}, D.~A., {Pogge}, R.~W., {Skillman}, E.~D., {et~al.} 2020, \apj, 893, 96

\bibitem[{{Berg} {et~al.}(2015){Berg}, {Skillman}, {Croxall}, {Pogge},
  {Moustakas}, \& {Johnson-Groh}}]{Berg+2015}
{Berg}, D.~A., {Skillman}, E.~D., {Croxall}, K.~V., {et~al.} 2015, \apj, 806,
  16

\bibitem[{{Boquien} {et~al.}(2015){Boquien}, {Calzetti}, {Aalto}, {Boselli},
  {Braine}, {Buat}, {Combes}, {Israel}, {Kramer}, {Lord}, {Rela{\~n}o},
  {Rosolowsky}, {Stacey}, {Tabatabaei}, {van der Tak}, {van der Werf},
  {Verley}, \& {Xilouris}}]{Boquien+2015}
{Boquien}, M., {Calzetti}, D., {Aalto}, S., {et~al.} 2015, \aap, 578, A8

\bibitem[{{Boquien} {et~al.}(2016){Boquien}, {Kennicutt}, {Calzetti}, {Dale},
  {Galametz}, {Sauvage}, {Croxall}, {Draine}, {Kirkpatrick}, {Kumari}, {Hunt},
  {De Looze}, {Pellegrini}, {Rela{\~n}o}, {Smith}, \&
  {Tabatabaei}}]{Boquien+2016}
{Boquien}, M., {Kennicutt}, R., {Calzetti}, D., {et~al.} 2016, \aap, 591, A6

\bibitem[{{Boselli} {et~al.}(2004){Boselli}, {Lequeux}, \&
  {Gavazzi}}]{Boselli+2004}
{Boselli}, A., {Lequeux}, J., \& {Gavazzi}, G. 2004, \aap, 428, 409

\bibitem[{{Brown} \& {Gnedin}(2021)}]{Brown+2021}
{Brown}, G., \& {Gnedin}, O.~Y. 2021, \mnras, 508, 5935

\bibitem[{{Buat} \& {Deharveng}(1988)}]{Buat+1988}
{Buat}, V., \& {Deharveng}, J.~M. 1988, \aap, 195, 60

\bibitem[{{Buat} {et~al.}(1999){Buat}, {Donas}, {Milliard}, \&
  {Xu}}]{Buat+1999}
{Buat}, V., {Donas}, J., {Milliard}, B., \& {Xu}, C. 1999, \aap, 352, 371

\bibitem[{{Buat} \& {Xu}(1996)}]{Buat+1996}
{Buat}, V., \& {Xu}, C. 1996, \aap, 306, 61

\bibitem[{{Bushouse} {et~al.}(2022){Bushouse}, {Eisenhamer}, {Dencheva},
  {Davies}, {Greenfield}, {Morrison}, {Hodge}, {Simon}, {Grumm}, {Droettboom},
  {Slavich}, {Sosey}, {Pauly}, {Miller}, {Jedrzejewski}, {Hack}, {Davis},
  {Crawford}, {Law}, {Gordon}, {Regan}, {Cara}, {MacDonald}, {Bradley},
  {Shanahan}, {Jamieson}, {Teodoro}, \& {Williams}}]{Bushouse+2022}
{Bushouse}, H., {Eisenhamer}, J., {Dencheva}, N., {et~al.} 2022,
  doi:10.5281/zenodo.7487203

\bibitem[{{Calapa} {et~al.}(2014){Calapa}, {Calzetti}, {Draine}, {Boquien},
  {Kramer}, {Xilouris}, {Verley}, {Braine}, {Rela{\~n}o}, {van der Werf},
  {Israel}, {Hermelo}, \& {Albrecht}}]{Calapa+2014}
{Calapa}, M.~D., {Calzetti}, D., {Draine}, B.~T., {et~al.} 2014, \apj, 784, 130

\bibitem[{{Calzetti}(2001)}]{Calzetti2001}
{Calzetti}, D. 2001, \pasp, 113, 1449

\bibitem[{{Calzetti}(2013)}]{Calzetti2013}
---. 2013, {Star Formation Rate Indicators}, ed. J.~{Falc{\'o}n-Barroso} \&
  J.~H. {Knapen}, 419

\bibitem[{{Calzetti} {et~al.}(1997){Calzetti}, {Meurer}, {Bohlin}, {Garnett},
  {Kinney}, {Leitherer}, \& {Storchi-Bergmann}}]{Calzetti+1997}
{Calzetti}, D., {Meurer}, G.~R., {Bohlin}, R.~C., {et~al.} 1997, \aj, 114, 1834

\bibitem[{{Calzetti} {et~al.}(2005){Calzetti}, {Kennicutt}, {Bianchi},
  {Thilker}, {Dale}, {Engelbracht}, {Leitherer}, {Meyer}, {Sosey}, {Mutchler},
  {Regan}, {Thornley}, {Armus}, {Bendo}, {Boissier}, {Boselli}, {Draine},
  {Gordon}, {Helou}, {Hollenbach}, {Kewley}, {Madore}, {Martin}, {Murphy},
  {Rieke}, {Rieke}, {Roussel}, {Sheth}, {Smith}, {Walter}, {White}, {Yi},
  {Scoville}, {Polletta}, \& {Lindler}}]{Calzetti+2005}
{Calzetti}, D., {Kennicutt}, R.~C., J., {Bianchi}, L., {et~al.} 2005, \apj,
  633, 871

\bibitem[{{Calzetti} {et~al.}(2007){Calzetti}, {Kennicutt}, {Engelbracht},
  {Leitherer}, {Draine}, {Kewley}, {Moustakas}, {Sosey}, {Dale}, {Gordon},
  {Helou}, {Hollenbach}, {Armus}, {Bendo}, {Bot}, {Buckalew}, {Jarrett}, {Li},
  {Meyer}, {Murphy}, {Prescott}, {Regan}, {Rieke}, {Roussel}, {Sheth}, {Smith},
  {Thornley}, \& {Walter}}]{Calzetti+2007}
{Calzetti}, D., {Kennicutt}, R.~C., {Engelbracht}, C.~W., {et~al.} 2007, \apj,
  666, 870

\bibitem[{{Calzetti} {et~al.}(2010){Calzetti}, {Wu}, {Hong}, {Kennicutt},
  {Lee}, {Dale}, {Engelbracht}, {van Zee}, {Draine}, {Hao}, {Gordon},
  {Moustakas}, {Murphy}, {Regan}, {Begum}, {Block}, {Dalcanton}, {Funes}, {Gil
  de Paz}, {Johnson}, {Sakai}, {Skillman}, {Walter}, {Weisz}, {Williams}, \&
  {Wu}}]{Calzetti+2010}
{Calzetti}, D., {Wu}, S.~Y., {Hong}, S., {et~al.} 2010, \apj, 714, 1256

\bibitem[{{Calzetti} {et~al.}(2015){Calzetti}, {Lee}, {Sabbi}, {Adamo},
  {Smith}, {Andrews}, {Ubeda}, {Bright}, {Thilker}, {Aloisi}, {Brown},
  {Chandar}, {Christian}, {Cignoni}, {Clayton}, {da Silva}, {de Mink}, {Dobbs},
  {Elmegreen}, {Elmegreen}, {Evans}, {Fumagalli}, {Gallagher}, {Gouliermis},
  {Grebel}, {Herrero}, {Hunter}, {Johnson}, {Kennicutt}, {Kim}, {Krumholz},
  {Lennon}, {Levay}, {Martin}, {Nair}, {Nota}, {{\"O}stlin}, {Pellerin},
  {Prieto}, {Regan}, {Ryon}, {Schaerer}, {Schiminovich}, {Tosi}, {Van Dyk},
  {Walterbos}, {Whitmore}, \& {Wofford}}]{Calzetti+2015a}
{Calzetti}, D., {Lee}, J.~C., {Sabbi}, E., {et~al.} 2015, \aj, 149, 51

\bibitem[{{Cervi{\~n}o} {et~al.}(2002){Cervi{\~n}o}, {Valls-Gabaud},
  {Luridiana}, \& {Mas-Hesse}}]{Cervino+2002}
{Cervi{\~n}o}, M., {Valls-Gabaud}, D., {Luridiana}, V., \& {Mas-Hesse}, J.~M.
  2002, \aap, 381, 51

\bibitem[{{Cox} {et~al.}(1986){Cox}, {Kruegel}, \& {Mezger}}]{Cox+1986}
{Cox}, P., {Kruegel}, E., \& {Mezger}, P.~G. 1986, \aap, 155, 380

\bibitem[{{Crocker} {et~al.}(2013){Crocker}, {Calzetti}, {Thilker}, {Aniano},
  {Draine}, {Hunt}, {Kennicutt}, {Sandstrom}, \& {Smith}}]{Crocker+2013}
{Crocker}, A.~F., {Calzetti}, D., {Thilker}, D.~A., {et~al.} 2013, \apj, 762,
  79

\bibitem[{{Dale} {et~al.}(2007){Dale}, {Gil de Paz}, {Gordon}, {Hanson},
  {Armus}, {Bendo}, {Bianchi}, {Block}, {Boissier}, {Boselli}, {Buckalew},
  {Buat}, {Burgarella}, {Calzetti}, {Cannon}, {Engelbracht}, {Helou},
  {Hollenbach}, {Jarrett}, {Kennicutt}, {Leitherer}, {Li}, {Madore}, {Martin},
  {Meyer}, {Murphy}, {Regan}, {Roussel}, {Smith}, {Sosey}, {Thilker}, \&
  {Walter}}]{Dale+2007}
{Dale}, D.~A., {Gil de Paz}, A., {Gordon}, K.~D., {et~al.} 2007, \apj, 655, 863

\bibitem[{{Dale} {et~al.}(2012){Dale}, {Aniano}, {Engelbracht}, {Hinz},
  {Krause}, {Montiel}, {Roussel}, {Appleton}, {Armus}, {Beir{\~a}o}, {Bolatto},
  {Brandl}, {Calzetti}, {Crocker}, {Croxall}, {Draine}, {Galametz}, {Gordon},
  {Groves}, {Hao}, {Helou}, {Hunt}, {Johnson}, {Kennicutt}, {Koda}, {Leroy},
  {Li}, {Meidt}, {Miller}, {Murphy}, {Rahman}, {Rix}, {Sandstrom}, {Sauvage},
  {Schinnerer}, {Skibba}, {Smith}, {Tabatabaei}, {Walter}, {Wilson}, {Wolfire},
  \& {Zibetti}}]{Dale+2012}
{Dale}, D.~A., {Aniano}, G., {Engelbracht}, C.~W., {et~al.} 2012, \apj, 745, 95

\bibitem[{{de la Fuente Marcos} \& {de la Fuente
  Marcos}(2009)}]{Delafuente+2009}
{de la Fuente Marcos}, R., \& {de la Fuente Marcos}, C. 2009, \apj, 700, 436

\bibitem[{{de Vaucouleurs} {et~al.}(1991){de Vaucouleurs}, {de Vaucouleurs},
  {Corwin}, {Buta}, {Paturel}, \& {Fouque}}]{deVaucouleurs+1991}
{de Vaucouleurs}, G., {de Vaucouleurs}, A., {Corwin}, Herold~G., J., {et~al.}
  1991, {Third Reference Catalogue of Bright Galaxies}

\bibitem[{{Della Bruna} {et~al.}(2021){Della Bruna}, {Adamo}, {Lee}, {Smith},
  {Krumholz}, {Bik}, {Calzetti}, {Fox}, {Fumagalli}, {Grasha}, {Messa},
  {{\"O}stlin}, {Walterbos}, \& {Wofford}}]{DellaBruna+2021}
{Della Bruna}, L., {Adamo}, A., {Lee}, J.~C., {et~al.} 2021, \aap, 650, A103

\bibitem[{{Desert} {et~al.}(1990){Desert}, {Boulanger}, \&
  {Puget}}]{Desert+1990}
{Desert}, F.~X., {Boulanger}, F., \& {Puget}, J.~L. 1990, \aap, 237, 215

\bibitem[{{Draine}(2011)}]{Draine+2011}
{Draine}, B.~T. 2011, \apj, 732, 100

\bibitem[{{Draine} \& {Li}(2001)}]{Draine+2001}
{Draine}, B.~T., \& {Li}, A. 2001, \apj, 551, 807

\bibitem[{{Draine} \& {Li}(2007)}]{DraineLi2007}
---. 2007, \apj, 657, 810

\bibitem[{{Draine} {et~al.}(2021){Draine}, {Li}, {Hensley}, {Hunt},
  {Sandstrom}, \& {Smith}}]{Draine+2021}
{Draine}, B.~T., {Li}, A., {Hensley}, B.~S., {et~al.} 2021, \apj, 917, 3

\bibitem[{{Draine} {et~al.}(2007){Draine}, {Dale}, {Bendo}, {Gordon}, {Smith},
  {Armus}, {Engelbracht}, {Helou}, {Kennicutt}, {Li}, {Roussel}, {Walter},
  {Calzetti}, {Moustakas}, {Murphy}, {Rieke}, {Bot}, {Hollenbach}, {Sheth}, \&
  {Teplitz}}]{Draine+2007}
{Draine}, B.~T., {Dale}, D.~A., {Bendo}, G., {et~al.} 2007, \apj, 663, 866

\bibitem[{{Efremov} \& {Elmegreen}(1998)}]{Efremov+1998}
{Efremov}, Y.~N., \& {Elmegreen}, B.~G. 1998, \mnras, 299, 588

\bibitem[{{Ekstr{\"o}m} {et~al.}(2012){Ekstr{\"o}m}, {Georgy}, {Eggenberger},
  {Meynet}, {Mowlavi}, {Wyttenbach}, {Granada}, {Decressin}, {Hirschi},
  {Frischknecht}, {Charbonnel}, \& {Maeder}}]{Ekstrom+2012}
{Ekstr{\"o}m}, S., {Georgy}, C., {Eggenberger}, P., {et~al.} 2012, \aap, 537,
  A146

\bibitem[{{Elmegreen} \& {Hunter}(2010)}]{Elmegreen+2010}
{Elmegreen}, B.~G., \& {Hunter}, D.~A. 2010, \apj, 712, 604

\bibitem[{{Emsellem} {et~al.}(2022){Emsellem}, {Schinnerer}, {Santoro},
  {Belfiore}, {Pessa}, {McElroy}, {Blanc}, {Congiu}, {Groves}, {Ho}, {Kreckel},
  {Razza}, {Sanchez-Blazquez}, {Egorov}, {Faesi}, {Klessen}, {Leroy}, {Meidt},
  {Querejeta}, {Rosolowsky}, {Scheuermann}, {Anand}, {Barnes},
  {Be{\v{s}}li{\'c}}, {Bigiel}, {Boquien}, {Cao}, {Chevance}, {Dale},
  {Eibensteiner}, {Glover}, {Grasha}, {Henshaw}, {Hughes}, {Koch}, {Kruijssen},
  {Lee}, {Liu}, {Pan}, {Pety}, {Saito}, {Sandstrom}, {Schruba}, {Sun},
  {Thilker}, {Usero}, {Watkins}, \& {Williams}}]{Emsellem+2022}
{Emsellem}, E., {Schinnerer}, E., {Santoro}, F., {et~al.} 2022, \aap, 659, A191

\bibitem[{{Fahrion} \& {De Marchi}(2023)}]{Fahrion+2023}
{Fahrion}, K., \& {De Marchi}, G. 2023, \aap, 671, L14

\bibitem[{{Ferguson} {et~al.}(1996){Ferguson}, {Wyse}, {Gallagher}, \&
  {Hunter}}]{Ferguson+1996}
{Ferguson}, A. M.~N., {Wyse}, R. F.~G., {Gallagher}, J.~S., I., \& {Hunter},
  D.~A. 1996, \aj, 111, 2265

\bibitem[{{Figueira} {et~al.}(2022){Figueira}, {Pollo}, {Ma{\l}ek}, {Buat},
  {Boquien}, {Pistis}, {Cassar{\`a}}, {Vergani}, {Hamed}, \&
  {Salim}}]{Figueira+2022}
{Figueira}, M., {Pollo}, A., {Ma{\l}ek}, K., {et~al.} 2022, \aap, 667, A29

\bibitem[{{F{\"o}rster Schreiber} {et~al.}(2004){F{\"o}rster Schreiber},
  {Roussel}, {Sauvage}, \& {Charmandaris}}]{Forster+2004}
{F{\"o}rster Schreiber}, N.~M., {Roussel}, H., {Sauvage}, M., \&
  {Charmandaris}, V. 2004, \aap, 419, 501

\bibitem[{{Fumagalli} {et~al.}(2011){Fumagalli}, {da Silva}, \&
  {Krumholz}}]{Fumagalli+2011}
{Fumagalli}, M., {da Silva}, R.~L., \& {Krumholz}, M.~R. 2011, \apjl, 741, L26

\bibitem[{{Galliano} {et~al.}(2018){Galliano}, {Galametz}, \&
  {Jones}}]{Galliano+2018}
{Galliano}, F., {Galametz}, M., \& {Jones}, A.~P. 2018, \araa, 56, 673

\bibitem[{{Garn} \& {Best}(2010)}]{Garn+2010}
{Garn}, T., \& {Best}, P.~N. 2010, \mnras, 409, 421

\bibitem[{{Girardi} {et~al.}(2000){Girardi}, {Bressan}, {Bertelli}, \&
  {Chiosi}}]{Girardi+2000}
{Girardi}, L., {Bressan}, A., {Bertelli}, G., \& {Chiosi}, C. 2000, \aaps, 141,
  371

\bibitem[{{Gonzaga} {et~al.}(2012){Gonzaga}, {Hack}, {Fruchter}, \&
  {Mack}}]{Gonzaga+2012}
{Gonzaga}, S., {Hack}, W., {Fruchter}, A., \& {Mack}, J. 2012, {The DrizzlePac
  Handbook}

\bibitem[{{Grasha} {et~al.}(2017){Grasha}, {Elmegreen}, {Calzetti}, {Adamo},
  {Aloisi}, {Bright}, {Cook}, {Dale}, {Fumagalli}, {Gallagher}, {Gouliermis},
  {Grebel}, {Kahre}, {Kim}, {Krumholz}, {Lee}, {Messa}, {Ryon}, \&
  {Ubeda}}]{Grasha+2017}
{Grasha}, K., {Elmegreen}, B.~G., {Calzetti}, D., {et~al.} 2017, \apj, 842, 25

\bibitem[{{Greenberg}(1968)}]{Greenberg1968}
{Greenberg}, J.~M. 1968, in Nebulae and Interstellar Matter, ed. B.~M.
  {Middlehurst} \& L.~H. {Aller}, 221

\bibitem[{{Greenfield} \& {Miller}(2016)}]{Greenfield+2016}
{Greenfield}, P., \& {Miller}, T. 2016, Astronomy and Computing, 16, 41

\bibitem[{{Gregg} {et~al.}(2022){Gregg}, {Calzetti}, \& {Heyer}}]{Gregg+2022}
{Gregg}, B., {Calzetti}, D., \& {Heyer}, M. 2022, \apj, 928, 120

\bibitem[{{Hao} {et~al.}(2011){Hao}, {Kennicutt}, {Johnson}, {Calzetti},
  {Dale}, \& {Moustakas}}]{Hao+2011}
{Hao}, C.-N., {Kennicutt}, R.~C., {Johnson}, B.~D., {et~al.} 2011, \apj, 741,
  124

\bibitem[{{Helou}(1986)}]{Helou1986}
{Helou}, G. 1986, \apjl, 311, L33

\bibitem[{{Helou} {et~al.}(2004){Helou}, {Roussel}, {Appleton}, {Frayer},
  {Stolovy}, {Storrie-Lombardi}, {Hurt}, {Lowrance}, {Makovoz}, {Masci},
  {Surace}, {Gordon}, {Alonso-Herrero}, {Engelbracht}, {Misselt}, {Rieke},
  {Rieke}, {Willner}, {Pahre}, {Ashby}, {Fazio}, \& {Smith}}]{Helou+2004}
{Helou}, G., {Roussel}, H., {Appleton}, P., {et~al.} 2004, \apjs, 154, 253

\bibitem[{{Hirashita} {et~al.}(2003){Hirashita}, {Buat}, \&
  {Inoue}}]{Hirashita+2003}
{Hirashita}, H., {Buat}, V., \& {Inoue}, A.~K. 2003, \aap, 410, 83

\bibitem[{{Hoopes} \& {Walterbos}(2003)}]{Hoopes+2003}
{Hoopes}, C.~G., \& {Walterbos}, R. A.~M. 2003, \apj, 586, 902

\bibitem[{{Hoopes} {et~al.}(1996){Hoopes}, {Walterbos}, \&
  {Greenwalt}}]{Hoopes+1996}
{Hoopes}, C.~G., {Walterbos}, R. A.~M., \& {Greenwalt}, B.~E. 1996, \aj, 112,
  1429

\bibitem[{{Hopkins} {et~al.}(2001){Hopkins}, {Connolly}, {Haarsma}, \&
  {Cram}}]{Hopkins+2001}
{Hopkins}, A.~M., {Connolly}, A.~J., {Haarsma}, D.~B., \& {Cram}, L.~E. 2001,
  \aj, 122, 288

\bibitem[{{Hopkins} {et~al.}(2003){Hopkins}, {Miller}, {Nichol}, {Connolly},
  {Bernardi}, {G{\'o}mez}, {Goto}, {Tremonti}, {Brinkmann}, {Ivezi{\'c}}, \&
  {Lamb}}]{Hopkins+2003}
{Hopkins}, A.~M., {Miller}, C.~J., {Nichol}, R.~C., {et~al.} 2003, \apj, 599,
  971

\bibitem[{{Jang} \& {Lee}(2014)}]{Jang+2014}
{Jang}, I.~S., \& {Lee}, M.~G. 2014, \apj, 792, 52

\bibitem[{{Joye} \& {Mandel}(2003)}]{Joye+2003}
{Joye}, W.~A., \& {Mandel}, E. 2003, in Astronomical Society of the Pacific
  Conference Series, Vol. 295, Astronomical Data Analysis Software and Systems
  XII, ed. H.~E. {Payne}, R.~I. {Jedrzejewski}, \& R.~N. {Hook}, 489

\bibitem[{{Kennicutt}(1983)}]{Kennicutt+1983}
{Kennicutt}, R.~C., J. 1983, \apj, 272, 54

\bibitem[{{Kennicutt}(1998)}]{KennicuttARAA1998}
{Kennicutt}, Robert~C., J. 1998, \araa, 36, 189

\bibitem[{{Kennicutt} {et~al.}(2008){Kennicutt}, {Lee}, {Funes}, {J.}, {Sakai},
  \& {Akiyama}}]{Kennicutt+2008}
{Kennicutt}, Robert~C., J., {Lee}, J.~C., {Funes}, J.~G., {et~al.} 2008, \apjs,
  178, 247

\bibitem[{{Kennicutt} {et~al.}(2007){Kennicutt}, {Calzetti}, {Walter}, {Helou},
  {Hollenbach}, {Armus}, {Bendo}, {Dale}, {Draine}, {Engelbracht}, {Gordon},
  {Prescott}, {Regan}, {Thornley}, {Bot}, {Brinks}, {de Blok}, {de Mello},
  {Meyer}, {Moustakas}, {Murphy}, {Sheth}, \& {Smith}}]{Kennicutt+2007}
{Kennicutt}, Robert~C., J., {Calzetti}, D., {Walter}, F., {et~al.} 2007, \apj,
  671, 333

\bibitem[{{Kennicutt} {et~al.}(2009){Kennicutt}, {Hao}, {Calzetti},
  {Moustakas}, {Dale}, {Bendo}, {Engelbracht}, {Johnson}, \&
  {Lee}}]{Kennicutt+2009}
{Kennicutt}, Robert~C., J., {Hao}, C.-N., {Calzetti}, D., {et~al.} 2009, \apj,
  703, 1672

\bibitem[{{Kennicutt} \& {Evans}(2012)}]{KennicuttEvans2012}
{Kennicutt}, R.~C., \& {Evans}, N.~J. 2012, \araa, 50, 531

\bibitem[{{Kroupa}(2001)}]{Kroupa2001}
{Kroupa}, P. 2001, \mnras, 322, 231

\bibitem[{{Krumholz} \& {Matzner}(2009)}]{Krumholz+2009}
{Krumholz}, M.~R., \& {Matzner}, C.~D. 2009, \apj, 703, 1352

\bibitem[{{Lang} {et~al.}(2020){Lang}, {Meidt}, {Rosolowsky}, {Nofech},
  {Schinnerer}, {Leroy}, {Emsellem}, {Pessa}, {Glover}, {Groves}, {Hughes},
  {Kruijssen}, {Querejeta}, {Schruba}, {Bigiel}, {Blanc}, {Chevance},
  {Colombo}, {Faesi}, {Henshaw}, {Herrera}, {Liu}, {Pety}, {Puschnig}, {Saito},
  {Sun}, \& {Usero}}]{Lang+2020}
{Lang}, P., {Meidt}, S.~E., {Rosolowsky}, E., {et~al.} 2020, \apj, 897, 122

\bibitem[{{Lee} {et~al.}(2023){Lee}, {Sandstrom}, {Leroy}, {Thilker},
  {Schinnerer}, {Rosolowsky}, {Larson}, {Egorov}, {Williams}, {Schmidt},
  {Emsellem}, {Anand}, {Barnes}, {Belfiore}, {Be{\v{s}}li{\'c}}, {Bigiel},
  {Blanc}, {Bolatto}, {Boquien}, {den Brok}, {Cao}, {Chandar}, {Chastenet},
  {Chevance}, {Chiang}, {Congiu}, {Dale}, {Deger}, {Eibensteiner}, {Faesi},
  {Glover}, {Grasha}, {Groves}, {Hassani}, {Henny}, {Henshaw}, {Hoyer},
  {Hughes}, {Jeffreson}, {Jim{\'e}nez-Donaire}, {Kim}, {Kim}, {Klessen},
  {Koch}, {Kreckel}, {Kruijssen}, {Li}, {Liu}, {Lopez}, {Maschmann}, {Chen},
  {Meidt}, {Murphy}, {Neumann}, {Neumayer}, {Pan}, {Pessa}, {Pety},
  {Querejeta}, {Pinna}, {Rodr{\'\i}guez}, {Saito}, {S{\'a}nchez-Bl{\'a}zquez},
  {Santoro}, {Sardone}, {Smith}, {Sormani}, {Scheuermann}, {Stuber}, {Sutter},
  {Sun}, {Teng}, {Tre{\ss}}, {Usero}, {Watkins}, {Whitmore}, \&
  {Razza}}]{Lee+2023}
{Lee}, J.~C., {Sandstrom}, K.~M., {Leroy}, A.~K., {et~al.} 2023, \apjl, 944,
  L17

\bibitem[{{Leger} \& {Puget}(1984)}]{Leger+1984}
{Leger}, A., \& {Puget}, J.~L. 1984, \aap, 137, L5

\bibitem[{{Leitherer} {et~al.}(1999){Leitherer}, {Schaerer}, {Goldader},
  {Delgado}, {Robert}, {Kune}, {de Mello}, {Devost}, \&
  {Heckman}}]{Leitherer+1999}
{Leitherer}, C., {Schaerer}, D., {Goldader}, J.~D., {et~al.} 1999, \apjs, 123,
  3

\bibitem[{{Leroy} {et~al.}(2012){Leroy}, {Bigiel}, {de Blok}, {Boissier},
  {Bolatto}, {Brinks}, {Madore}, {Munoz-Mateos}, {Murphy}, {Sandstrom},
  {Schruba}, \& {Walter}}]{Leroy+2012}
{Leroy}, A.~K., {Bigiel}, F., {de Blok}, W.~J.~G., {et~al.} 2012, \aj, 144, 3

\bibitem[{{Leroy} {et~al.}(2023){Leroy}, {Sandstrom}, {Rosolowsky}, {Belfiore},
  {Bolatto}, {Cao}, {Koch}, {Schinnerer}, {Barnes}, {Be{\v{s}}li{\'c}},
  {Bigiel}, {Blanc}, {Chastenet}, {Chen}, {Chevance}, {Chown}, {Congiu},
  {Dale}, {Egorov}, {Emsellem}, {Eibensteiner}, {Faesi}, {Glover}, {Grasha},
  {Groves}, {Hassani}, {Henshaw}, {Hughes}, {Jim{\'e}nez-Donaire}, {Kim},
  {Klessen}, {Kreckel}, {Kruijssen}, {Larson}, {Lee}, {Levy}, {Liu}, {Lopez},
  {Meidt}, {Murphy}, {Neumann}, {Pessa}, {Pety}, {Saito}, {Sardone}, {Sun},
  {Thilker}, {Usero}, {Watkins}, {Whitcomb}, \& {Williams}}]{Leroy+2023}
{Leroy}, A.~K., {Sandstrom}, K., {Rosolowsky}, E., {et~al.} 2023, \apjl, 944,
  L9

\bibitem[{{Li} {et~al.}(2005){Li}, {Wang}, {Zhou}, {Dong}, \&
  {Cheng}}]{Li+2005}
{Li}, C., {Wang}, T.-G., {Zhou}, H.-Y., {Dong}, X.-B., \& {Cheng}, F.-Z. 2005,
  \aj, 129, 669

\bibitem[{{Li} {et~al.}(2013){Li}, {Crocker}, {Calzetti}, {Wilson},
  {Kennicutt}, {Murphy}, {Brandl}, {Draine}, {Galametz}, {Johnson}, {Armus},
  {Gordon}, {Croxall}, {Dale}, {Engelbracht}, {Groves}, {Hao}, {Helou}, {Hinz},
  {Hunt}, {Krause}, {Roussel}, {Sauvage}, \& {Smith}}]{Li+2013}
{Li}, Y., {Crocker}, A.~F., {Calzetti}, D., {et~al.} 2013, \apj, 768, 180

\bibitem[{{Liu} {et~al.}(2011){Liu}, {Koda}, {Calzetti}, {Fukuhara}, \&
  {Momose}}]{Liu+2011}
{Liu}, G., {Koda}, J., {Calzetti}, D., {Fukuhara}, M., \& {Momose}, R. 2011,
  \apj, 735, 63

\bibitem[{{Lonsdale Persson} \& {Helou}(1987)}]{LonsdaleHelou1987}
{Lonsdale Persson}, C.~J., \& {Helou}, G. 1987, \apj, 314, 513

\bibitem[{{Magnelli} {et~al.}(2014){Magnelli}, {Lutz}, {Saintonge}, {Berta},
  {Santini}, {Symeonidis}, {Altieri}, {Andreani}, {Aussel}, {B{\'e}thermin},
  {Bock}, {Bongiovanni}, {Cepa}, {Cimatti}, {Conley}, {Daddi}, {Elbaz},
  {F{\"o}rster Schreiber}, {Genzel}, {Ivison}, {Le Floc'h}, {Magdis},
  {Maiolino}, {Nordon}, {Oliver}, {Page}, {P{\'e}rez Garc{\'\i}a}, {Poglitsch},
  {Popesso}, {Pozzi}, {Riguccini}, {Rodighiero}, {Rosario}, {Roseboom},
  {Sanchez-Portal}, {Scott}, {Sturm}, {Tacconi}, {Valtchanov}, {Wang}, \&
  {Wuyts}}]{Magnelli+2014}
{Magnelli}, B., {Lutz}, D., {Saintonge}, A., {et~al.} 2014, \aap, 561, A86

\bibitem[{{McCall} {et~al.}(1985){McCall}, {Rybski}, \&
  {Shields}}]{McCall+1985}
{McCall}, M.~L., {Rybski}, P.~M., \& {Shields}, G.~A. 1985, \apjs, 57, 1

\bibitem[{{McQuinn} {et~al.}(2017){McQuinn}, {Skillman}, {Dolphin}, {Berg}, \&
  {Kennicutt}}]{McQuinn+2017}
{McQuinn}, K. B.~W., {Skillman}, E.~D., {Dolphin}, A.~E., {Berg}, D., \&
  {Kennicutt}, R. 2017, \aj, 154, 51

\bibitem[{{Messa} {et~al.}(2021){Messa}, {Calzetti}, {Adamo}, {Grasha},
  {Johnson}, {Sabbi}, {Smith}, {Bajaj}, {Finn}, \& {Lin}}]{Messa+2021}
{Messa}, M., {Calzetti}, D., {Adamo}, A., {et~al.} 2021, \apj, 909, 121

\bibitem[{{Meurer} {et~al.}(1999){Meurer}, {Heckman}, \&
  {Calzetti}}]{Meurer+1999}
{Meurer}, G.~R., {Heckman}, T.~M., \& {Calzetti}, D. 1999, \apj, 521, 64

\bibitem[{{Oey} {et~al.}(2007){Oey}, {Meurer}, {Yelda}, {Furst},
  {Caballero-Nieves}, {Hanish}, {Levesque}, {Thilker}, {Walth},
  {Bland-Hawthorn}, {Dopita}, {Ferguson}, {Heckman}, {Doyle}, {Drinkwater},
  {Freeman}, {Kennicutt}, {Kilborn}, {Knezek}, {Koribalski}, {Meyer}, {Putman},
  {Ryan-Weber}, {Smith}, {Staveley-Smith}, {Webster}, {Werk}, \&
  {Zwaan}}]{Oey+2007}
{Oey}, M.~S., {Meurer}, G.~R., {Yelda}, S., {et~al.} 2007, \apj, 661, 801

\bibitem[{{Osterbrock} \& {Ferland}(2006)}]{Osterbrock+2006}
{Osterbrock}, D.~E., \& {Ferland}, G.~J. 2006, {Astrophysics of gaseous nebulae
  and active galactic nuclei}

\bibitem[{{Pellegrini} {et~al.}(2012){Pellegrini}, {Oey}, {Winkler}, {Points},
  {Smith}, {Jaskot}, \& {Zastrow}}]{Pellegrini+2012}
{Pellegrini}, E.~W., {Oey}, M.~S., {Winkler}, P.~F., {et~al.} 2012, \apj, 755,
  40

\bibitem[{{P{\'e}rez-Gonz{\'a}lez} {et~al.}(2006){P{\'e}rez-Gonz{\'a}lez},
  {Kennicutt}, {Gordon}, {Misselt}, {Gil de Paz}, {Engelbracht}, {Rieke},
  {Bendo}, {Bianchi}, {Boissier}, {Calzetti}, {Dale}, {Draine}, {Jarrett},
  {Hollenbach}, \& {Prescott}}]{Perez+2006}
{P{\'e}rez-Gonz{\'a}lez}, P.~G., {Kennicutt}, Robert~C., J., {Gordon}, K.~D.,
  {et~al.} 2006, \apj, 648, 987

\bibitem[{{Rela{\~n}o} {et~al.}(2007){Rela{\~n}o}, {Lisenfeld},
  {P{\'e}rez-Gonz{\'a}lez}, {V{\'\i}lchez}, \& {Battaner}}]{Relano+2007}
{Rela{\~n}o}, M., {Lisenfeld}, U., {P{\'e}rez-Gonz{\'a}lez}, P.~G.,
  {V{\'\i}lchez}, J.~M., \& {Battaner}, E. 2007, \apjl, 667, L141

\bibitem[{{Reynolds}(1984)}]{Reynolds+1984}
{Reynolds}, R.~J. 1984, \apj, 282, 191

\bibitem[{{Reynolds}(1990)}]{Reynolds+1990}
---. 1990, \apjl, 349, L17

\bibitem[{{Rieke} {et~al.}(2009){Rieke}, {Alonso-Herrero}, {Weiner},
  {P{\'e}rez-Gonz{\'a}lez}, {Blaylock}, {Donley}, \& {Marcillac}}]{Rieke+2009}
{Rieke}, G.~H., {Alonso-Herrero}, A., {Weiner}, B.~J., {et~al.} 2009, \apj,
  692, 556

\bibitem[{{Rieke} {et~al.}(2004){Rieke}, {Young}, {Engelbracht}, {Kelly},
  {Low}, {Haller}, {Beeman}, {Gordon}, {Stansberry}, {Misselt}, {Cadien},
  {Morrison}, {Rivlis}, {Latter}, {Noriega-Crespo}, {Padgett}, {Stapelfeldt},
  {Hines}, {Egami}, {Muzerolle}, {Alonso-Herrero}, {Blaylock}, {Dole}, {Hinz},
  {Le Floc'h}, {Papovich}, {P{\'e}rez-Gonz{\'a}lez}, {Smith}, {Su}, {Bennett},
  {Frayer}, {Henderson}, {Lu}, {Masci}, {Pesenson}, {Rebull}, {Rho}, {Keene},
  {Stolovy}, {Wachter}, {Wheaton}, {Werner}, \& {Richards}}]{Rieke+2004}
{Rieke}, G.~H., {Young}, E.~T., {Engelbracht}, C.~W., {et~al.} 2004, \apjs,
  154, 25

\bibitem[{{Rieke} {et~al.}(2015){Rieke}, {Ressler}, {Morrison}, {Bergeron},
  {Bouchet}, {Garc{\'\i}a-Mar{\'\i}n}, {Greene}, {Regan}, {Sukhatme}, \&
  {Walker}}]{Rieke+2015}
{Rieke}, G.~H., {Ressler}, M.~E., {Morrison}, J.~E., {et~al.} 2015, \pasp, 127,
  665

\bibitem[{{Rieke} {et~al.}(2005){Rieke}, {Kelly}, \& {Horner}}]{Rieke+2005}
{Rieke}, M.~J., {Kelly}, D., \& {Horner}, S. 2005, in Society of Photo-Optical
  Instrumentation Engineers (SPIE) Conference Series, Vol. 5904, Cryogenic
  Optical Systems and Instruments XI, ed. J.~B. {Heaney} \& L.~G. {Burriesci},
  1--8

\bibitem[{{Rosa-Gonz{\'a}lez} {et~al.}(2002){Rosa-Gonz{\'a}lez}, {Terlevich},
  \& {Terlevich}}]{RosaGonzalez+2002}
{Rosa-Gonz{\'a}lez}, D., {Terlevich}, E., \& {Terlevich}, R. 2002, \mnras, 332,
  283

\bibitem[{{Roussel} {et~al.}(2001){Roussel}, {Sauvage}, {Vigroux}, \&
  {Bosma}}]{Roussel+2001}
{Roussel}, H., {Sauvage}, M., {Vigroux}, L., \& {Bosma}, A. 2001, \aap, 372,
  427

\bibitem[{{Rowan-Robinson} \& {Crawford}(1989)}]{RowanRobinson+1989}
{Rowan-Robinson}, M., \& {Crawford}, J. 1989, \mnras, 238, 523

\bibitem[{{Ryon} {et~al.}(2015){Ryon}, {Bastian}, {Adamo}, {Konstantopoulos},
  {Gallagher}, {Larsen}, {Hollyhead}, {Silva-Villa}, \& {Smith}}]{Ryon+2015}
{Ryon}, J.~E., {Bastian}, N., {Adamo}, A., {et~al.} 2015, \mnras, 452, 525

\bibitem[{{Ryon} {et~al.}(2017){Ryon}, {Gallagher}, {Smith}, {Adamo},
  {Calzetti}, {Bright}, {Cignoni}, {Cook}, {Dale}, {Elmegreen}, {Fumagalli},
  {Gouliermis}, {Grasha}, {Grebel}, {Kim}, {Messa}, {Thilker}, \&
  {Ubeda}}]{Ryon+2017}
{Ryon}, J.~E., {Gallagher}, J.~S., {Smith}, L.~J., {et~al.} 2017, \apj, 841, 92

\bibitem[{{Sabbi} {et~al.}(2018){Sabbi}, {Calzetti}, {Ubeda}, {Adamo},
  {Cignoni}, {Thilker}, {Aloisi}, {Elmegreen}, {Elmegreen}, {Gouliermis},
  {Grebel}, {Messa}, {Smith}, {Tosi}, {Dolphin}, {Andrews}, {Ashworth},
  {Bright}, {Brown}, {Chandar}, {Christian}, {Clayton}, {Cook}, {Dale}, {de
  Mink}, {Dobbs}, {Evans}, {Fumagalli}, {Gallagher}, {Grasha}, {Herrero},
  {Hunter}, {Johnson}, {Kahre}, {Kennicutt}, {Kim}, {Krumholz}, {Lee},
  {Lennon}, {Martin}, {Nair}, {Nota}, {{\"O}stlin}, {Pellerin}, {Prieto},
  {Regan}, {Ryon}, {Sacchi}, {Schaerer}, {Schiminovich}, {Shabani}, {Van Dyk},
  {Walterbos}, {Whitmore}, \& {Wofford}}]{Sabbi+2018}
{Sabbi}, E., {Calzetti}, D., {Ubeda}, L., {et~al.} 2018, \apjs, 235, 23

\bibitem[{{Sauvage} \& {Thuan}(1992)}]{Sauvage+1992}
{Sauvage}, M., \& {Thuan}, T.~X. 1992, \apjl, 396, L69

\bibitem[{{Schlafly} \& {Finkbeiner}(2011)}]{Schlafly+2011}
{Schlafly}, E.~F., \& {Finkbeiner}, D.~P. 2011, \apj, 737, 103

\bibitem[{{Sellgren} {et~al.}(1983){Sellgren}, {Werner}, \&
  {Dinerstein}}]{Sellgren+1983}
{Sellgren}, K., {Werner}, M.~W., \& {Dinerstein}, H.~L. 1983, \apjl, 271, L13

\bibitem[{{Sirianni} {et~al.}(2005){Sirianni}, {Jee}, {Ben{\'\i}tez},
  {Blakeslee}, {Martel}, {Meurer}, {Clampin}, {De Marchi}, {Ford}, {Gilliland},
  {Hartig}, {Illingworth}, {Mack}, \& {McCann}}]{Sirianni+2005}
{Sirianni}, M., {Jee}, M.~J., {Ben{\'\i}tez}, N., {et~al.} 2005, \pasp, 117,
  1049

\bibitem[{{Smith} {et~al.}(2012){Smith}, {Dunne}, {da Cunha}, {Rowlands},
  {Maddox}, {Gomez}, {Bonfield}, {Charlot}, {Driver}, {Popescu}, {Tuffs},
  {Dunlop}, {Jarvis}, {Seymour}, {Symeonidis}, {Baes}, {Bourne}, {Clements},
  {Cooray}, {De Zotti}, {Dye}, {Eales}, {Scott}, {Verma}, {van der Werf},
  {Andrae}, {Auld}, {Buttiglione}, {Cava}, {Dariush}, {Fritz}, {Hopwood},
  {Ibar}, {Ivison}, {Kelvin}, {Madore}, {Pohlen}, {Rigby}, {Robotham},
  {Seibert}, \& {Temi}}]{SmithDunne+2012}
{Smith}, D.~J.~B., {Dunne}, L., {da Cunha}, E., {et~al.} 2012, \mnras, 427, 703

\bibitem[{{Smith} {et~al.}(2007){Smith}, {Draine}, {Dale}, {Moustakas},
  {Kennicutt}, {Helou}, {Armus}, {Roussel}, {Sheth}, {Bendo}, {Buckalew},
  {Calzetti}, {Engelbracht}, {Gordon}, {Hollenbach}, {Li}, {Malhotra},
  {Murphy}, \& {Walter}}]{Smith+2007}
{Smith}, J.~D.~T., {Draine}, B.~T., {Dale}, D.~A., {et~al.} 2007, \apj, 656,
  770

\bibitem[{{Tody}(1986)}]{Tody1986}
{Tody}, D. 1986, in Society of Photo-Optical Instrumentation Engineers (SPIE)
  Conference Series, Vol. 627, Instrumentation in astronomy VI, ed. D.~L.
  {Crawford}, 733

\bibitem[{{Tody}(1993)}]{Tody1993}
{Tody}, D. 1993, in Astronomical Society of the Pacific Conference Series,
  Vol.~52, Astronomical Data Analysis Software and Systems II, ed. R.~J.
  {Hanisch}, R.~J.~V. {Brissenden}, \& J.~{Barnes}, 173

\bibitem[{{Treyer} {et~al.}(2010){Treyer}, {Schiminovich}, {Johnson}, {O'Dowd},
  {Martin}, {Wyder}, {Charlot}, {Heckman}, {Martins}, {Seibert}, \& {van der
  Hulst}}]{Treyer+2010}
{Treyer}, M., {Schiminovich}, D., {Johnson}, B.~D., {et~al.} 2010, \apj, 719,
  1191

\bibitem[{{V{\'a}zquez} \& {Leitherer}(2005)}]{Vazquez+2005}
{V{\'a}zquez}, G.~A., \& {Leitherer}, C. 2005, \apj, 621, 695

\bibitem[{{Voges} \& {Walterbos}(2006)}]{Voges+2006}
{Voges}, E.~S., \& {Walterbos}, R.~A.~M. 2006, \apjl, 644, L29

\bibitem[{{Walterbos} \& {Greenawalt}(1996)}]{Walterbos+1996}
{Walterbos}, R. A.~M., \& {Greenawalt}, B. 1996, \apj, 460, 696

\bibitem[{{Wang} \& {Heckman}(1996)}]{Wang+1996}
{Wang}, B., \& {Heckman}, T.~M. 1996, \apj, 457, 645

\bibitem[{{Whitmore} {et~al.}(2011){Whitmore}, {Chandar}, {Kim}, {Kaleida},
  {Mutchler}, {Stankiewicz}, {Calzetti}, {Saha}, {O'Connell}, {Balick}, {Bond},
  {Carollo}, {Disney}, {Dopita}, {Frogel}, {Hall}, {Holtzman}, {Kimble},
  {McCarthy}, {Paresce}, {Silk}, {Trauger}, {Walker}, {Windhorst}, \&
  {Young}}]{Whitmore+2011}
{Whitmore}, B.~C., {Chandar}, R., {Kim}, H., {et~al.} 2011, \apj, 729, 78

\bibitem[{{Williams} {et~al.}(2024){Williams}, {Lee}, {Larson}, {Leroy},
  {Sandstrom}, {Schinnerer}, {Thilker}, {Belfiore}, {Egorov}, {Rosolowsky},
  {Sutter}, {DePasquale}, {Pagan}, {Anand}, {Barnes}, {Bigiel}, {Boquien},
  {Cao}, {Chastenet}, {Chevance}, {Chown}, {Dale}, {Eibensteiner}, {Emsellem},
  {Faesi}, {Glover}, {Grasha}, {Hannon}, {Hassani}, {Henshaw},
  {Jim{\'e}nez-Donaire}, {Kim}, {Klessen}, {Koch}, {Li}, {Liu}, {Meidt},
  {M{\'e}ndez-Delgado}, {Murphy}, {Neumann}, {Neumann}, {Neumayer}, {Oakes},
  {Pathak}, {Pety}, {Pinna}, {Querejeta}, {Ramambason}, {Romanelli}, {Sormani},
  {Stuber}, {Sun}, {Teng}, {Usero}, {Watkins}, \& {Weinbeck}}]{Williams+2024}
{Williams}, T.~G., {Lee}, J.~C., {Larson}, K.~L., {et~al.} 2024, arXiv
  e-prints, arXiv:2401.15142

\bibitem[{{Wu} {et~al.}(2005){Wu}, {Cao}, {Hao}, {Liu}, {Wang}, {Xia}, {Deng},
  \& {Young}}]{Wu+2005}
{Wu}, H., {Cao}, C., {Hao}, C.-N., {et~al.} 2005, \apjl, 632, L79

\bibitem[{{Xiao} {et~al.}(2012){Xiao}, {Wang}, {Wang}, {Zhou}, {Lu}, \&
  {Dong}}]{Xiao+2012}
{Xiao}, T., {Wang}, T., {Wang}, H., {et~al.} 2012, \mnras, 421, 486

\bibitem[{{Zhang} {et~al.}(2017){Zhang}, {Yan}, {Bundy}, {Bershady}, {Haffner},
  {Walterbos}, {Maiolino}, {Tremonti}, {Thomas}, {Drory}, {Jones}, {Belfiore},
  {S{\'a}nchez}, {Diamond-Stanic}, {Bizyaev}, {Nitschelm}, {Andrews},
  {Brinkmann}, {Brownstein}, {Cheung}, {Li}, {Law}, {Roman Lopes}, {Oravetz},
  {Pan}, {Storchi Bergmann}, \& {Simmons}}]{Zhang+2017}
{Zhang}, K., {Yan}, R., {Bundy}, K., {et~al.} 2017, \mnras, 466, 3217

\bibitem[{{Zhu} {et~al.}(2008){Zhu}, {Wu}, {Cao}, \& {Li}}]{Zhu+2008}
{Zhu}, Y.-N., {Wu}, H., {Cao}, C., \& {Li}, H.-N. 2008, \apj, 686, 155

\end{thebibliography}

\end{document}